\documentclass[12pt]{article}
\makeatletter \@addtoreset{equation}{section} \makeatother
\renewcommand{\theequation}{\thesection.\arabic{equation}}
\newcommand{\ba}{\begin{array}}
\newcommand{\ea}{\end{array}}
\newcommand{\beq}{\begin{equation}}
\newcommand{\eeq}{\end{equation}}
\newcommand{\bea}{\begin{eqnarray}}
\newcommand{\eea}{\end{eqnarray}}

\def\bce{\begin{center}}
\def\ece{\end{center}}

\def\nonu{\nonumber}

\def\pa{\partial}
\def\al{\alpha}
\def\be{\beta}

\def\de{\delta}

\def\la{\lambda}

\def\eps6{{\displaystyle \mathop{\epsilon}^{6}}{}}
\def\g6{{\displaystyle \mathop{g}^{6}}{}}

\def\nab6{{\displaystyle \mathop{\nabla}^{6}}{}}

\def\0{{\sst{(0)}}}
\def\1{{\sst{(1)}}}
\def\2{{\sst{(2)}}}
\def\3{{\sst{(3)}}}
\def\4{{\sst{(4)}}}
\def\5{{\sst{(5)}}}
\def\6{{\sst{(6)}}}
\def\7{{\sst{(7)}}}
\def\8{{\sst{(8)}}}

\def\ba{\begin{array}}
\def\ea{\end{array}}
\def\beq{\begin{equation}}
\def\eeq{\end{equation}}
\def\be{\begin{equation}}
\def\ee{\end{equation}}

\def\la{\lambda}
\def\eps{\epsilon}

\def\ba{\begin{array}}
\def\ea{\end{array}}
\def\beq{\begin{equation}}
\def\eeq{\end{equation}}
\def\be{\begin{equation}}
\def\ee{\end{equation}}

\def\la{\lambda}
\def\eps{\epsilon}

\def\eps6{{\displaystyle \mathop{\epsilon}^{6}}{}}

\def\nab6{{\displaystyle \mathop{\nabla}^{6}}{}}

\newcommand{\bean}{\begin{eqnarray*}}
\newcommand{\eean}{\end{eqnarray*}}

\begin{document}
\thispagestyle{empty} \addtocounter{page}{-1}
   \begin{flushright}
\end{flushright}

\vspace*{1.3cm}
  
\centerline{ \Large \bf   
  The Operator Product Expansions
 } 
\vspace*{0.3cm}
\centerline{ \Large \bf 
 in the ${\cal N}=4$ Orthogonal Wolf Space Coset Model} 
\vspace*{1.5cm}
\centerline{{\bf Changhyun Ahn, Man Hea Kim} 
and {\bf Jinsub Paeng}}
\vspace*{1.0cm} 
\centerline{\it 
Department of Physics, Kyungpook National University, Taegu
41566, Korea} 
\vspace*{0.8cm} 
\centerline{\tt ahn, manhea, jdp2r@knu.ac.kr
} 
\vskip2cm

\centerline{\bf Abstract}
\vspace*{0.5cm}

Some of the operator product expansions (OPEs) between the
lowest $SO(4)$ singlet higher spin-$2$ multiplet of spins
$(2, \frac{5}{2}, \frac{5}{2}, \frac{5}{2}, \frac{5}{2},
3, 3, 3, 3, 3, 3, \frac{7}{2}, \frac{7}{2}, \frac{7}{2}, \frac{7}{2},
4)$
in an extension of the large ${\cal N}=4$ (non)linear superconformal
algebra were constructed in the ${\cal N}=4$ superconformal coset
$\frac{SO(N+4)}{SO(N) \times SO(4)}$
theory with $N=4$ previously.
In this paper, by rewriting the above OPEs with $N=5$,
the remaining undetermined OPEs are completely determined.
There exist additional $SO(4)$ singlet higher spin-$2$ multiplet,
six $SO(4)$ adjoint higher spin-$3$ multiplets,
four $SO(4)$ vector higher spin-$\frac{7}{2}$ multiplets,
$SO(4)$ singlet higher spin-$4$ multiplet
and
four $SO(4)$ vector higher spin-$\frac{9}{2}$ multiplets
in the right hand side of these OPEs.
Furthermore, by introducing the arbitrary coefficients
in front of the composite fields in the right hand sides
of the above complete 136 OPEs,
the complete structures of the above OPEs
are obtained by using various Jacobi
identities for generic $N$. 
Finally, we describe them as one single ${\cal N}=4$ super
OPE between the above lowest
$SO(4)$ singlet higher spin-$2$ multiplet in the
${\cal N}=4$ superspace.

\baselineskip=18pt
\newpage
\renewcommand{\theequation}
{\arabic{section}\mbox{.}\arabic{equation}}

\tableofcontents

\section{Introduction}

The large ${\cal N}=4$ holography \cite{GG1305}
connects the unitary Wolf space coset conformal field theory
in two dimensions and the matrix extended higher spin theory
on $AdS_3$ space.
The (large) ${\cal N}=4$ supersymmetry plays an important role 
in this holography.
One of the reasons why we want to understand
the Wolf space coset construction is that this
coset construction is
a generalization of the free field construction.
We obtain the latter by taking large level (the second order pole
term of the OPE between spin-$1$ currents)  $k$ limit of the former.
In other words, the large level $k$ limit corresponds to
the vanishing of 't Hooft-like coupling constant \cite{GG1406}.
In order to
observe the behavior of finite 't Hooft-like coupling constant,
the Wolf space coset construction is necessary to describe
its nontrivial structure fully. 
Then we obtain the finite $(N,k)$ behavior
in the Wolf space coset construction and this
will provide some hints for
the higher spin theory on $AdS_3$ space at the
quantum level.
According to the results of \cite{ST},
there exist different types of Wolf space cosets.
 One of them
 is given by orthogonal Wolf space coset we are interested in
 \footnote{There exists
  the large ${\cal N}=4$ holography corresponding to a symplectic
  Wolf space coset conformal field theory \cite{Ferreira} and its
  $AdS_3$ Vasiliev higher spin theory \cite{PV9806,PV9812}.}.
See also the relevant work in \cite{GG1512}.
 
 Contrary to the unitary Wolf space coset model
 \cite{AK1509,Ahn1311,Ahn1408,Ahn1504}, 
the orthogonal Wolf space coset model contains the lowest
higher spin current of spin-$2$ rather than spin-$1$ \cite{AP}.
This will make some calculations be rather involved.
So far, the complete OPEs between the lowest
$16$ higher spin currents are not known.
In \cite{AKP}, 
the lowest higher spin-$2$ current living in the
lowest ${\cal N}=4$ higher spin-$2$ multiplet (that is, there are $16$
higher spin currents) in terms of
orthogonal Wolf space coset fields for generic $N$ was found.
Then it is straightforward to obtain
the remaining $15$ higher spin currents from
this higher spin-$2$ current by using the four supersymmetry
generators of the large ${\cal N}=4$ (non)linear
superconformal algebra for fixed low values of $N$.
For fixed $N=4$, the three kinds of higher spin-$3$
currents were obtained from the OPEs between  
the higher spin currents of the above the
lowest ${\cal N}=4$ higher spin-$2$ multiplet.
It was not clear how they appear in different ${\cal N}=4$ multiplets
at that time. 
We should look at the OPEs between the ${\cal N}=4$ stress energy
tensor and the possible ${\cal N}=4$ multiplets by allowing
the $SO({\cal N}=4)$
nonsinglet property to these ${\cal N}=4$ multiplets
\footnote{
  The ${\cal N}=3$ supersymmetric example, where the
  $SO({\cal N}=3)$ nonsinglet structure plays an important role,
  can be found in
  \cite{AK1607} in the context of the ${\cal N}=3$ Kazama-Suzuki
model. See also the nonsinglet structure in \cite{GPZ}.}.

In this paper, we reconsider 
the complete OPEs between the lowest
$16$ higher spin currents in the ${\cal N}=4$ orthogonal Wolf space
coset model.
After determining the complete OPEs for the particular $N=4$
where all the higher spin currents can be written in terms of
orthogonal Wolf space coset fields,
we discuss the $N=5$ case by adding more fields.
The main idea to this purpose is that by using the fundamental
orthogonal Wolf space coset fields, we compute the various OPEs.
When the new higher spin primary fields arise in the right hand side
of the OPEs, then we should reorganize them under the $SO({\cal N}=4)$
symmetry and observe how they transform under the $SO(4)$
symmetry
\footnote{
  We will use the notation $SO(4)$ for
  $SO({\cal N}=4)$ for simplicity.}. For the $SO(4)$ singlet
${\cal N}=4$ multiplet in the unitary case,
we can construct all the new higher spin currents
living this multiplet once the lowest new higher spin current
is determined with the help of the four supersymmetry generators. 

However, if we have  a single
 $SO(4)$ nonsinglet
${\cal N}=4$ multiplet which will appear in the
right hand side of the our OPEs,
then there are several elements
on this multiplet.
Each element transforms nontrivially
under the $16$ currents of the large ${\cal N}=4$ linear
superconformal algebra.
The reason why we need to have this transformation is
that we should calculate the OPEs between the $16$ currents
and the higher spin currents living in the
$SO(4)$ nonsinglet
${\cal N}=4$ multiplet
in order to
use the Jacobi identities.
More explicitly, the several lowest higher spin currents
can be determined by the six spin-$1$ currents of the above
$16$ currents after computing the OPEs between them and reading
off the first order poles.
Once these lowest higher spin currents
are obtained completely, then we can repeat 
the procedure for the singlet case because we can act
the four supersymmetry generators on each lowest higher spin current.

The most difficult part of the present work is to write down
all the possible
orthogonal
Wolf space composite fields appearing
in the right hand sides of the OPEs
in terms of the known (higher spin) currents. 
As the spin at the specific pole increases,
the number of composite fields becomes large.
When the new higher spin current appears, 
the situation is more involved.
Even for $N=5$, when the spin at the particular pole is large, 
then it is not obvious to observe how to express
that pole in terms of known (higher spin) currents and
a new primary higher spin current.
Due to the many independent terms, it is not possible to
solve the linear equations for the undetermined coefficients
coming from the vanishing of the sum of
the particular pole (written in terms of
coset fields), the possible known composite
(higher spin) current terms with arbitrary coefficients
and a new higher spin current.
We can calculate the OPE between the spin-$\frac{1}{2}$
currents and the above particular pole term
given in terms of coset fields. Then the first order pole
of this OPE has spin which is less than the original spin by
$\frac{1}{2}$. This will make some computations easier.
At the same time,
we can calculate the OPE between  the spin-$\frac{1}{2}$
currents and the above sum of known composite terms and a
new higher spin current.
By comparing these two expressions, we can reduce  
the number of unknown coefficients and moreover,
by using other conditions from other
spin-$1, \frac{3}{2}$ and $2$ currents from the above ${\cal N}=4$
primary, we can eventually determine all the coefficients if there
is no new higher spin current. If not, the new higher spin current
can be written in terms of the known composite terms as well as
the extra terms which can be written in terms of coset fields.

Instead of considering all the possible $136(=\sum_{i=1}^{16} i)$
OPEs (which arise
from the OPEs between $16$ higher spin currents),
we focus on the $16$ OPEs among them (the OPEs between
$16$ higher spin currents and the lowest higher spin-$2$ current
or their reversed OPEs)
because the remaining $120$
OPEs can be extracted from the ${\cal N}=4$ supersymmetry.
That is, once the above $16$ OPEs are obtained
(this implies that we can write down the corresponding
${\cal N}=4$ OPE explicitly by putting these
five kinds of OPEs into the expansion of fermionic
coordinates), then the ${\cal N}=4$
superspace description allows us to write down the above $120$ OPEs
automatically
by multiplying
various super derivatives and putting the fermionic coordinates
to zero in this OPE successively.
Then after inserting the arbitrary coefficients which will depend on
$(N,k)$ explicitly (and possible other structure constants)
in front of all the composite fields arising
in the right hand side of $136$ OPEs,
the Jacobi identities can be used.
Eventually, the complete OPEs can be determined and we will present
them in a single OPE in ${\cal N}=4$ superspace.

In section $2$,
the ${\cal N}=4$ orthogonal Wolf space coset model is reviewed
\footnote{There are related works in
  \cite{Ahn1311,Ahn1408,AK1411,Ahn1504}
  on this Wolf space coset
  model. There are also previous works on the orthogonal coset
  models in
  \cite{AP1902,Ahn1701,Ahn1106,GV1106,Ahn1202,CHR1209,CGKV,AP1301,AP1310}
  along the line of \cite{GG1011,GG1205,GG1207}.}.

In section $3$,
based on the findings in \cite{AKP} which is valid for $N=4$,
the new observations will be added.

In section $4$,
based on the results of section $3$, we consider $N=5$ case.
We will find various new higher spin currents (some of them
are not present for $N=4$ case). The $136$ OPEs will be obtained
eventually.

In Appendices, some of the detailed expressions described
in previous sections are given.

The Thielemans package \cite{Thielemans} is used.

An ancillary (mathematica) file $\tt{ancillary.nb}$,
where the complete OPEs with the explicit structure
constants appearing in Appendices $B$ and $C$ are given,
is included.

\section{Review of ${\cal N}=4$ orthogonal Wolf space coset model}

We consider the Wolf space coset in the `supersymmetric' version
with groups $G=SO(N+4)$ and $H=SO(N) \times SO(4)$
as follows \footnote{After we divide $SU(2) \times U(1)$ in the coset
  of \cite{ST},
we obtain the Wolf space coset.}:
\bea
\mbox{Wolf}= \frac{G}{H} = 
\frac{SO(N+4)}{SO(N) \times SO(4)}.
\label{cosetWolf}
\eea
The group indices 
are denoted by
\bea 
G \quad \mbox{indices} &:& a, b, c, \cdots=1,2, \cdots, 
\frac{1}{4} (N+4)(N+3),
1^{\ast}, 2^{\ast}, \cdots, \left( \frac{1}{4} (N+4)(N+3)  \right)^{\ast},
\nonu \\
\frac{G}{H} \quad \mbox{indices} &:& \bar{a},\bar{b},\bar{c},\cdots=
1,2, \cdots, 2N, 1^{\ast}, 2^{\ast}, \cdots, 2N^{\ast}.
\label{abnotation}
\eea
In the bosonic version, there exist $4N$ free fermions
living in the extra $SO(4N)$ group in the numerator of the
coset at level $1$.

The ${\cal N}=1$ affine Kac-Moody  
algebra can be determined by the adjoint spin-$1$
current and the spin-$\frac{1}{2}$ current of group
$G=SO(N+4)$.
By adding the quadratic term in the fermions to the
above spin-$1$ current, the operator product expansions
between the `modified' spin-$1$ current $V^a(z)$
and the spin-$\frac{1}{2}$
current $Q^a(z)$ are described as
\bea
V^a(z) \, V^b(w) & = & \frac{1}{(z-w)^2} \, k \, g^{ab}
-\frac{1}{(z-w)} \, f^{ab}_{\,\,\,\,\,\,c} \, V^c(w) 
+\cdots,
\nonu \\
Q^a(z) \, Q^b(w) & = & -\frac{1}{(z-w)} \, (k+N+2) \, g^{ab} + \cdots.
\label{opevq}
\eea
The level $k$ is a positive integer.
The metric can be obtained from
$g_{ab} = \frac{1}{2\, c_g} \, f_{ac}^{\,\,\,d} \, f_{bd}^{\,\,\,c}$
where $c_g$ is the dual Coxeter number of the
Lie algebra $G=SO(N+4)$. That is, $c_g=(N+2)$. 
The metric $g_{ab}$ is given by the generators of
$SO(N+4)$ in the complex basis, $g_{ab} = \frac{1}{2} \,
\mbox{Tr}(T_a \, T_b)$.
The commutation relation of generators
is given by $[T_a, T_b]=f_{ab}^{\,\,\,c} \,
T_c$.

For given $(N+4) \times (N+4)$ matrix,
the above $4N$ coset indices (\ref{abnotation})
can be associated with
the following locations with asterisk
\bea
\left(\begin{array}{rrrrr|rrrr}
&&&&& {\ast} & {\ast} & {\ast} & {\ast} \\
&&&&& {\ast} & {\ast} & {\ast} & {\ast} \\
&& &&& \vdots & \vdots & \vdots & \vdots \\
&&&&& {\ast} & {\ast} & {\ast} & {\ast} \\
&&&&& {\ast} & {\ast} & {\ast} & {\ast} \\ \hline 
{\ast} & {\ast} & \cdots & {\ast} & {\ast} &&&& \\
{\ast} & {\ast} & \cdots & {\ast} & {\ast} &&&& \\ 
{\ast} & {\ast} & \cdots & {\ast} & {\ast} &&&& \\ 
{\ast} & {\ast} & \cdots & {\ast} & {\ast} &&&& \\
\end{array}\right)_{(N+4) \times (N+4)}.
\label{Matrix}
\eea
That is, the generators with $2N$ coset indices
have two nonzero elements located at the above
$N \times 4$ and $4 \times N$ off diagonal matrices
in (\ref{Matrix}).
The remaining $2N$ coset generators can be obtained
from the above coset generators by transposing.
Note that the size of two block diagonals
at $N=4$ is equal to each other. 

\subsection{The $11$ currents of
  ${\cal N}=4$ nonlinear superconformal algebra}

The four supersymmetry currents of spin-$\frac{3}{2}$,
the six spin-$1$ currents of $\hat{SU}(2)_k \times \hat{SU}(2)_N$,
and the spin-$2$ stress energy tensor \cite{GS,Van,GPTV,GK}
can be described as follows:
\bea
\hat{G}^{0}(z) &  = &   \frac{i}{(k+N+2)}  \, Q_{\bar{a}} \, V^{\bar{a}}(z),
\qquad
\hat{G}^{i}(z)  =  \frac{i}{(k+N+2)} 
\, h^{i}_{\bar{a} \bar{b}} \, Q^{\bar{a}} \, V^{\bar{b}}(z), \qquad
i =1,2,3,
\nonu \\
\hat{A}_{i}(z) &  = & 
(-1)^{i+1} \frac{1}{4N} \, f^{\bar{a} \bar{b}}_{\,\,\,\,\,\, c} \, h^i_{\bar{a} \bar{b}} \, V^c(z), 
\qquad
\hat{B}_{i}(z)  =  
-\frac{1}{4(k+N+2)} \, h^i_{\bar{a} \bar{b}} \, Q^{\bar{a}} \, Q^{\bar{b}}(z),
\nonu \\
\hat{T}(z)  & = & 
\frac{1}{2(k+N+2)^2} \left[ (k+N+2) \, V_{\bar{a}} \, V^{\bar{a}} 
+k \, Q_{\bar{a}} \, \pa \, Q^{\bar{a}} 
+f_{\bar{a} \bar{b} c} \, Q^{\bar{a}} \, Q^{\bar{b}} \, V^c  \right] (z)
\nonu \\
&- & 
\frac{1}{(k+N+2)} \sum_{i=1}^3
( (-1)^{i} \hat{A}_{i} + \hat{B}_{i}  )^2 (z).
\label{closedform}
\eea
Here the three almost complex structures
are given by
\bea
h^1_{\bar{a} \bar{b}} = 
\left(
\begin{array}{cccc}
0 & 1  & 0 & 0 \\
-1 & 0 & 0 & 0 \\
0 & 0 & 0 & 1 \\
0 & 0 & -1 & 0 \\
\end{array}
\right), 
h^2_{\bar{a} \bar{b}} = 
\left(
\begin{array}{cccc}
0 &i  & 0 & 0 \\
-i & 0 & 0 & 0 \\
0 & 0 & 0 & -i \\
0 & 0 & i & 0 \\
\end{array}
\right), 
h^{3}_{\bar{a} \bar{b}}
=
\left(
\begin{array}{cccc}
0 & 0  & i & 0 \\
0 & 0 & 0 & i \\
-i & 0 & 0 & 0 \\
0 & -i & 0 & 0 \\
\end{array}
\right)
\label{himatrix},
\eea
where each entry is $N \times N$ matrix.
Note that we introduce $h^{0}_{\bar{a} \bar{b}} \equiv
g_{\bar{a} \bar{b}}$.
The only coset indices associated with
(\ref{Matrix}) appear in the fermionic fields.

We introduce the above $11$ currents in different basis
in order to describe them in the ${\cal N}=4$ superspace as follows:
\bea
\hat{T}(z) & \rightarrow &  \hat{L}(z), \label{1703}
\\
  \hat{G}^0(z) & \rightarrow & \hat{G}_{ss}^2(z), \qquad
 \hat{ G}^1(z) \rightarrow \hat{G}_{ss}^3(z), \qquad
   \hat{G}^2(z) \rightarrow -\hat{G}_{ss}^4(z), \qquad
   \hat{G}^3(z) \rightarrow \hat{G}_{ss}^1(z), \nonu \\
   \hat{A}^{\pm 1}(z) & \rightarrow & \frac{i}{2}
   ( \hat{T}^{14} \mp T^{23})(z), \,\,
  \hat{A}^{\pm 2}(z)  \rightarrow  \frac{i}{2} ( \hat{T}^{13} \pm \hat{T}^{24})(z), \,\,
\hat{A}^{\pm 3}(z)  \rightarrow  \pm
\frac{i}{2} ( \hat{T}^{12} \mp \hat{T}^{34})(z).
  \nonu
\eea
We use $\hat{A}_1(z) =-\hat{A}^{+1}(z)$, $
\hat{A}_2(z) =\hat{A}^{+2}(z)$, $\hat{A}_3(z) =-\hat{A}^{+3}(z)$
and $\hat{B}_i(z) = \hat{A}^{-i}(z)$.
Then the $11$ currents in this new basis
satisfy Appendix $(A.1)$ of \cite{AKK} by using (\ref{opevq}).

\subsection{The $16$ currents of
  ${\cal N}=4$ linear superconformal algebra}

In the linear version \cite{STV,STVS,Schoutensnpb,ST,Saulina}
of ${\cal N}=4$ superconformal algebra,
there are also four spin-$\frac{1}{2}$ currents and
one spin-$1$ current
\bea
{ F}_{11}(z)&=& \frac{i }{\sqrt{2}} Q^{(2N+3)}(z),
\qquad
{ F}_{22}(z)=-\frac{i }{\sqrt{2}} Q^{(2N+3)^\ast}(z),
\nonu \\
{ F}_{12}(z)&=& \frac{(1-i) }{2} Q^{(2N+2)^\ast}(z),
\qquad
{ F}_{21}(z)= \frac{(1+i) }{2} Q^{(2N+2)}(z),
\nonu \\
{ U}(z)&=& 
\frac{(1+i)}{2 \sqrt{2}} V^{(2N+2)} (z)+ \frac{(-1+i)}{2 \sqrt{2}} V^{(2N+2)^*}(z)
+\frac{i}{(N+k+2)} Q^{(2N+1)} Q^{(2N+1)^\ast}(z)
\nonu \\
&-& \frac{i}{2(N+k+2)} (  \sum_{a=1}^N Q^{a} Q^{a^\ast}
- \sum_{a=N+1}^{2N} Q^{a} Q^{a^\ast} )(z).
\label{1andhalfexp}
\eea

We use the following transformation in order to describe in the
${\cal N}=4$ superspace
\bea
\frac{i}{\sqrt{2}}(F_{12}+F_{21})(z) & \rightarrow& \Gamma^{1}(z),
\qquad
 -\frac{1}{\sqrt{2}}(F_{12}+F_{21})(z) \rightarrow \Gamma^{2}(z),
\nonu\\
-\frac{i}{\sqrt{2}}(F_{11}+F_{22})(z) &\rightarrow& \Gamma^{3}(z),
\qquad
 -\frac{1}{\sqrt{2}}(F_{11}-F_{22})(z) \rightarrow \Gamma^{4}(z),
\qquad
U(z) \rightarrow U(z).
\label{gammaru}
\eea

Furthermore,  the six spin-$1$ currents of $\hat{SU}(2)_{k+1}
\times \hat{SU}(2)_{N+1}$, four supersymmetry
currents of spin-$\frac{3}{2}$ and the spin-$2$ stress energy tensor
can be described as follows \cite{GS}:
\bea
T^{ij}(z) & = & \hat{T}^{ij}(z)
+\frac{2\,i}{(2+k+N)}\,\Gamma^{i}\Gamma^{j}(z),
\nonu\\
G^{i}(z) & =& \hat{G}_{ss}^{i}(z)+\frac{2\,i}{(2+k+N)}\,
U\Gamma^{i}(z)+\varepsilon_{ijkl}\,
\Bigg[\,
\frac{4\,i}{3(2+k+N)^{2}}\,\Gamma^{j}\Gamma^{k}\Gamma^{l}
-\frac{1}{(2+k+N)}\,T^{jk}\Gamma^{l}
\,\Bigg](z),
\nonu\\
L(z) &=&  \hat{L}(z)-
\frac{1}{(2+k+N)}\Big(\,UU-\partial \Gamma^{i} \Gamma^{i}\,\Big)(z).
\label{eleven}
\eea
Here we should use (\ref{closedform}), (\ref{himatrix}), (\ref{1703})
and (\ref{1andhalfexp}).
The ${\cal N}=4$ linear superconformal algebra
can be realized from (\ref{gammaru}) and (\ref{eleven})
in the orthogonal Wolf space coset model (\ref{cosetWolf}).
The $16$ currents are given by
$L(z)$, $G^i(z)$, $T^{ij}(z)$, $\Gamma^i(z)$ and $U(z)$ which can be
written in terms of $Q^{\bar{a}}(z)$ and $V^a(z)$.

\subsection{The $16$ lowest higher spin currents}

The lowest higher spin-$2$ current
in the ${\cal N}=4$ orthogonal Wolf space is found in \cite{AKP}
and it is given by
\bea
\Phi_0^{(2)}(z)&=&c_1  \, V_{\bar{a}} V^{\bar{a}}(z)+ c_2 \,  \sum_{a':so(N)} 
V_{a'} V^{a'}(z)+ 
c_3 \,  \sum_{a'':so(4)} V_{a''} V^{a''}(z)+
c_4 \, \sum_{i=1}^3 \hat{A}_i \hat{A}_i(z) \nonu \\
& + & c_5 \,
\sum_{i=1}^3 \hat{B}_i \hat{B}_i(z)
+  c_6 \, Q_{\bar{a}} \pa Q^{\bar{a}}(z)
+ c_7 \, \sum_{\mu =0}^3 h^{\mu}_{\bar{a} \bar{b}} h^{\mu}_{\bar{c} \bar{d}}
f^{\bar{a} \bar{c}}_{\,\,\,\,\,\, e} Q^{\bar{b}} Q^{\bar{d}} V^e(z),
\label{T2ansatz}
\eea
where the coefficients are \footnote{The coefficients
$c_2$ and $c_3$ in (\ref{t2coeff}) are the same for $N=4$.}
\bea
c_1  & = & -\frac{(2 k^2 N+k^2+4 k N^2+6 k N+2 k+11
   N^2-2 N-24)}{2 (k-1) N (k+N+2)^2}, 
   \nonu \\
   c_2 & = & \frac{6 (2 k N+3 k+3 N+4)}{(k-1) N
   (k+N+2)^2},
\qquad
c_3  =  \frac{3 (k+N-2) (2 k N+3 k+3 N+4)}{2
   (k-1) (k+2) (k+N+2)^2}, 
   \nonu \\
c_4 & = & \frac{2 (N+2) (2 k+N)}{(k+2) (k+N+2)^2},
\qquad
c_5  =  \frac{2 k (2 k+N)}{N (k+N+2)^2},
\nonu \\
c_6 & = & \frac{k (N+2) (2 k+N)}{N (k+N+2)^3},
\qquad
c_7  =  \frac{(N+2) (2 k+N)}{4 N (k+N+2)^3}.
\label{t2coeff}
\eea

The OPE between this higher spin-$2$ current and itself
is described as
\bea
\Phi_{0}^{(2)}(z)\,\Phi_{0}^{(2)}(w)
& = & \frac{1}{(z-w)^{4}}\,\,
c^{0}
+\frac{1}{(z-w)^{2}}\,
Q^{(2)}_{0}(w)
+\frac{1}{(z-w)}\,
\frac{1}{2}\,\partial Q^{(2)}_{0}(w) 
+\cdots,
\label{spin2spin2}
\eea
where the central term is
\bea
&& c_{0}^{0,4} =
\label{central22}
\\
&& \frac{3k(2k+N)(2kN+3k+3N+4)(2k^2N+k^2+4kN^2+6kN+2k+11N^2-2N-24)}{(k-1)(k+2)N(k+N+2)^3}.
\nonu
\eea
The quasi primary field
$Q^{(2)}_{0}(w)$ depends on
the higher spin-$2$ current itself
$\Phi^{(2)}_{0}(w)$ and the spin-$\frac{1}{2}, 1, 2$
currents of the
${\cal N}=4$ linear superconformal algebra.
The explicit form is given Appendix $B$.
The $16$ higher spin currents can be 
combined into one single ${\cal N}=4$ super field
as follows:
\bea
 {\bf \Phi}^{(s=2)}(Z) \equiv \Bigg(
\Phi_{0}^{(2)}(z),
\Phi_{\frac{1}{2}}^{(2),i}(z),
\Phi_{1}^{(2),ij}(z),
\Phi_{\frac{3}{2}}^{(2),i}(z),
\Phi_{2}^{(2)}(z) \Bigg).
\label{Phiexp}
\eea
In this paper, we construct $136$ OPEs between the higher spin currents
in (\ref{Phiexp}) explicitly \footnote{
  In addition to (\ref{Phiexp}),
  we will consider the following ${\cal N}=4$ multiplets 
  \bea
 {\bf X}^{(2)} & \equiv & \Bigg(
X_{0}^{(2)},
X_{\frac{1}{2}}^{(2),i},
X_{1}^{(2),ij},
X_{\frac{3}{2}}^{(2),i},
X_{2}^{(2)} \Bigg), \,\,
 {\bf \Phi}^{(3),\alpha} \equiv \Bigg(
\Phi_{0}^{(3),\alpha},
\Phi_{\frac{1}{2}}^{(3),i,\alpha},
\Phi_{1}^{(2),ij,\alpha},
\Phi_{\frac{3}{2}}^{(3),i,\alpha},
\Phi_{2}^{(3),\alpha} \Bigg),
\nonu \\
 {\bf \Phi}^{(\frac{7}{2}),\mu} & \equiv & \Bigg(
\Phi_{0}^{(\frac{7}{2}),\mu},
\Phi_{\frac{1}{2}}^{(\frac{7}{2}),i,\mu},
\Phi_{1}^{(2),ij,\mu},
\Phi_{\frac{3}{2}}^{(\frac{7}{2}),i,\mu},
\Phi_{2}^{(\frac{7}{2}),\mu} \Bigg),
\,\,
 {\bf \Phi}^{(\frac{9}{2}),\mu}  \equiv  \Bigg(
\Phi_{0}^{(\frac{9}{2}),\mu},
\Phi_{\frac{1}{2}}^{(\frac{9}{2}),i,\mu},
\Phi_{1}^{(2),ij,\mu},
\Phi_{\frac{3}{2}}^{(\frac{9}{2}),i,\mu},
\Phi_{2}^{(\frac{9}{2}),\mu} \Bigg),
\nonu \\
 {\bf \Phi}^{(4)} & \equiv & \Bigg(
\Phi_{0}^{(4)},
\Phi_{\frac{1}{2}}^{(4),i},
\Phi_{1}^{(4),ij},
\Phi_{\frac{3}{2}}^{(4),i},
\Phi_{2}^{(4)} \Bigg).
\nonu
\eea
Note that there are $SO(4)$ nonsinglet representations
denoted by $\alpha$ and $\mu$. Sometimes the index $\mu$ is replaced
by the index $i$ because it is a $SO(4)$ vector index.}.

\section{The OPE for $N=4$}

In this section, we continue to
calculate the OPEs between the lowest higher spin-$2$ multiplet
for $N=4$. Some of the OPEs were found in \cite{AKP}.
We would like to obtain the general structure of these
OPEs which will give us some hints for the general $N$ in next section.


\subsection{The known facts}

For fixed $N=4$, it was straightforward to
calculate the various higher spin currents
in ${\cal N}=4$ orthogonal Wolf space starting from the
above higher spin-$2$ current in (\ref{T2ansatz}).
One of the main results in \cite{AKP} was to
obtain the new three higher spin-$3$ currents which live in
different higher spin multiplet. It was not clear
how they appear in the right hand side of the whole $136$ OPEs.
They will turn out to be
the lowest components of three
$SO(4)$ vector ${\cal N}=4$ higher spin-$3$ multiplets
in next subsection. 

\subsection{The complete OPEs in components and ${\cal N}=4$
superspace}

In order to observe the symmetry behind the presence of
the new three higher spin-$3$ currents, we should go into
the ${\cal N}=4$ superspace.
It is known that the above $16$ currents can be combined into
the following single ${\cal N}=4$ super field \cite{Schoutensnpb}
\bea
{\bf J}(Z) & = & 
- \Delta(z) + i \theta^{j} \Gamma^{j}(z)-
\frac{i}{2}\theta^{4-jk} T^{jk}(z)-\theta^{4-j} (G^{j}-2
\alpha i \partial \Gamma^{j})(z)+\theta^{4-0}  
(2L-2\alpha\partial^2 \Delta)(z)
\nonu \\
&\equiv &
- \Delta(z) + i \theta^{j} \Gamma^{j}(z)-
\frac{i}{2}\theta^{4-jk} T^{jk}(z)-\theta^{4-j} \, \tilde{G}^{j}(z)+
\theta^{4-0} \,  
2 \, \tilde{L}(z).
\label{J4}
\eea
Here we use  the notation $\theta^{4-0}$ for the product
of fermionic coordinates $\theta^{4-0} \equiv \theta^1 \,
\theta^2 \, \theta^3 \, \theta^4$ and we have $ U(z) \equiv -\pa \,
\Delta(z)$. 
The parameter $\alpha$ appears in the above
and is given by
\bea
 { \alpha} & \equiv &
    \frac{1}{2}\frac{(k^{+}-k^{-})}{(k^{+}+k^{-})}, \qquad k^+
    \equiv k+1, \qquad k^- \equiv N+1.
    \label{alpha}
    \eea
    Then the explicit realization for these $16$
    currents described in previous section 
    can be inserted into the above single ${\cal N}=4$ $SO(4)$
    singlet super field.

    It is known that the ${\cal N}=4$ higher spin multiplet,
    which transforms nontrivially under the $SO(4)$ (the index
    $\alpha$ stands for this representation which is nothing to do with
    (\ref{alpha})),
    of (conformal) (super)spin $s$
    has the following OPE with the above
    ${\cal N}=4$ stress energy tensor
    as follows \cite{Schoutensnpb}:
    \bea
{\bf J }(Z_{1}) \, {\bf \Phi}^{(s),\alpha}(Z_{2}) & = & 
\frac{\theta_{12}^{4-0}}{z_{12}^{2}} \, 2s\, {\bf \Phi}^{(s),\alpha}(Z_{2})+
\frac{\theta_{12}^{4-i}}{z_{12}} \, D^{i}  {\bf \Phi}^{(s),\alpha}(Z_{2})+
\frac{\theta_{12}^{4-0}}{z_{12}} \, 2 \, \partial {\bf \Phi}^{(s),\alpha}(Z_{2})
\nonu \\
&- &
\frac{i}{2} \:\frac{\theta_{12}^{4-ij}}{z_{12}}({ T}^{ij})^{\alpha \beta}\: {\bf \Phi}^{(s),\beta}(Z_{2})
+\cdots.
\label{JPhi}
\eea
Note that for the $SO(4)$ singlet higher spin multiplet
the last term in (\ref{JPhi}) will disappear.
We will see two kinds of $T^{ij}$ which span the representation
of the $SO(4)$ Lie algebra in this OPE of this paper.

By using the $16$ component fields for fixed $\alpha$ 
as in 
\bea
 {\bf \Phi}^{(s),\alpha}(Z) \equiv \Bigg(
\Phi_{0}^{(s),\alpha}(z),
\Phi_{\frac{1}{2}}^{(s),i,\alpha}(z),
\Phi_{1}^{(s),ij,\alpha}(z),
\Phi_{\frac{3}{2}}^{(s),i,\alpha}(z),
\Phi_{2}^{(s),\alpha}(z) \Bigg),
\label{Phialpha}
\eea
the various complicated component results of (\ref{JPhi})
are presented in Appendix $A$.

We can show that the three higher spin-$3$ currents
($P^{(3)}(z)$, $\tilde{Q}^{(3)}_{-}(z)$ and $\tilde{R}^{(3)}_{+}(z)$)
found in \cite{AKP} with proper change of basis
can be written in terms of
the three lowest components of the
three $SO(3)$ vector  ${\bf \Phi}^{(s=3),\alpha}(Z)$
where $\alpha=1,2,3$ and
this higher spin-$3$ multiplet transforms as in (\ref{JPhi})
with
\bea
{ T}^{12} & = & \left(\begin{array}{cccc}
0 & i & 0 \\
-i & 0 & 0 \\
0 & 0 & 0
\end{array}\right)= -T^{34},
\qquad
    { T}^{13}=\left(\begin{array}{cccc}
0 & 0 & i \\
0 & 0 & 0 \\
-i & 0 & 0
 \end{array}\right) = -T^{24},
 \nonu \\
{ T}^{23} & = & \left(\begin{array}{cccc}
0 & 0 & 0 \\
0 & 0 & -i \\
0 & i & 0
\end{array}\right) = -T^{14}.
\label{threeT1}
\eea
We can check that
the above three $SO(3)$ (inside of $SO(4)$)
generators in (\ref{threeT1})
satisfy $[T^i,T^j]=  \varepsilon^{ijk} \, T^k$ where $T^i
\equiv -\frac{i}{2} \, \varepsilon^{ijk}\, T^{jk}$
\footnote{For example,
  the lowest component  $\Phi_0^{(s=3),\alpha}$ has the nontrivial OPEs
  with the spin-$1$ currents $A^{+i}$ associated with
  $\hat{SU}(2)_{k+1}$ and has trivial OPEs with
  the spin-$1$ currents $A^{-i}$ associated with $\hat{SU}(2)_{N+1}$.
  That is, $\Phi_0^{(s=3),\alpha}$ transforms in the representation
  $({\bf 3}, {\bf 1} )$ under the $\hat{SU}(2)_{k+1} \times
 \hat{SU}(2)_{N+1} $. }. 
Each higher spin-$3$ multiplet
${\bf \Phi}^{(s=3),\alpha}(Z)$
has $16$ components of higher spin
currents according to (\ref{Phialpha}).
Totally, we have $48$ higher spin currents.

Furthermore, there exist the $SO(4)$ singlet
higher spin-$4$ multiplet
${\bf \Phi}^{(s=4)}(Z)$ and the four $SO(4)$ vector
higher spin-$\frac{9}{2}$ multiplets
${\bf \Phi}^{(s=\frac{9}{2}),i}(Z)$ with $i=1,2,3,4$.
They have their component fields as in (\ref{Phialpha}).

It turns out that 
there will be a problem to generalize the OPEs for general $N$
without introducing the new primary fields which will be discussed
in next section. 

\section{The OPE for $N \geq 5$}


We would like to construct the OPEs
between the $16$ higher spin currents
for generic $N$ in component approach and in ${\cal N}=4$
superspace.

\subsection{What happens for $N=5$?}

It is natural to take the OPEs found in previous section and
introduce the arbitrary coefficients in front of
the composite fields appearing in the right hand side of the OPEs.
It is straightforward to apply the Jacobi identity in order to
determine these coefficients completely.
It turns out that there are no
consistent solutions unless we introduce the new primary fields.
Therefore, we focus on the case of $N=5$ in order to
understand the algebraic structures more clearly.

\subsection{The  new higher spin current
of spin $2$}

Let us consider the OPE between the
last component and the first component of (\ref{Phiexp}).
How we can obtain the last component in terms of orthogonal
Wolf space
coset fields 
from the first component?
According to the OPEs in Appendix $A$ with a singlet $\alpha$,
we obtain
the second component using the OPE between $G^i(z)$
and the first component $\Phi_0^{(2)}(w)$.
Then we can calculate the OPE between $G^i(z)$ and
the second component $\Phi_{\frac{1}{2}}^{(2),j}(w)$
with $i \neq j$. Then the third component can be
determined.
The fourth component can be obtained
by the OPE between 
$G^i(z)$ and
the third component $\Phi_{1}^{(2),jk}(w)$ with $i=k$.
Finally, the last component can be determined
by 
 the OPE between 
$G^i(z)$ and
 the fourth component $\Phi_{\frac{3}{2}}^{(2),j}(w)$
 with $i=j$.

The reason for describing this particular OPE rather than
others
is that the structure of the right hand side of this OPE
will be simple because these two higher spin currents
are $SO(4)$ singlets rather than nonsinglets, although
we have found the presence of this new higher spin-$2$ current
in other OPEs.

Let us emphasize that the last component of (\ref{Phiexp})
is not a quasi primary field. See also Appendix $A$.
As in unitary case \cite{AK1509}, we subtract the additional terms
from the last component of (\ref{Phiexp}) in order to
make it to be primary field \footnote{ That is, we have
$\Phi_{2}^{(s=2)}(z) \equiv  \widetilde\Phi_{2}^{(2)}(z)-
  p_{1}\:\partial^{2}\Phi_{0}^{(2)}(z)-p_{2}\:L\,\Phi_{0}^{(2)}(z)$
  where $ \widetilde\Phi_{2}^{(2)}(z)$ is a primary field under the
  stress energy tensor.
  The coefficients $p_1$ and $p_2$ were given in \cite{AK1509} or in
  (\ref{p1p2p3p4}). }.
It turns out that we have the following OPE with implicit notation
\bea
\Phi_{2}^{(2)}(z)\,\Phi_{0}^{(2)}(w)
& = & \frac{1}{(z-w)^{6}}\,
8 \, \alpha \, c_{0}^{0,4}
+\frac{1}{(z-w)^{5}}\,
Q^{(1)}_{2}(w)
+\frac{1}{(z-w)^{4}}\,\Bigg[\,
\frac{3}{2}\,\partial Q^{(1)}_{2}
+Q^{(2)}_{2}
\,\Bigg](w)
\nonu\\
& + &
\frac{1}{(z-w)^{3}}\,\Bigg[\,
\partial^{2} Q^{(1)}_{2}
+\partial Q^{(2)}_{2}
+Q^{(3)}_{2}
\,\Bigg](w)
\nonu\\
& + &
\frac{1}{(z-w)^{2}}\,\Bigg[\,
\frac{5}{12}\,\partial^{3} Q^{(1)}_{2}
+\frac{1}{2}\, \partial^{2} Q^{(2)}_{2}
+\frac{5}{6}\,\partial Q^{(3)}_{2}
+Q^{(4)}_{2}
\,\Bigg](w)
\nonu\\
& + &
\frac{1}{(z-w)}\,\Bigg[\,
\frac{1}{8}\,\partial^{4} Q^{(1)}_{2}
+\frac{1}{6}\,\partial^{3} Q^{(2)}_{2}
+\frac{5}{14}\,\partial^{2} Q^{(3)}_{2}
+\frac{3}{4}\,\partial Q^{(4)}_{2}
+ Q^{(5)}_{2}
\,\Bigg](w)
\label{OPE4-1}
\\
& + &
p_{1}\,\sum\limits _{n=3}^{4}
\frac{1}{(z-w)^{n}} \, \Big\{
\partial^{2} \Phi_{0}^{(2)} \,\Phi_{0}^{(2)}
\Big\}_{n}(w)+
p_{2}\,\sum\limits _{m=2}^{4}
\frac{1}{(z-w)^{m}}\, \Big\{
(L \Phi_{0}^{(2)})\,\Phi_{0}^{(2)}
\Big\}_{m}(w)
\nonu \\
& + & \cdots.
\nonu
\eea
The central term is proportional to
the previous central term in (\ref{central22}) together with
(\ref{alpha}) \footnote{
The last line of (\ref{OPE4-1}) with specific notations for
the singular terms \cite{BS}
comes from the subtracted terms as described above.}. 
Note that the above central term comes from
these $p_1$ and $p_2$ terms.

We will use the following quasi primary fields
with their spins, $SO(4)$ indices $i,j$ and the subscript
indicating the number of fermionic coordinates
\bea
&& Q_0^{(2)}(z) \, ; \,
Q_{\frac{1}{2}}^{(\frac{3}{2}),i}(z), \, Q_{\frac{1}{2}}^{(\frac{5}{2}),i}(z),
\, Q_{\frac{1}{2}}^{(\frac{7}{2}),i}(z) \, ; \,
Q_{1}^{(1),ij}(z), \, Q_{1}^{(2),ij}(z), \, Q_{1}^{(3),ij}(z),
 \, Q_{1}^{(4),ij}(z) \, ;
\label{quasicomponent}
 \\
 &&
 Q_{\frac{3}{2}}^{(\frac{1}{2}),ij}(z), \, Q_{\frac{3}{2}}^{(\frac{3}{2}),ij}(z),
 \, Q_{\frac{3}{2}}^{(\frac{5}{2}),ij}(z), \,  Q_{\frac{3}{2}}^{(\frac{7}{2}),ij}(z),
 \, Q_{\frac{3}{2}}^{(\frac{9}{2}),ij}(z) \, ; \,
 Q_2^{(1)}(z),  \, Q_2^{(2)}(z), \, Q_2^{(3)}(z),
 \, Q_2^{(4)}(z), \, Q_2^{(5)}(z).
 \nonu
\eea
Note that the spin is given by the number inside the bracket and 
we do not add the subscript for the spin,
contrary to the notation of
(\ref{Phialpha}) \footnote{
The corresponding ${\cal N}=4$ super fields
can be denoted by the boldface later.}.

First of all, the fifth order pole has
the spin-$1$ current $U(w)$ of the ${\cal N}=4$
linear superconformal algebra from Appendix $B$.
The next fourth order pole contains 
the descendant field $\pa U(w)$ with the known
coefficient and other terms.

We observe that there exists
a new primary higher spin field of spin-$2$ denoted by
$X_0^{(2)}(w)$
which cannot be written in terms of the known
composite fields of the currents and higher spin currents
as in Appendix $B$.
That is,
\bea
Q_2^{(2)}(w) & = & w_{1,2} \, \Phi_0^{(2)}(w)+
w_{2,2} \, X_0^{(2)}(w) + \cdots,
\label{Q22}
\eea
where other remaining terms
are given in Appendix $B$.
See also Appendix $E$ for explicit form for the
$X_0^{(2)}(w)$ for $N=5$.
By considering the condition that
the fourth order pole of the OPE
between $\Phi_0^{(2)}(z)$ and $X_0^{(2)}(w)$
should vanish, we can determine   the structure constant
appearing the $\Phi_0^{(2)}(w)$ of (\ref{Q22}).
For $N=5$, we obtain this particular structure constant as 
\bea
w_{1,2} \equiv
C_{(4)(2)}^{(2)}\Bigg|_{N=5} & = & -\frac{36}{5 (k-1) (k+2) (k+7)^2
  (28 k+61) (11 k^2+132 k+241)}
\nonu \\
& \times & (7227 k^7+201718 k^6+2017067 k^5+8606534 k^4+
13128257 k^3 \nonu \\
& - & 11460814 k^2-54096247 k-42478238).
\label{Cstructure}
\eea
At the moment, it is rather difficult to determine
the $N$ generalization of (\ref{Cstructure}) because
although we can expect the $N$ dependence for the denominator
of (\ref{Cstructure}) by increasing the $N$ values,
the numerical values appearing in front of $k$-th power in the
numerator are functions of $N$. Even if we can try to
calculate (\ref{Cstructure})  for seven (which is the maximum
power of $k$) $N$ values where $N=5,8,9,12,13,16$ and $17$, 
it will take too much time to extract the higher spin-$4$
$\Phi_2^{(2)}(z)$.
In this paper, the above structure constant for generic $N$
is not determined.

The next third order pole can be expressed in terms
of the descendant fields and other known composite fields
where there are two higher spin dependent terms
$U \,\Phi_0^{(2)}(w)$ and $\Gamma^i \,
\Phi_{\frac{1}{2}}^{(2), i}(w)$. The four component fields
of spin-$\frac{5}{2}$
in (\ref{Phiexp}) arise in this pole.

The second and first order poles appearing in the third
and fourth lines of (\ref{OPE4-1}) will be described later
subsection. We will observe that there will be a  primary
higher spin-$4$ current.

We can easily see that the singular terms
appearing in the last two terms of (\ref{OPE4-1})
can be rewritten as follows \footnote{ \label{somerelation}
For the first term, there is a relation
$
\Big\{ \partial^{2} \Phi_{0}^{(2)} \,\Phi_{0}^{(2)}
\Big\}_{n+2}(w)
=
n(n+1)\, \Big\{ \Phi_{0}^{(2)} \,\Phi_{0}^{(2)}
\Big\}_{n}(w)$  with $ n=1,2$.
That is,
we have
$6 \, Q_0^{(2)}(w)$ for the fourth order pole ($n=2$)
and $\pa  \, Q_0^{(2)}(w)$ for the third order pole ($n=1$) from
(\ref{spin2spin2}).}.

Moreover, for the second term, by introducing
\bea
\Big\{
(L \Phi_{0}^{(2)})\,\Phi_{0}^{(2)}
\Big\}_{n+2} \equiv E^{(4-n)}_{2}, \qquad n=0,1,2,
\label{Edef}
\eea
the following relations for three in (\ref{Edef})
can be obtained from the OPE
$(L \Phi_{0}^{(2)})(z) \,\Phi_{0}^{(2)}(w)$, where
the previous relation (\ref{spin2spin2}) is used,
in terms of $Q_0^{(2)}(w)$, $L(w)$ and $\Phi_0^{(2)}(w)$,
\bea
E^{(2)}_{2}(w)& = &
(\,
4 Q^{(2)}_{0}+c^{0,4}_{0}L
\,)(w),
\nonu\\
E^{(3)}_{2}(w)& = &
(\,
\frac{5}{2} \, \pa \, Q^{(2)}_{0}+c^{0,4}_{0} \, \pa \, L
\,)(w),
\nonu\\
E^{(4)}_{2}(w)& = &
(\,
\frac{1}{2}\,\partial^{2} Q^{(2)}_{0}
+ L  Q^{(2)}_{0}
+\frac{1}{2}\,\partial^{2} L
+2\,\Phi^{(2)}_{0}\Phi^{(2)}_{0}
\,)(w),
\label{Es}
\eea
where 
$c^{0,4}_{0}$ is the central term of the OPE between the
lowest higher spin-$2$ current in (\ref{central22}).

Therefore, we can present the above OPE (\ref{OPE4-1}),
together with (\ref{Es}), 
in complete form as follows:
\bea
\Phi_{2}^{(2)}(z)\,\Phi_{0}^{(2)}(w)
& = & \frac{1}{(z-w)^{6}}\,
8 \, \alpha \, c_{0}^{0,4}
+\frac{1}{(z-w)^{5}}\,
Q^{(1)}_{2}(w)
\nonu\\
& + & \frac{1}{(z-w)^{4}}\,\Bigg[\,
\frac{3}{2}\,\partial Q^{(1)}_{2}
+Q^{(2)}_{2}-6\,p_{1}\,Q^{(2)}_{0}
-p_2 \, E^{(2)}_{2}
\,\Bigg](w)
\nonu\\
& + &
\frac{1}{(z-w)^{3}}\,\Bigg[\,
\partial^{2} Q^{(1)}_{2}
+\partial Q^{(2)}_{2}
+Q^{(3)}_{2} -p_1 \, \pa \, Q_0^{(2)}
- p_2 \, E^{(3)}_{2}
\,\Bigg](w)
\nonu\\
& + &
\frac{1}{(z-w)^{2}}\,\Bigg[\,
\frac{5}{12}\,\partial^{3} Q^{(1)}_{2}
+\frac{1}{2}\, \partial^{2} Q^{(2)}_{2}
+\frac{5}{6}\,\partial Q^{(3)}_{2}
+Q^{(4)}_{2}
-p_2 \, E^{(4)}_{2}
\,\Bigg](w)
\nonu\\
& + &
\frac{1}{(z-w)}\,\Bigg[\,
\frac{1}{8}\,\partial^{4} Q^{(1)}_{2}
+\frac{1}{6}\,\partial^{3} Q^{(2)}_{2}
+\frac{5}{14}\,\partial^{2} Q^{(3)}_{2}
+\frac{3}{4}\,\partial Q^{(4)}_{2}
+ Q^{(5)}_{2}
\,\Bigg](w)
\nonu\\
&+& \cdots,
\label{OPE4-1-1}
\eea
where the relations (\ref{alpha}), (\ref{central22}),
(\ref{Es}) and (\ref{p1p2p3p4})
are used.
All the singlet quasi primary fields
appearing in this OPE are obtained and we present the
partial expressions given in Appendix $B$. 

\subsection{The  new higher spin currents
of spin $3$}

Let us describe the OPE between the higher spin-$3$ (primary) current
transforming as the $SO(4)$ adjoint representation
and the higher spin-$2$ current.
We observe the following OPE
\bea
\Phi_{1}^{(2),ij}(z)\,\Phi_{0}^{(2)}(w)
& = & \frac{1}{(z-w)^{4}}\,
Q^{(1),ij}_{1}(w)
+\frac{1}{(z-w)^{3}}\,\Bigg[\,
\partial  Q^{(1),ij}_{1}+ Q^{(2),ij}_{1}
\,\Bigg](w)
\nonu \\ 
& + &
\frac{1}{(z-w)^{2}}\,\Bigg[\,
\frac{1}{2}\,\partial^{2}  Q^{(1),ij}_{1}
+\frac{3}{4}\,\partial  Q^{(2),ij}_{1}
+ Q^{(3),ij}_{1}
\,\Bigg](w)
\nonu \\ 
& + &
\frac{1}{(z-w)}\,\Bigg[\,
\frac{1}{6}\,\partial^{3}  Q^{(1),ij}_{1} 
+\frac{3}{10}\,\partial^{2} Q^{(2),ij}_{1}
+\frac{2}{3}\,\partial  Q^{(3),ij}_{1}
+ Q^{(4),ij}_{1}\,
\Bigg](w)
\nonu \\ 
& - &
\sum\limits _{n=1}^{4}\frac{1}{(z-w)^{n}} \, (i\leftrightarrow j)
+\cdots.
\label{OPE3-1}
\eea
Note that the higher spin-$3$ currents
$\Phi_{1}^{(2),ij}(z)$
are antisymmetric under the interchange of the index $i$
and the index $j$. The last line of (\ref{OPE3-1})
implies that we should take the three lines with
$ i \leftrightarrow j$ with minus sign.
The quasi primary fields appearing in the fourth and third
order poles, which are written in terms of the
known composite fields are given in Appendix $B$.

After subtracting the descendant fields, the second order pole
contains the six $SO(4)$ adjoint  higher spin-$3$
currents $\Phi_0^{(3),\alpha}(w)$ with adjoint $\alpha$
(that is, there will be $96$ higher spin currents
in these six higher spin multiplets ${\bf \Phi}^{(3),\alpha}(Z_2)$)
\footnote{
One way to observe the presence of the higher spin-$2$ current
$X_0^{(2)}(w)$ described in previous subsection is as follows.
We can calculate the OPE between $G^i(z)$ and the quasi
primary fields $Q_1^{(3),jk}(w)$ of spin $3$
and focus on the second order pole which has spin $\frac{5}{2}$.
Then we can compute the OPE between
$G^i(z)$ and this second order pole and look at the
particular second order pole which has spin-$2$. 
We can check whether this spin-$2$ field can be written in terms of
the known composite fields, as usual. It turns out that
we observe that
there should be $X_0^{(2)}(w)$-dependent terms.}.

By introducing the following six generators $M^{\alpha}$ which are
$4 \times 4$ matrices 
with $\alpha =1, 2, \cdots, 6$ of $SO(4)$
\bea
    { M^{1}} & \equiv &
    L_{1}=
    \left(\begin{array}{cccc}
0 & 0 & 0 & 0\\
0 & 0 & -i & 0\\
0 & i & 0 & 0\\
0 & 0 & 0 & 0
\end{array}\right),
         { M^{2}} \equiv
         L_{2}=
         \left(\begin{array}{cccc}
0 & 0 & i & 0\\
0 & 0 & 0 & 0\\
-i & 0 & 0 & 0\\
0 & 0 & 0 & 0
\end{array}\right),
              { M^{3}} \equiv
              L_{3}=
              \left(\begin{array}{cccc}
0 & -i & 0 & 0\\
i & 0 & 0 & 0\\
0 & 0 & 0 & 0\\
0 & 0 & 0 & 0
\end{array}\right),
              \nonu \\
                    { M^{4}} & \equiv &
                    K_{1}=
                    \left(\begin{array}{cccc}
0 & 0 & 0 & -i\\
0 & 0 & 0 & 0\\
0 & 0 & 0 & 0\\
i & 0 & 0 & 0
\end{array}\right),
                         { M^{5}}
                         \equiv K_{2}
                         =\left(\begin{array}{cccc}
0 & 0 & 0 & 0\\
0 & 0 & 0 & -i\\
0 & 0 & 0 & 0\\
0 & i & 0 & 0
\end{array}\right),
                         { M^{6}}
                         \equiv K_{3}
                         =\left(\begin{array}{cccc}
0 & 0 & 0 & 0\\
0 & 0 & 0 & 0\\
0 & 0 & 0 & -i\\
0 & 0 & i & 0
\end{array}\right),
                         \nonu \\
                         \label{Malpha}
\eea
we can write down 
\bea
Q_1^{(3),ij}(w) = w_{0,3} \, (M^{\alpha})^{ij} \, \Phi_0^{(3),\alpha}(w) +
\cdots,
\label{Q3alpha}
\eea
as the one in Appendix $B$ \footnote{
  \label{othercomp}
  For given
  the higher spin current $\Phi_0^{(3),\alpha}(w)$ for fixed $\alpha$,
  the other component
 $\Phi_0^{(3),\beta}(w)$ with $\beta \neq \alpha$
  can be obtained, for example,
  from the OPE $T^{ij}(z) \, \Phi_0^{(3),\alpha}(w) = \frac{1}{(z-w)}
  \, (T^{ij})^{\alpha \beta} \,\Phi_0^{(3),\beta}(w)+ \cdots$
  in Appendix $A$.
  From the explicit form of $6\times 6$ nondiagonal
  matrix (\ref{6x6}),
the remaining higher spin currents can be determined completely.
}.
There are commutation relations
$[L_i, L_j] = i \, \varepsilon_{ijk} \, L_k$,
$[K_i, K_j] = i \, \varepsilon_{ijk} \, L_k$,
and
$[L_i, K_j] = i \, \varepsilon_{ijk} \, K_k$ \cite{Beveren}.
The relative coefficients of (\ref{Q3alpha}) can be determined by
using the defining nontrivial OPEs of $L(z) \, \Phi_0^{(3),\alpha}(w)$,
$G^i(z) \, \Phi_0^{(3),\alpha}(w)$ and $T^{ij}(z) \,
\Phi_0^{(3),\alpha}(w)$ in Appendix $A$.

By recalling that from the relation (\ref{JPhi}) or
Appendix $A$, the OPE $T^{ij}(z) \, \Phi_0^{(3),\alpha}(w)$
between the spin-$1$ currents of
the ${\cal N}=4$ linear superconformal algebra and
the lowest higher spin-$3$ currents contains
the nontrivial singular terms $(T^{ij})^{\alpha \beta} \,
\Phi_0^{(3),\beta}(w)$
where the generators $T^{ij}$ in the $SO(4)$ adjoint representation
are given by
\bea
{ T}^{12} & = & \left(\begin{array}{cccccc}
0 & -i & 0 & 0 & 0 & 0 \\
i & 0 & 0 & 0 & 0 & 0 \\
0 & 0 & 0 & 0 & 0 & 0 \\
0 & 0 & 0 & 0 & -i & 0 \\
0 & 0 & 0 & i & 0 & 0 \\
0 & 0 & 0 & 0 & 0 & 0 
\end{array}\right), \qquad
{ T}^{13}=\left(\begin{array}{cccccc}
0 & 0 & -i & 0 & 0 & 0 \\
0 & 0 & 0 & 0 & 0 & 0 \\
i & 0 & 0 & 0 & 0 & 0 \\
0 & 0 & 0 & 0 & 0 & -i \\
0 & 0 & 0 & 0 & 0 & 0 \\
0 & 0 & 0 & i & 0 & 0 
\end{array}\right),
\nonu \\
{ T}^{14} & = & \left(\begin{array}{cccccc}
0 & 0 & 0 & 0 & 0 & 0 \\
0 & 0 & 0 & 0 & 0 & -i \\
0 & 0 & 0 & 0 & i & 0 \\
0 & 0 & 0 & 0 & 0 & 0 \\
0 & 0 & -i & 0 & 0 & 0 \\
0 & i & 0 & 0 & 0 & 0 
\end{array}\right), \qquad
{ T}^{23}=\left(\begin{array}{cccccc}
0 & 0 & 0 & 0 & 0 & 0 \\
0 & 0 & -i & 0 & 0 & 0 \\
0 & i & 0 & 0 & 0 & 0 \\
0 & 0 & 0 & 0 & 0 & 0 \\
0 & 0 & 0 & 0 & 0 & -i \\
0 & 0 & 0 & 0 & i & 0 
\end{array}\right), 
\nonu \\
{ T}^{24} & = & \left(\begin{array}{cccccc}
0 & 0 & 0 & 0 & 0 & i \\
0 & 0 & 0 & 0 & 0 & 0 \\
0 & 0 & 0 & -i & 0 & 0 \\
0 & 0 & i & 0 & 0 & 0 \\
0 & 0 & 0 & 0 & 0 & 0 \\
-i & 0 & 0 & 0 & 0 & 0 
\end{array}\right), \qquad
{ T}^{34}=\left(\begin{array}{cccccc}
0 & 0 & 0 & 0 & -i & 0 \\
0 & 0 & 0 & i & 0 & 0 \\
0 & 0 & 0 & 0 & 0 & 0 \\
0 & -i & 0 & 0 & 0 & 0 \\
i & 0 & 0 & 0 & 0 & 0 \\
0 & 0 & 0 & 0 & 0 & 0 
\end{array}\right),
\label{6x6}
\eea
where they satisfy
$[T^{ij}, T^{kl}]= i (\de^{ik} \, T^{jl}-\de^{il} \, T^{jk}-
\de^{jk} \, T^{il} + \de^{jl} \, T^{ik})$.

Sometimes we need to have the (quasi)primary fields
in order to express the OPE with known coefficients
appearing in the descendant fields. Then
we should introduce each primary field at
the third, fourth and fifth components of (\ref{Phialpha}).
The point here is to consider the possible terms with correct spins,
$SO(4)$ vector index and  $SO(4)$ adjoint index.
It turns out that we obtain
the following decomposition where the primary field has a tilde
\bea
{\bf \Phi}^{(s),\alpha}(Z)
&=&
\Phi_{0}^{(s),\alpha}(z)+\theta^{i}\:
\Phi_{\frac{1}{2}}^{(s),i,\alpha}(z)
+\frac{1}{2} \, \theta^{4-ij} \:\Phi_{1}^{(s),ij,\alpha}(z)
+
\theta^{4-i}\,
\Phi_{\frac{3}{2}}^{(s),i,\alpha}(z)
+
\theta^{4-0}\,
\Phi_{2}^{(s),\alpha}(z)
\nonu \\
& \equiv&
\Phi_{0}^{(s),\alpha}(z)+\theta^{i}\:
\Phi_{\frac{1}{2}}^{(s),i,\alpha}(z)
+\frac{1}{2}\theta^{4-ij}\Bigg[\widetilde\Phi_{1}^{(s),ij,\alpha}
  +\frac{i}{ s}\,
  ({ T}_R^{ij})^{\alpha \beta}\,
\partial \Phi_{0}^{(s),\beta}
\Bigg](z)
\nonu\\
&+&
\theta^{4-i}\,\Bigg[
\,\widetilde\Phi_{\frac{3}{2}}^{(s),i,\alpha}+
\frac{2\,\alpha}{(2s+1)}\,\partial\Phi_{\frac{1}{2}}^{(s),i,\alpha}
-\frac{2i}{(2\, s+1)}
 \, ({ T}_R^{ij})^{\alpha \beta}\,
\partial \Phi_{\frac{1}{2}}^{(s),j,\beta}\,
\Bigg](z)
\nonumber\\
&+&
\theta^{4-0}\,\Bigg[\,
\widetilde\Phi_{2}^{(s),\alpha}-
p_{1}\:\partial^{2}\Phi_{0}^{(s),\alpha}-p_{2}\:L\,\Phi_{0}^{(s),\alpha}
+\frac{i}{2(2\, s+1)}\,({ T}_L^{ij})^{\alpha \beta}
\,
\partial \Phi_{1}^{(s),ij,\beta}
\nonu\\ 
&+&
({ T}_L^{ij} { T}_R^{ij})^{\alpha \beta}\,(
p_{3}\:\partial^{2}\Phi_{0}^{(s),\beta}
+p_{4}\:L\,\Phi_{0}^{(s),\beta})
\,
\Bigg](z),
\label{Phialphaprimary}
\eea
where the coefficients in the last component
depend on $N,k$ and $s$ (together with (\ref{alpha}))
and are given by
\bea
&&
p_{1}  =  -\frac{2\alpha\,(3+3k+3N+3kN+26s+13ks+13Ns)}{(3+3k+3N+3kN-4s+ks+Ns+6kNs+16s^{2}+8ks^{2}+8Ns^{2})},
\nonu\\
&&
p_{2}  =  \frac{12(k-N)s(1+s)}{(3+3k+3N+3kN-4s+ks+Ns+6kNs+16s^{2}+8ks^{2}+8Ns^{2})},
\nonu\\
&&
p_{3}= \frac{(-15 - 6 k - 6 N + 3 k N + 8 s + 4 k s + 4 N s)}
{2 (1 + s) (3 + 3 k + 3 N + 3 k N - 4 s + k s + N s + 6 k N s + 16 s^2 + 8 k s^2 + 8 N s^2)},
\nonu\\
&&
p_{4}= \frac{3 (2 + k + N)}{(3 + 3 k + 3 N + 3 k N - 4 s + k s + N s + 6 k N s + 16 s^2 + 8 k s^2 + 8 N s^2)}.
\label{p1p2p3p4}
\eea
Note that
the values $p_1$ and $p_2$ also appear in the 
corresponding higher spin-$s$ multiplet in the unitary Wolf space
coset model \cite{AK1509}. The expression (\ref{Phialphaprimary})
holds for $SO(4)$ nonsinglet higher spin multiplets
in the unitary Wolf space coset.
Moreover, the following quantities with (\ref{alpha})
are introduced
\bea
\widetilde{{ T}}^{ij} & \equiv &
\frac{1}{2}\:\varepsilon_{ijkl}\:{ T}^{kl}, \qquad
{ T}^{ij}_{L}  \equiv 
    \frac{1}{2} \, { T}^{ij}+{ \alpha}\,{ \widetilde{T}}^{ij},
\qquad
    { T}^{ij}_{R}  \equiv 
    { \alpha}\,{ T}^{ij}+\frac{1}{2}\,{ \widetilde{T}}^{ij}.
\label{threeT}
\eea
For the $SO(4)$ vector representation $\alpha$,
we will see similar construction  in next
subsection. 

Then we can check that the $SO(4)$ adjoint
higher spin multiplet (\ref{Phialphaprimary}) satisfies
the relation (\ref{JPhi}) with (\ref{6x6}).
Its component relations are given in Appendix $A$. 

The first order pole of (\ref{OPE3-1})
contains other higher spin currents in various way.
For example, the other components of the ${\cal N}=4$ multiplets
${\bf \Phi}^{(3),\alpha}(Z)$ can arise.

\subsection{The  new higher spin currents
of spin $\frac{7}{2}$}

We consider the OPE
between the four higher spin-$\frac{5}{2}$ currents
and the
higher spin-$2$ current.
It turns out that 
we obtain
\bea
\Phi_{\frac{1}{2}}^{(2),i}(z)\,\Phi_{0}^{(2)}(w)
& = & \frac{1}{(z-w)^{3}}\,
Q^{(\frac{3}{2}),i}_{\frac{1}{2}}(w)
+\frac{1}{(z-w)^{2}}\,\Bigg[\,
\frac{2}{3}\,\partial  
Q^{(\frac{3}{2}),i}_{\frac{1}{2}}
+ Q^{(\frac{5}{2}),i}_{\frac{1}{2}}
\,\Bigg](w)
\nonu \\ 
& + &
\frac{1}{(z-w)}\,\Bigg[\,
\frac{1}{4}\,\partial^{2}  Q^{(\frac{3}{2}),i}_{\frac{1}{2}} +
\frac{3}{5}\,\partial  Q^{(\frac{5}{2}),i}_{\frac{1}{2}}
+ Q^{(\frac{7}{2}),i}_{\frac{1}{2}}\,
\Bigg](w)+\cdots.
\label{fivehalf-1}
\eea
The quasi primary fields appearing in the above poles
of (\ref{fivehalf-1})
are given in Appendix $B$.
In particular, the quasi primary field
$ Q^{(\frac{5}{2}),i}_{\frac{1}{2}}(w)$ contains
$ \Phi^{(2),i}_{\frac{1}{2}}(w)$ which is the second component
of the lowest higher spin-$2$ multiplet in (\ref{Phiexp}).
In the first order pole, there exist four
new primary fields $ \Phi^{(\frac{7}{2}),i}_{0}(w)$
\footnote{We can obtain each component by following the procedure
described in the footnote \ref{othercomp}.}
as well as
the composite fields containing the $SO(4)$ adjoint
$\Phi_0^{(3),\alpha}(w)$ (and 
$\Phi_{\frac{1}{2}}^{(3),i, \alpha}(w)$), the $SO(4)$ singlet
$\Phi_0^{(2)}(w)$ ($\Phi_{\frac{1}{2}}^{(2),i}(w)$, $\Phi_{1}^{(2),ij}(w)$ 
and $\Phi_{\frac{3}{2}}^{(2),i}(w)$) and 
the other $SO(4)$ singlet $X_0^{(2)}(w)$ ($X_{\frac{1}{2}}^{(2),i}(w)$,
$X_{1}^{(2),ij}(w)$ and $X_{\frac{3}{2}}^{(2),i}(w)$).
In other words, we have
\bea
Q_{\frac{1}{2}}^{(\frac{7}{2}),i}(w) = w_{0,\frac{7}{2}} \,
\Phi_0^{(\frac{7}{2}),i}(w)
+ \cdots.
\label{Q7half}
\eea
The abbreviated part is given in Appendix $B$. 

The ${\cal N}=4$
four $SO(4)$ vector higher spin-$\frac{7}{2}$ multiplets
transform under the stress energy tensor
as follows \cite{Schoutensnpb}. The OPE looks like (\ref{JPhi})
with $s =\frac{7}{2}$:
\bea
{\bf J }(Z_{1}) \, {\bf \Phi}^{(s),\mu}(Z_{2}) & = & 
\frac{\theta_{12}^{4}}{z_{12}^{2}} \, 2s\, {\bf \Phi}^{(s),\mu}(Z_{2})+
\frac{\theta_{12}^{4-i}}{z_{12}} \, D^{i}  {\bf \Phi}^{(s),\mu}(Z_{2})+
\frac{\theta_{12}^{4}}{z_{12}} \, 2 \, \partial {\bf \Phi}^{(s),\mu}(Z_{2})
\nonu \\
&- &
\frac{i}{2} \:\frac{\theta_{12}^{4-ij}}{z_{12}}({ T}^{ij})^{\mu \nu}\: {\bf \Phi}^{(s),\nu}(Z_{2})
+\cdots,
\label{eq0}
\eea
where the $T^{ij}$ matrix is the generator of
the $SO(4)$ vector representation (\ref{Malpha})
\bea
{ T}^{12} & = & \left(\begin{array}{cccc}
0 & -i & 0 & 0\\
i & 0 & 0 & 0\\
0 & 0 & 0 & 0\\
0 & 0 & 0 & 0
\end{array}\right), \,
{ T}^{13}=\left(\begin{array}{cccc}
0 & 0 & -i & 0\\
0 & 0 & 0 & 0\\
i & 0 & 0 & 0\\
0 & 0 & 0 & 0
\end{array}\right), \,
{ T}^{14}=\left(\begin{array}{cccc}
0 & 0 & 0 & -i\\
0 & 0 & 0 & 0\\
0 & 0 & 0 & 0\\
i & 0 & 0 & 0
\end{array}\right),
\nonu \\
{ T}^{23} & = & \left(\begin{array}{cccc}
0 & 0 & 0 & 0\\
0 & 0 & -i & 0\\
0 & i & 0 & 0\\
0 & 0 & 0 & 0
\end{array}\right), \,
{ T}^{24}=\left(\begin{array}{cccc}
0 & 0 & 0 & 0\\
0 & 0 & 0 & -i\\
0 & 0 & 0 & 0\\
0 & i & 0 & 0
\end{array}\right), \,
{ T}^{34}=\left(\begin{array}{cccc}
0 & 0 & 0 & 0\\
0 & 0 & 0 & 0\\
0 & 0 & 0 & -i\\
0 & 0 & i & 0
\end{array}\right),
\label{4x4}
\eea
where the commutators $[T^{ij}, T^{kl}]$ satisfy the previous relations
described in (\ref{6x6}).
The corresponding component OPEs of (\ref{eq0})
can be obtained from Appendix $A$ by considering
the $T^{ij}$ matrix as the ones in (\ref{4x4}).
Moreover, we should use the corresponding primary fields
(if we need them)
according to (\ref{Phialphaprimary}) by substituting the
$4\times 4$ matrices in (\ref{4x4}).

\subsection{The  higher spin current
of spin $4$}

In the second order pole of (\ref{OPE4-1}), the quasi primary
field $Q_2^{(4)}(w)$ contains
the $SO(4)$ singlet higher spin-$4$ current $\Phi_0^{(4)}(w)$.
The other part of the quasi primary
field $Q_2^{(4)}(w)$ is given in Appendix $B$.
Note that the composite field $X_0^{(2)} \, X_0^{(2)}(w)$ (as well as
other dependent terms)
is absorbed in  $\Phi_0^{(4)}(w)$ such that
$\Phi_0^{(4)}(w)$ should transform as the primary field
under the stress energy tensor.
If the composite field $X_0^{(2)} \, X_0^{(2)}(w)$ is not included
in the  $SO(4)$ singlet higher spin-$4$ current, then
we should calculate the OPE $X_0^{(2)}(z) \, X_0^{(2)}(w)$ (and their
${\cal N}=4$ version)
in order to use its normal ordered product for the Jacobi identity.
We observe that all the components of the ${\cal N}=4$ multiplets
${\bf \Phi}^{(2)}(Z)$ and ${\bf X}^{(2)}(Z)$
appear in the quasi primary field
$Q_2^{(4)}(w)$.

Moreover, once we combine the above higher spin-$4$ current   
with $\Phi_0^{(2)} \, \Phi_0^{(2)}(w)$ term as well as other 
terms as in 
\bea
\tilde{\Phi}_0^{(4)}(w)
& = & w_{11,4}\,\Phi^{(2)}_0\Phi^{(2)}_0(w)
+(M^{\alpha})^{ij}(\, 
c_{1}\,\Phi_{1}^{(3),ij,\alpha}
+\cdots
+\varepsilon^{ijkl}c_{8}\,\Gamma^{k}\Gamma^{l}\Phi_{0}^{(3),\alpha}\,)(w)
+  \delta_{\mu}^{i}\, 
c_{9}\,\Phi_{\frac{1}{2}}^{(\frac{7}{2}),i,\mu}(w)
\nonu \\
& + & ( c_{10}\,\Phi_{2}^{(2)}
+c_{11}\,\tilde{L}\Phi_{0}^{(2)}
+\cdots
+c_{32}\,\varepsilon^{ijkl}\, T^{ij}T^{kl}\Phi_{0}^{(2)})(w)
+c_{33}\, X_{2}^{(2)}(w)
+c_{34}\,\tilde{L} X_{0}^{(2)}(w)
\nonu
\\
&
+ &  \cdots + c_{55}\,\varepsilon^{ijkl}\, T^{ij}T^{kl}X_{0}^{(2)}(w)
+c_{56}\,\tilde{L}\tilde{L}(w)
+c_{57}\,\tilde{L}UU(w)
\nonu \\
& + & \cdots+c_{155}\,
\varepsilon^{ijkl}\,\Gamma^{i}
\Gamma^{j}\partial\Gamma^{k}\partial\Gamma^{l}(w),
\label{finalspin4}
\eea
then the structure constant $w_{11,4} \equiv C_{(4)(2)}^{(4)}$
does not appear
in the remaining OPEs.
The coefficients $c_{10}, \cdots, c_{32}$ in (\ref{finalspin4})
also depend on the structure constant $C_{(4)(2)}^{(2)}$
appeared in the subsection $4.2$.
We have checked that
this feature arises also in the unitary case \cite{AK1509}.

The final first order pole of (\ref{OPE4-1}) can be obtained
and it turns out that there is no new primary field.
All the terms after subtracting the descendant fields can be
written in terms of the known composite fields (including the higher
spin-$\frac{9}{2}$ currents which will be described in
next subsection). We expect that
the ${\cal N}=4$ higher spin-$5$ multiplets (we have not found in this
paper)
will appear by considering
the other OPEs between the ${\cal N}=4$ multiplets we have found in
this paper
\footnote{
\label{nonlinear}
  Let us emphasize that in this case (together with
  the higher spin-$\frac{9}{2}$ current case which will be described
  in next subsection), due to the
  too many number of composite fields, we should go into the
  nonlinear version where there are no spin-$\frac{1}{2}$ and
  spin-$1$ currents. Without these currents,
  the possible composite terms are reduced significantly.
  We can obtain the higher spin currents in the
  nonlinear version, by following the work of \cite{BCG}, in terms of
  the ones we have found in the linear version (and vice versa).
  Then by using the
  OPEs between them we can rewrite the right hand sides of these OPEs
  in terms of the fields in the nonlinear version. After that, we can
  go into the linear version (by changing the composite terms
  in the linear basis) together with the known field
  contents of the composite terms. }.

\subsection{The   higher spin currents
of spin $\frac{9}{2}$}

We consider the OPE
between the four higher spin-$\frac{7}{2}$ currents
and the
higher spin-$2$ current.
It turns out that 
we obtain 
\bea
\Phi_{\frac{3}{2}}^{(2),i}(z)\,\Phi_{0}^{(2)}(w)
& = & \frac{1}{(z-w)^{5}}\,
Q^{(\frac{1}{2}),i}_{\frac{3}{2}}(w)
+\frac{1}{(z-w)^{4}}\,\Bigg[\,
2\,\partial Q^{(\frac{1}{2}),i}_{\frac{3}{2}}+Q^{(\frac{3}{2}),i}_{\frac{3}{2}}
\,\Bigg](w)
\nonu \\ 
&+&\frac{1}{(z-w)^{3}}\,\Bigg[\,
\frac{3}{2}\,\partial^{2} Q^{(\frac{1}{2}),i}_{\frac{3}{2}}
+\partial Q^{(\frac{3}{2}),i}_{\frac{3}{2}}
+Q^{(\frac{5}{2}),i}_{\frac{3}{2}}
\,\Bigg](w)
\nonu \\ 
& + &
\frac{1}{(z-w)^{2}}\,\Bigg[\,
\frac{2}{3}\,\partial^{3} Q^{(\frac{1}{2}),i}_{\frac{3}{2}}
+\frac{1}{2}\,\partial^{2} Q^{(\frac{3}{2}),i}_{\frac{3}{2}}
+\frac{4}{5}\,\partial  Q^{(\frac{5}{2}),i}_{\frac{3}{2}}
+ Q^{(\frac{7}{2}),i}_{\frac{3}{2}}
\,\Bigg](w)
\nonu \\ 
& + &
\frac{1}{(z-w)}\,\Bigg[\,
\frac{5}{24}\,\partial^{4} Q^{(\frac{1}{2}),i}_{\frac{3}{2}}
+\frac{1}{6}\,\partial^{3} Q^{(\frac{3}{2}),i}_{\frac{3}{2}}
+\frac{1}{3}\,\partial^{2} Q^{(\frac{5}{2}),i}_{\frac{3}{2}}
+\frac{5}{7}\,\partial Q^{(\frac{7}{2}),i}_{\frac{3}{2}}
+Q^{(\frac{9}{2}),i}_{\frac{3}{2}}
\,
\Bigg](w)
\nonu \\ 
& + &
\frac{2\,\alpha}{5}\,\sum\limits _{n=2}^{4}
\frac{1}{(z-w)^{n}} \, \Big\{
\partial \Phi_{\frac{1}{2}}^{(2),i} \,\Phi_{0}^{(2)}
\Big\}_{n}(w)
+\cdots.
\label{OPE7half-1}
\eea
Again, the OPE (\ref{OPE7half-1}) consists of two parts.
The first four lines comes from the corresponding
four primary higher spin-$\frac{7}{2}$ currents
$\tilde{\Phi}_{\frac{3}{2}}^{(2),i}(z)$
and the last line
comes from the additional term described in (\ref{Phialphaprimary})
\footnote{By decoupling of four spin-$\frac{1}{2}$ currents
and spin-$1$ current of the ${\cal N}=4$ linear
superconformal algebra, we obtain the $11$ currents of
the ${\cal N}=4$ nonlinear superconformal algebra.
The explicit expressions are given in (\ref{closedform}).
Then we can determine the remaining higher spin currents
in the nonlinear version
starting from the lowest higher spin-$2$
current (\ref{T2ansatz}) as  in the footnote \ref{nonlinear}.
After obtaining the composite fields in the nonlinear version,
we can use them in the linear version. Note that we saw the higher
spin-$4, \frac{9}{2}$ currents for $N=4$ in previous section.}.

By using the relation (by using the notation of \cite{BS})
\bea
\Big\{ \partial \Phi_{\frac{1}{2}}^{(2),i} \,\Phi_{0}^{(2)}
\Big\}_{n+1}(w)
=
-n\, \Big\{ \Phi_{\frac{1}{2}}^{(2),i} \,\Phi_{0}^{(2)}
\Big\}_{n}(w),
\qquad  n=1,2,3,
\nonu
\eea
where the previous relation (\ref{fivehalf-1}) can be used,
we can present the above OPE as follows:
\bea
\Phi_{\frac{3}{2}}^{(2),i}(z)\,\Phi_{0}^{(2)}(w)
& = & \frac{1}{(z-w)^{5}}\,
Q^{(\frac{1}{2}),i}_{\frac{3}{2}}(w)
+\frac{1}{(z-w)^{4}}\,\Bigg[\,
  2\,\partial Q^{(\frac{1}{2}),i}_{\frac{3}{2}}+Q^{(\frac{3}{2}),i}_{\frac{3}{2}}
  + R^{(\frac{3}{2}),i}_{\frac{3}{2}}
\,\Bigg](w)
\nonu \\ 
&+&\frac{1}{(z-w)^{3}}\,\Bigg[\,
\frac{3}{2}\,\partial^{2} Q^{(\frac{1}{2}),i}_{\frac{3}{2}}
+\partial Q^{(\frac{3}{2}),i}_{\frac{3}{2}}
+Q^{(\frac{5}{2}),i}_{\frac{3}{2}} +R^{(\frac{5}{2}),i}_{\frac{3}{2}}
\,\Bigg](w)
\nonu \\ 
& + &
\frac{1}{(z-w)^{2}}\,\Bigg[\,
\frac{2}{3}\,\partial^{3} Q^{(\frac{1}{2}),i}_{\frac{3}{2}}
+\frac{1}{2}\,\partial^{2} Q^{(\frac{3}{2}),i}_{\frac{3}{2}}
+\frac{4}{5}\,\partial  Q^{(\frac{5}{2}),i}_{\frac{3}{2}}
+ Q^{(\frac{7}{2}),i}_{\frac{3}{2}} +R^{(\frac{7}{2}),i}_{\frac{3}{2}}
\,\Bigg](w)
\nonu \\ 
& + &
\frac{1}{(z-w)}\,\Bigg[\,
\frac{5}{24}\,\partial^{4} Q^{(\frac{1}{2}),i}_{\frac{3}{2}}
+\frac{1}{6}\,\partial^{3} Q^{(\frac{3}{2}),i}_{\frac{3}{2}}
+\frac{1}{3}\,\partial^{2} Q^{(\frac{5}{2}),i}_{\frac{3}{2}}
+\frac{5}{7}\,\partial Q^{(\frac{7}{2}),i}_{\frac{3}{2}}
+Q^{(\frac{9}{2}),i}_{\frac{3}{2}}
\,
\Bigg](w) \nonu \\
&+ & \cdots,
\label{OPE7half-2}
\eea
where we introduce 
\bea
R_{\frac{3}{2}}^{(\frac{7}{2}),i}(w) & \equiv &
-\frac{2  \, \alpha}{5}\, (\frac{1}{4} \, \pa^2 \,
Q_{\frac{1}{2}}^{(\frac{3}{2}),i} +\frac{3}{5} \, \pa \,
Q_{\frac{1}{2}}^{(\frac{5}{2}),i} + Q_{\frac{1}{2}}^{(\frac{7}{2}),i})(w),
\nonu \\
R_{\frac{3}{2}}^{(\frac{5}{2}),i}(w) & \equiv &
-\frac{4 \, \alpha}{5} \, (\frac{2}{3} \, \pa \,
Q_{\frac{1}{2}}^{(\frac{3}{2}),i} + Q_{\frac{1}{2}}^{(\frac{5}{2}),i})(w),
\nonu \\
R_{\frac{3}{2}}^{(\frac{3}{2}),i}(w) & \equiv &
-\frac{6 \, \alpha}{5} \, Q_{\frac{1}{2}}^{(\frac{3}{2}),i}(w),
\label{bigR}
\eea
from the last line of (\ref{OPE7half-1}).
We can see that the fifth and fourth order terms of
(\ref{OPE7half-2}) consist of the composite fields from the
$16$ currents as in Appendix $B$.
The third and second order poles of (\ref{OPE7half-2})
have the higher spin currents (found before) dependent terms.
The first order pole has the following form
\bea
Q_{\frac{3}{2}}^{(\frac{9}{2}),i}(w) & = &
w_{0,\frac{9}{2}} \, \Phi_0^{(\frac{9}{2}),i}(w) + \cdots,
\label{Q9half}
\eea
where the other remaining terms are given in Appendix $B$.
Each component of the higher spin-$\frac{9}{2}$
current can be obtained by following the procedure
described in the footnote \ref{othercomp}.
We see that all kinds of higher spin currents appear in this
quasi primary field.
We can use (\ref{eq0}) together with
the above $4\times 4$ matrices in (\ref{4x4}) in order to
obtain the OPEs between this higher spin multiplet and
the ${\cal N}=4$ stress energy tensor.
Its component results are given in Appendix $A$.

\subsection{ The fundamental $16$ OPEs}

Therefore, the fundamental $16$ OPEs (five kinds of OPEs)
are given by (\ref{spin2spin2}),
(\ref{OPE4-1-1}), (\ref{OPE3-1}), (\ref{fivehalf-1}) and
(\ref{OPE7half-2}). The structure constants are written in terms of
$N$, $k$ and $C_{(4)(2)}^{(2)}$ which will be given in the
${\tt ancillary.nb}$.
These will determine the remaining $120$ OPEs by using the
${\cal N}=4$ supersymmetry soon.

We summarize the higher spin currents appearing in various
quasi primary fields (\ref{quasicomponent}) of
these OPEs in Table $1$.
Some of the higher spin currents are not present in this Table $1$
and they will arise in the remaining $120$ OPEs. 
For simplicity, we do not include the dependence of
$16$ currents of
the large ${\cal N}=4$ linear superconformal algebra.

\begin{table}[ht]
\centering 
\begin{tabular}{|c|c| } 
\hline 
Quasi primaries & Higher spin currents  \\
[1ex] 
\hline
$Q^{(\frac{5}{2}),i}_{\frac{1}{2}}(z)$  & $\Phi^{(2),j}_{\frac{1}{2}} $
\\ 
[1.5ex]
\hline
$Q^{(\frac{7}{2}),i}_{\frac{1}{2}}(z)$ &  $\Phi^{(\frac{7}{2}),j}_{0};\, \,
\Phi^{(3),j,\alpha}_{\frac{1}{2}},
\Phi^{(3),\alpha}_{0}; \,\,
\Phi^{(2),j}_{\frac{3}{2}},
\Phi^{(2),jk}_{1},
\Phi^{(2),j}_{\frac{1}{2}},
\Phi^{(2)}_{0}; \,\,
X^{(2),j}_{\frac{3}{2}},
X^{(2),jk}_{1},
X^{(2),j}_{\frac{1}{2}},
X^{(2)}_{0}$ \\
[1.5ex]
\hline
$Q^{(3),ij}_{1}(z)$ & $\Phi^{(3),\alpha}_{0};\,\,
\Phi^{(2),kl}_{1},
\Phi^{(2),k}_{\frac{1}{2}},
\Phi^{(2)}_{0};\,\,
X^{(2),kl}_{1},
X^{(2),k}_{\frac{1}{2}},
X^{(2)}_{0}$  \\
[1.5ex]
\hline
$Q^{(4),ij}_{1}(z) $ & $\Phi^{(3),kl,\alpha}_{1},
\Phi^{(3),k,\alpha}_{\frac{1}{2}},
\Phi^{(3),\alpha}_{0};\,\,
\Phi^{(2),k}_{\frac{3}{2}},
\Phi^{(2),kl}_{1},
\Phi^{(2),k}_{\frac{1}{2}},
\Phi^{(2)}_{0};\,\,
X^{(2),k}_{\frac{3}{2}},
X^{(2),kl}_{1},
X^{(2),k}_{\frac{1}{2}},
X^{(2)}_{0}$  \\
[1.5ex]
\hline
$Q^{(\frac{5}{2}),i}_{\frac{3}{2}}(z)$ & $\Phi^{(2),j}_{\frac{1}{2}},
\Phi^{(2)}_{0};\,\,
X^{(2),j}_{\frac{1}{2}}$  \\
[1.5ex]
\hline
$Q^{(\frac{7}{2}),i}_{\frac{3}{2}}(z)$ & $\Phi^{(\frac{7}{2}),j}_{0}; \,\,
\Phi^{(3),j,\alpha}_{\frac{1}{2}},
\Phi^{(3),\alpha}_{0};\,\,
\Phi^{(2),j}_{\frac{3}{2}},
\Phi^{(2),jk}_{1},
\Phi^{(2),j}_{\frac{1}{2}},
\Phi^{(2)}_{0};\,\,
X^{(2),j}_{\frac{3}{2}},
X^{(2),jk}_{1},
X^{(2),j}_{\frac{1}{2}},
X^{(2)}_{0} $  \\
[1.5ex]
\hline
$Q^{(\frac{9}{2}),i}_{\frac{3}{2}}(z)$ & $\Phi^{(\frac{9}{2}),j}_{0};\,\,
\Phi^{(4),j}_{\frac{1}{2}};
\,\,
\Phi^{(\frac{7}{2}),jk,l}_{1},
\Phi^{(\frac{7}{2}),j,k}_{\frac{1}{2}},
\Phi^{(\frac{7}{2}),j}_{0}; \,\,
\Phi^{(3),j,\alpha}_{\frac{3}{2}},
\Phi^{(3),jk,\alpha}_{1},
\Phi^{(3),j,\alpha}_{\frac{1}{2}},
\Phi^{(3),\alpha}_{0};$  \\
$ $ & $ \Phi^{(2),j}_{\frac{3}{2}},
\Phi^{(2),jk}_{1},
\Phi^{(2),j}_{\frac{1}{2}},
\Phi^{(2)}_{0}; \,\,
X^{(2),j}_{\frac{3}{2}},
X^{(2),jk}_{1},
X^{(2),j}_{\frac{1}{2}},
X^{(2)}_{0}$  \\
[1.5ex]
\hline
$ Q^{(2)}_{2}(z)$ & $\Phi^{(2)}_{0}; \,\,
X^{(2)}_{0}$  \\
[1.5ex]
\hline
$Q^{(3)}_{2}(z)$ & $\Phi^{(2),i}_{\frac{1}{2}}; \,\,
\Phi^{(2)}_{0}$ \\
[1.5ex]
\hline
$Q^{(4)}_{2}(z)$ & $\Phi_0^{(4)}; \,\, \Phi^{(\frac{7}{2}),i,\mu}_{\frac{1}{2}},
\Phi^{(\frac{7}{2}),i}_{0};\,\,
\Phi^{(3),jk,\alpha}_{1},
\Phi^{(3),j,\alpha}_{\frac{1}{2}},
\Phi^{(3),\alpha}_{0};
\Phi^{(2)}_{2},
\Phi^{(2),j}_{\frac{3}{2}},
\Phi^{(2),jk}_{1},
\Phi^{(2),j}_{\frac{1}{2}},
\Phi^{(2)}_{0};$ \\
$$ & $ X^{(2)}_{2},
X^{(2),j}_{\frac{3}{2}},
X^{(2),jk}_{1},
X^{(2),j}_{\frac{1}{2}},
X^{(2)}_{0}$ \\
[1.5ex]
\hline
$Q^{(5)}_{2}(z)$ & $\Phi^{(\frac{9}{2}),i,j}_{\frac{1}{2}};\,\,
\Phi^{(\frac{7}{2}),i,j}_{\frac{3}{2}},
\Phi^{(\frac{7}{2}),ij,k}_{1},
\Phi^{(\frac{7}{2}),i,j}_{\frac{1}{2}},
\Phi^{(\frac{7}{2}),i}_{0};\,\,
\Phi^{(3),ij,\alpha}_{1},
\Phi^{(3),i,\alpha}_{\frac{1}{2}},
\Phi^{(3),\alpha}_{0};$ \\
$$ & $\Phi^{(2)}_{2},
\Phi^{(2),i}_{\frac{3}{2}},
\Phi^{(2),ij}_{1},
\Phi^{(2),i}_{\frac{1}{2}},
\Phi^{(2)}_{0};\,\,
X^{(2)}_{2},
X^{(2),i}_{\frac{3}{2}},
X^{(2),ij}_{1},
X^{(2),i}_{\frac{1}{2}},
X^{(2)}_{0}$ \\
[1.5ex] 
\hline 
\end{tabular}
\caption{The structure constant
  $w_{1,2} \equiv C_{(4)(2)}^{(2)}$ appears 
  in front of
  the above quasi primaries appearing in the
  OPEs except $Q_{\frac{1}{2}}^{(\frac{5}{2}),i}(z)$
  and $Q_2^{(3)}(z)$, which do not have the components of
  ${\cal N}=4$ multiplet ${\bf X}^{(2)}(Z)$.
  The quasi primaries, which are not in this list, have the
  composite fields consisting of $16$ currents.
  Although the $X_0^{(2)}(z)$ dependence in $Q_2^{(3)}(z)$ does not appear,
  its dependence  appears in
  $\pa \, Q_2^{(2)}(z)$ in (\ref{OPE4-1-1}).
  Similarly,  the $\Phi_0^{(4)}(z)$ dependence in $Q_2^{(5)}(z)$
  does not appear and 
  its dependence appears via
  $\pa \, Q_2^{(4)}(z)$ in (\ref{OPE4-1-1}). 
} 
\end{table}

\subsection{One single ${\cal N}=4$ super OPE}

From the fundamental $16$ OPEs
(\ref{spin2spin2}), (\ref{fivehalf-1}), (\ref{OPE3-1}),
(\ref{OPE7half-2}) and (\ref{OPE4-1-1}) (that is, five
different kinds of OPEs),
we can generalize them in ${\cal N}=4$ superspace
by taking \cite{AK1509}
\bea
U(w) & \rightarrow &  \pa \, {\bf J}(Z_2),
\nonu \\
\Gamma^i(w) & \rightarrow & -i \, D^i {\bf J}(Z_2)
\equiv -i \, {\bf J}^i(Z_2),  
\nonu \\
T^{ij}(w) & \rightarrow & -\frac{i}{2!} \, \varepsilon^{ijkl} \,
D^k D^l {\bf J}(Z_2) \equiv -\frac{i}{2!} \,  \varepsilon^{ijkl} \,
{\bf J}^{kl}(Z_2),
\nonu \\
\tilde{G}^i(w) & \rightarrow & \frac{1}{3!} \,
\varepsilon^{ijkl} \, D^j D^k D^l {\bf J}(Z_2)
\equiv \frac{1}{3!} \,\varepsilon^{ijkl}  \, {\bf J}^{jkl}(Z_2),
\nonu \\
\tilde{L}(w)  & \rightarrow & \frac{1}{2 \cdot 4!} \,
\varepsilon^{ijkl} \, D^{i} D^j D^k D^l {\bf J}(Z_2)
\equiv \frac{1}{2 \cdot 4!} \,\varepsilon^{ijkl}  \, {\bf J}^{ijkl}(Z_2),
\nonu \\
\Phi_{0}^{(s),\alpha}(w) & \rightarrow &  
    {\bf \Phi}^{(s),\alpha}(Z_2),
    \nonu \\
\Phi_{\frac{1}{2}}^{(s),i,\alpha}(w) & \rightarrow &  D^i
    {\bf \Phi}^{(s),\alpha}(Z_2),  
\nonu \\
\Phi_{1}^{(s),ij,\alpha}(w) & \rightarrow & -\frac{1}{2!} \,
 \varepsilon^{ijkl} \, D^k D^l {\bf \Phi}^{(s),\alpha}(Z_2),
\nonu \\
\Phi_{\frac{3}{2}}^{(s),i,\alpha}(w) & \rightarrow & -\frac{1}{3!} \,
\varepsilon^{ijkl} \, D^j D^k D^l {\bf \Phi}^{(s),\alpha}(Z_2),
\nonu \\
\Phi_{2}^{(s),\alpha}(w)  & \rightarrow & \frac{1}{ 4!} \,
\varepsilon^{ijkl} \, D^{i} D^j D^k D^l {\bf \Phi}^{(s),\alpha}(Z_2),
\qquad \alpha = \mbox{singlet, \,\, adjoint, \,\, vector},
\label{comptosuper}
\eea
where $\tilde{G}^i(w)$ and $\tilde{L}(w)$ are given in (\ref{J4})
and putting the relevant fermionic coordinates.
For the singlet, adjoint and vector representations,
we substitute the corresponding indices into the $\alpha$.
In doing this, there are additional terms arising from the
summation over the same indices.
We present the quasi primary fields in ${\cal N}=4$ superspace
in Appendix $C$. The total number of terms in Appendix $B$
and Appendix $C$ is little different from each other.

Then the single ${\cal N}=4$ super OPE
between the $SO(4)$ singlet higher spin-$2$ multiplet
can be summarized by (after rearranging (\ref{intermediate}))
\bea
&& {\bf \Phi}^{(2)}(Z_{1})\,{\bf \Phi}^{(2)}(Z_{2}) =
\frac{1}{z_{12}^{4}}\, c^{0,4}_{0}
+\frac{\theta_{12}^{4-0}}{z_{12}^{6}}\,
8 \, \alpha \, c_{0}^{0,4}
+\frac{\theta_{12}^{4-i}}{z_{12}^{5}}\,
{\bf Q}^{(\frac{1}{2}),i}_{\frac{3}{2}}(Z_{2})
\nonu \\
&& +\frac{\theta_{12}^{4-0}}{z_{12}^{5}}\,
{\bf Q}^{(1)}_{2}(Z_{2})
+
\frac{\theta_{12}^{4-ij}}{z_{12}^{4}}\,
{\bf Q}^{(1),ij}_{1}(Z_{2})
 +\frac{\theta_{12}^{4-i}}{z_{12}^{4}}\,\Bigg[\,
2\,\partial {\bf Q}^{(\frac{1}{2}),i}_{\frac{3}{2}}+{\bf Q}^{(\frac{3}{2}),i}_{\frac{3}{2}}
+{\bf R}^{(\frac{3}{2}),i}_{\frac{3}{2}}
\,\Bigg](Z_{2})
 \nonu\\
&&
+\frac{\theta_{12}^{4-0}}{z_{12}^{4}}\,\Bigg[\,
\frac{3}{2}\,\partial {\bf Q}^{(1)}_{2}
 +{\bf Q}^{(2)}_{2}
+{\bf R}^{(2)}_{2}\,\Bigg](Z_{2})
+
\frac{\theta_{12}^{i}}{z_{12}^{3}}\,
{\bf Q}^{(\frac{3}{2}),i}_{\frac{1}{2}}(Z_{2})
 +
\frac{\theta_{12}^{4-ij}}{z_{12}^{3}}\,\Bigg[\,
\partial{\bf Q}^{(1),ij}_{1}
+{\bf Q}^{(2),ij}_{1}
\,\Bigg](Z_{2})
\nonu \\
&& +\frac{\theta_{12}^{4-i}}{z_{12}^{3}}\,\Bigg[\,
\frac{3}{2}\,\partial^{2} {\bf Q}^{(\frac{1}{2}),i}_{\frac{3}{2}}
+\partial {\bf Q}^{(\frac{3}{2}),i}_{\frac{3}{2}}
+{\bf Q}^{(\frac{5}{2}),i}_{\frac{3}{2}}
+{\bf R}^{(\frac{5}{2}),i}_{\frac{3}{2}}
\,\Bigg](Z_{2})
\nonu\\
&&
+\frac{\theta_{12}^{4-0}}{z_{12}^{3}}\,\Bigg[\,
\partial^{2} {\bf Q}^{(1)}_{2}
+\partial {\bf Q}^{(2)}_{2}
+{\bf Q}^{(3)}_{2}
+{\bf R}^{(3)}_{2}\,\Bigg](Z_{2})
+\frac{1}{z_{12}^{2}}\,
{\bf Q}^{(2)}_{0}(Z_{2})
\nonu \\
&&
+
\frac{\theta_{12}^{i}}{z_{12}^{2}}\,\Bigg[\,
\frac{2}{3}\,\partial  {\bf Q}^{(\frac{3}{2}),i}_{\frac{1}{2}}
 + {\bf Q}^{(\frac{5}{2}),i}_{\frac{1}{2}}
\,\Bigg](Z_{2})
+
\frac{\theta_{12}^{4-ij}}{z_{12}^{2}}\,\Bigg[\,
\frac{1}{2}\,\partial^{2} {\bf Q}^{(1),ij}_{1}
+\frac{3}{4}\,\partial {\bf Q}^{(2),ij}_{1}
+{\bf Q}^{(3),ij}_{1}
\,\Bigg](Z_{2})
\nonu\\
&&
+
\frac{\theta_{12}^{4-i}}{z_{12}^{2}}\,\Bigg[\,
\frac{2}{3}\,\partial^{3} {\bf Q}^{(\frac{1}{2}),i}_{\frac{3}{2}}
+\frac{1}{2}\,\partial^{2} {\bf Q}^{(\frac{3}{2}),i}_{\frac{3}{2}}
+\frac{4}{5}\,\partial {\bf Q}^{(\frac{5}{2}),i}_{\frac{3}{2}}
+ {\bf Q}^{(\frac{7}{2}),i}_{\frac{3}{2}}
+{\bf R}^{(\frac{7}{2}),i}_{\frac{3}{2}}
\,\Bigg](Z_{2})
\nonu\\
&&
+\frac{\theta_{12}^{4-0}}{z_{12}^{2}}\,\Bigg[\,
\frac{5}{12}\,\partial^{3} {\bf Q}^{(1)}_{2}
+\frac{1}{2}\, \partial {\bf Q}^{(2)}_{2}
+\frac{5}{6}\,{\bf Q}^{(3)}_{2}
+{\bf Q}^{(4)}_{2}
+{\bf R}^{(4)}_{2}
\,\Bigg](Z_{2})
\nonu\\
&&
+\frac{1}{z_{12}}\,
\frac{1}{2}\partial {\bf Q}^{(2)}_{0}(Z_{2})
+
\frac{\theta_{12}^{i}}{z_{12}}\,\Bigg[\,
\frac{1}{4}\,\partial^{2}  {\bf Q}^{(\frac{3}{2}),i}_{\frac{1}{2}} +
\frac{3}{5}\,\partial {\bf Q}^{(\frac{5}{2}),i}_{\frac{1}{2}}
+ {\bf Q}^{(\frac{7}{2}),i}_{\frac{1}{2}}
\,\Bigg](Z_{2})
\nonu\\
&&
\nonu \\ 
&&
+
\frac{\theta_{12}^{4-ij}}{z_{12}}\,\Bigg[\,
\frac{1}{6}\,\partial^{3} {\bf Q}^{(1),ij}_{1} 
+\frac{3}{10}\,\partial^{2} {\bf Q}^{(2),ij}_{1}
+\frac{2}{3}\,\partial {\bf Q}^{(3),ij}_{1}
+{\bf Q}^{(4),ij}_{1}
\,\Bigg](Z_{2})
\nonu \\ 
&&
+\frac{\theta_{12}^{4-i}}{z_{12}}\,\Bigg[\,
\frac{5}{24}\,\partial^{4} {\bf Q}^{(\frac{1}{2}),i}_{\frac{3}{2}}
+\frac{1}{6}\,\partial^{3} {\bf Q}^{(\frac{3}{2}),i}_{\frac{3}{2}}
+\frac{1}{3}\,\partial^{2} {\bf Q}^{(\frac{5}{2}),i}_{\frac{3}{2}}
+\frac{5}{7}\,\partial {\bf Q}^{(\frac{7}{2}),i}_{\frac{3}{2}}
+ {\bf Q}^{(\frac{9}{2}),i}_{\frac{3}{2}}
\,\Bigg](Z_{2})
\nonu \\ 
&&
+\frac{\theta_{12}^{4-0}}{z_{12}}\,\Bigg[\,
\frac{1}{8}\,\partial^{4} {\bf Q}^{(1)}_{2}
+\frac{1}{6}\,\partial^{3} {\bf Q}^{(2)}_{2}
+\frac{5}{14}\,\partial^{2} {\bf Q}^{(3)}_{2}
+\frac{3}{4}\,\partial {\bf Q}^{(4)}_{2}
+ {\bf Q}^{(5)}_{2}\,
\Bigg](Z_{2})
+\cdots.
\label{singleOPE}
\eea
The central term is presented in (\ref{central22}).
The maximum number of super spin is given by $5$
and the corresponding (composite)
higher spin currents appear in the last line
of (\ref{singleOPE}).
Due to the space of the paper, we cannot write down all the
operators in the right hand side of (\ref{singleOPE}).
The partial expressions of
quasi primary super fields corresponding to
the component fields in (\ref{quasicomponent})
are given in Appendix $C$ (together with ${\tt ancillary.nb}$).

Here we introduce the following quantities (${\cal N}=4$ expressions
of (\ref{bigR}), (\ref{Es}) and the one in the footnote
\ref{somerelation})
\footnote{
The  $p_1$ and $p_2$ are given in (\ref{p1p2p3p4})
and we also use, together with (\ref{central22}),
\bea
{\bf E}^{(2)}_{2}(Z_2)& \equiv &
(\,
4 {\bf Q}^{(2)}_{0}+ \frac{c^{0,4}_{0}}{2}{\bf J}^{4-0}
\,)(Z_2),
\qquad
{\bf E}^{(3)}_{2}(Z_2) \equiv 
(\,
\frac{5}{2} \, \pa \, {\bf Q}^{(2)}_{0}+
\frac{c^{0,4}_{0}}{2} \, \pa \, {\bf J}^{4-0}
\,)(Z_2),
\nonu\\
{\bf E}^{(4)}_{2}(Z_2) & \equiv &
(\,
\frac{1}{2}\,\partial^{2} {\bf Q}^{(2)}_{0}
+\frac{1}{4}\, {\bf J}^{4-0}  {\bf Q}^{(2)}_{0}
+\frac{1}{4}\,\partial^{2} {\bf J}^{4-0}
+2\,{\bf \Phi}^{(2)} {\bf\Phi}^{(2)}
\,)(Z_2),
\nonu
\eea
which are the supersymmetric extension of (\ref{Es}). Note that
there are some different numerical factors according to
(\ref{comptosuper}).}
with (\ref{alpha})
\bea
{\bf R}_{\frac{3}{2}}^{(\frac{7}{2}),i}(Z_2) & \equiv &
-\frac{2  \, \alpha}{5}\, (\frac{1}{4} \, \pa^2 \,
{\bf Q}_{\frac{1}{2}}^{(\frac{3}{2}),i} +\frac{3}{5} \, \pa \,
{\bf Q}_{\frac{1}{2}}^{(\frac{5}{2}),i} +
{\bf Q}_{\frac{1}{2}}^{(\frac{7}{2}),i})(Z_2),
\nonu \\
{\bf R}_{\frac{3}{2}}^{(\frac{5}{2}),i}(Z_2) & \equiv &
-\frac{4 \, \alpha}{5} \, (\frac{2}{3} \, \pa \,
{\bf Q}_{\frac{1}{2}}^{(\frac{3}{2}),i} +
{\bf Q}_{\frac{1}{2}}^{(\frac{5}{2}),i})(Z_2),
\nonu \\
{\bf R}_{\frac{3}{2}}^{(\frac{3}{2}),i}(Z_2) & \equiv &
-\frac{6 \, \alpha}{5} \, {\bf Q}_{\frac{1}{2}}^{(\frac{3}{2}),i}(Z_2),
\nonu \\
{\bf R}^{(4-n)}_{2}(Z_2)
& \equiv & - p_2 \,
{\bf E}^{(4-n)}_{2}(Z_2)-p_{1}\,n(n+1)\,{\bf Q}^{(4-n)}_{0}(Z_2),\qquad
n =0,1,2.
\nonu
\eea
Note that
${\bf R}_{\frac{3}{2}}^{(\frac{7}{2}),i}(Z_2)$,
${\bf R}_{\frac{3}{2}}^{(\frac{5}{2}),i}(Z_2)$,
${\bf R}_2^{(3)}(Z_2)$ and ${\bf R}_2^{(4)}(Z_2)$ are not quasi primary. 
From (\ref{Q22}), (\ref{Q3alpha}), (\ref{Q7half}),
(\ref{finalspin4}) and (\ref{Q9half}),
the new primary higher spin ${
  \cal N}=4$ multiplets in (\ref{singleOPE})
arise in the following quasi primary ones
\bea
{\bf Q}^{(2)}_{2}(Z_2) &  = &
c_{1}^{2,4}\,{\bf \Phi}^{(2)}(Z_2)
+c_{2}^{2,4}\,{\bf X}^{(2)}(Z_2) + \cdots, \nonu \\
 {\bf Q}^{(3),ij}_{1}(Z_2)  & = & 
c_{0}^{1,2}\,(M^{\alpha})^{ij}{\bf \Phi}^{(3),\alpha}(Z_2) + \cdots,
\nonu \\
 {\bf Q}^{(\frac{7}{2}),i}_{\frac{1}{2}}(Z_2) & = &
c_{0}^{\frac{1}{2},1}\,{\bf \Phi}^{(\frac{7}{2}),i}(Z_2)
+\cdots,
\nonu \\
{\bf Q}^{(4)}_{2}(Z_2)  & = & 
c_{0}^{2,2}\,{\bf \Phi}^{(4)}(Z_2) + \cdots,
\nonu \\
 {\bf Q}^{(\frac{9}{2}),i}_{\frac{3}{2}}(Z_2)  & = & 
c_{0}^{\frac{3}{2},1}\,{\bf \Phi}^{(\frac{9}{2}),i}(Z_2)
+\cdots,
\label{superQ}
\eea
where
some of the abbreviated parts are given in Appendix $C$. 
Its component relations can be found in Appendix $B$.
We can easily figure out the $SO(4)$ indices and
it is rather nontrivial to observe the second quasi primary
field of (\ref{superQ})
where $SO(4)$ adjoint index is contracted with the one in
the matrix $M$ (\ref{Malpha}).

Schematically, we can present the above OPE as follows:
\bea
    [{\bf \Phi}^{(2)} \cdot {\bf \Phi}^{(2)}] &=& [{\bf I}] +
   \theta^{4-0}\, ([{\bf  \Phi}^{(2)}] +  [{\bf  X}^{(2)}]) +
    \theta^{4-ij}\, (M^{\alpha})^{ij} \,
          [{\bf  \Phi}^{(3),\alpha}] + \theta^{i}\,
    [{\bf  \Phi}^{(\frac{7}{2}),i}]+
    \theta^{4-0}\, [{\bf  \tilde{\Phi}}^{(4)}] \nonu \\
    & + & \theta^{4-i}\,
      [{\bf  \Phi}^{(\frac{9}{2}),i}],
        \label{PhiPhi}
\eea
where $[{\bf I}]$ stands for the large
${\cal N}=4$ linear superconformal family of identity
operator and
the various composite fields consisting of ${\cal N}=4$ multiplet
${\bf J}$ up to the super spin $5$ can appear.
We also insert the $SO(4)$ vector indices $i, j$
and the $SO(4)$ matrix $M^{\alpha}$ in (\ref{Malpha}).
Note that in the component approach
described in previous subsections, the
$SO(4)$ indices are present in the two higher spin currents
of the left hand side of the given OPE.
In the ${\cal N}=4$ superspace description (\ref{PhiPhi}),
all the $SO(4)$ indices are contracted with the ones
in the fermionic coordinates.
In order to obtain the component results, we can
act various super derivatives both sides of the ${\cal N}=4$ OPE
to restore the $SO(4)$ indices.
Here the ${\cal N}=4$  multiplet ${\bf  \tilde{\Phi}}^{(4)}(Z)$
has its lowest component given in (\ref{finalspin4})
and contains the quadratic ${\cal N}=4$ multiplet
${\bf \Phi}^{(2)}(Z)$.
Compared with the ${\cal N}=3$ example \cite{AK1607},
there are more higher spin multiplets contracted with fermionic
coordinates. 

\subsection{The $136$
  OPEs between the $16$ lowest higher spin currents
for generic $N$}

We can calculate the remaining $136-16=120$ OPEs
from (\ref{singleOPE}) by taking the super derivatives
$D^i_1$ or $D^j_2$ both sides of
(\ref{singleOPE}) and putting $\theta_1^k=0=\theta^l_2$.
By introducing the various coefficients in front of
composite fields appearing in $136$ OPEs (the number of coefficients
is $2000$ or so and coefficients are denoted by
$w_{1,s}, \cdots,
w_{2043,s}$ in
Appendix $B$) and using the Jacobi
identities, we obtain
the $136$
  OPEs between the $16$ lowest higher spin currents
  (\ref{Phiexp}) for generic $N$.
  Note that this number of coefficients is huge
  compared to the unitary case in \cite{AK1509}.
Furthermore, after substituting these coefficients into
(\ref{singleOPE}) back,
we obtain the final single OPE with fixed coefficients
which depend on $N$ and $k$ (as well as $C_{(4)(2)}^{(2)}$)
\footnote{The field contents  appearing in the
  right hand side of (\ref{singleOPE}) are taken from
  the results for $N=5$ case. We also have checked that
  for $N=8,9$, the six $SO(4)$ adjoint higher spin-$3$ multiplets
  appear in the corresponding OPEs
  similarly and there are no new higher spin multiplets having
  super spins $2$ and $\frac{5}{2}$. If there exist the
  extra ${\cal N}=4$ higher spin multiplets in the right hand side of
  (\ref{singleOPE}) for large $N$, we expect that
  they will appear linearly without spoiling its algebraic structure.
  We may try to calculate the OPEs from the closed forms written
  in terms of the orthogonal Wolf space fields but this will be rather
  involved.}.
Because there are new $16$ ${\cal N}=4$ multiplets
(${\bf X}^{(2)}(Z)$, ${\bf \Phi}^{(3),\alpha}(Z)$,
${\bf \Phi}^{(\frac{7}{2}),\mu}(Z)$, ${\bf \Phi}^{(4)}(Z)$
and ${\bf \Phi}^{(\frac{9}{2}),\mu}(Z)$)
in the right hand side
of the $136$ OPEs, we have $16 \times 16=256$ higher spin
currents (in components)
totally. Under the $16$ currents of the large ${\cal N}=4$
superconformal algebra,
they transform nontrivially as in Appendix $A$.
It is an open problem to determine
the OPEs between these $256$ higher spin currents
(or the OPEs between the lowest $16$ higher spin currents
and those $256$ higher spin currents) systematically.

\section{Conclusions and outlook }

We have described one single ${\cal N}=4$ super
OPE (\ref{singleOPE}) between 
the lowest higher spin-$2$ multiplet in the
${\cal N}=4$ superspace.
As in the abstract, there exist several
${\cal N}=4$ higher spin multiplets
in the right hand side of this OPE. 

There are open problems we can consider in the future as follows:

$\bullet$ Higher spin algebra in the bulk theory

In \cite{EGR}, the free field construction at $\la =0$
by using the bosons and fermions is presented.
Maybe at this particular $\la=0$ case, the full higher spin algebra
can be described. In other words, the commutators and
anticommutators for the higher spin currents (including the $16$
currents) can be determined with complete structure constants.
The final goal is to obtain the higher spin algebra at finite
$\la$ which will provide the corresponding algebra in the
dual conformal field theory at the classical level.
Contrary to the unitary case,
the orthogonal case needs to obtain the appropriate truncation on
the matrix elements observed in \cite{EGR}.  

$\bullet$ Three-point functions

One way to check the dual relation between
the orthogonal Wolf space coset model and
the higher spin theory on $AdS_3$ space is to compute the
three-point functions of the two scalars and the higher spin
currents.
According to the results of this paper,
there are many higher spin currents from the
single OPE (\ref{singleOPE}).
It is an open problem to obtain the remaining $15$ higher spin
currents in terms of the orthogonal Wolf space coset fields
explicitly and to calculate the eigenvalues of the zero modes,
by following the procedure studied in
\cite{Ahn1805,Ahn1711,AK1506,AK1308,Ahn1111}.
Although this will be rather involved, once we obtain them, then
it is straightforward to compute the three-point functions
at finite $N$ and $k$.

$\bullet$ ${\cal N}=2$ superspace description

In principle, we can rewrite the above ${\cal N}=4$ superspace OPE
in terms of various $10$ OPEs in ${\cal N}=2$ superspace.
One merit for this description is that
contrary to the ${\cal N}=4$ superspace OPE we have described so far,
the ${\cal N}=2$ superspace description enables us to write down
in terms of quasi (super) primary fields completely.
In doing this, it is rather nontrivial
to obtain the correct component fields
for the $SO(4)$ nonsinglet ${\cal N}=4$ multiplets. 
The relevant work in this direction appeared in \cite{AK1607}. 

$\bullet$  The large $k$ limit

We can examine the behavior of large $k$ limit (for example,
see the work of \cite{FG}) from the 
results we have obtained in this paper. We take the large $k$ limit
in the structure constants appearing in the right hand sides of the
OPEs we have found.
We can read off the leading behavior of $k$ of the right hand sides. 
Even the $N=5$ results are enough to analyze this large $k$ limit.
We expect to observe the realization of vanishing of 't Hooft-like
coupling constant $\la =\frac{(N+1)}{(N+k+2)} \rightarrow 0$
for fixed $N$ \cite{GG1406}. It is also interesting to
observe whether there is an extension of the small ${\cal N}=4$
superconformal algebra along the line of \cite{GG1406} or not.

$\bullet $ The nonlinear version 

It is an open problem to
obtain the above single OPE (\ref{singleOPE})
in the context of nonlinear version which is an extension of
the large ${\cal N}=4$ nonlinear algebra.
Due to the fact that we do not know its ${\cal N}=4$ superspace
version, we need to present the whole $136$ OPEs.   
Although the lowest $16$ higher spin currents are primary
under the corresponding stress energy tensor, they do transform
nontrivially with respect to other $10$ currents.
In principle, because we do have the complete OPE
in the linear version, it is straightforward to obtain
them  in the nonlinear version although the careful
analysis should be done. In this paper, we applied some computations
in this nonlinear basis, although we did not present them
explicitly (some OPEs are rather complicated). 

\vspace{.7cm}

\centerline{\bf Acknowledgments}

CA acknowledges warm hospitality from 
the School of  Liberal Arts (and Institute of Convergence Fundamental
Studies), Seoul National University of Science and Technology.
This research was supported by Basic Science Research Program through
the National Research Foundation of Korea  
funded by the Ministry of Education  
(No. 2017R1D1A1A09079512).

\newpage

\appendix

\renewcommand{\theequation}{\Alph{section}\mbox{.}\arabic{equation}}

\section{The OPEs between the $16$ currents and the (non)singlet
  higher spin
  currents in the component approach }

We present the component results \cite{Schoutensnpb}
for the ${\cal N}=4$ primary
condition for the $SO(4)$ nonsinglet field in (\ref{JPhi})
\bea
L(z)\,\Phi_{2}^{(s),\alpha}(w) & = & 
-\frac{1}{(z-w)^{4}}\,12\: s\:{ \alpha}\,\Phi_{0}^{(s),\alpha}(w)
+  
\frac{1}{(z-w)^{3}}\Bigg[4\:{ \alpha}\,\partial\Phi_{0}^{(s),\alpha}
+i\, ({ T}^{ij}_{L})^{\alpha \beta}  \,\Phi_{1}^{(s),ij,\beta}
\Bigg](w)
\nonu\\
& + & 
\frac{1}{(z-w)^{2}}\,(s+2)\,\Phi_{2}^{(s),\alpha}(w)+\frac{1}{(z-w)}\,\partial\Phi_{2}^{(s),\alpha}(w)
+\cdots,
\nonu \\
L(z)\,\Phi_{\frac{3}{2}}^{(s),i,\alpha}(w) & = & 
\frac{1}{(z-w)^{3}}2\Bigg[\alpha\,\Phi_{\frac{1}{2}}^{(s),i,\alpha}-i\,({ T}^{ij}_{R})^{\alpha \beta}\,\Phi_{\frac{1}{2}}^{(s),j,\beta}
\Bigg](w)
+  
\frac{1}{(z-w)^{2}}\,(s+\frac{3}{2})\,\Phi_{\frac{3}{2}}^{(s),i,\alpha}(w)
\nonu \\
& + & \frac{1}{(z-w)}\,\partial\Phi_{\frac{3}{2}}^{(s),i,\alpha}(w)+\cdots,
\nonu \\
L(z)\,\Phi_{1}^{(s),ij,\alpha}(w)  & = & 
\frac{1}{(z-w)^{3}}\,2\,i\,({ T}^{ij}_{R})^{\alpha \beta}\,\Phi_{0}^{(s),\beta}(w)
+  \frac{1}{(z-w)^{2}}\,(s+1)\,\Phi_{1}^{(s),ij,\alpha}(w)
\nonu \\
& + & \frac{1}{(z-w)}\,\partial\Phi_{1}^{(s),ij,\alpha}(w)+\cdots,
\nonu \\
L(z) \, \Phi_{\frac{1}{2}}^{(s),i,\alpha}(w) & = &
\frac{1}{(z-w)^{2}}\, (s+\frac{1}{2})\, \Phi_{\frac{1}{2}}^{(s),i,\alpha}(w)
+\frac{1}{(z-w)} \, \partial\Phi_{\frac{1}{2}}^{(s),i,\alpha}(w)
+\cdots,
\nonu \\
L(z) \, \Phi_{0}^{(s),\alpha}(w) & = & 
\frac{1}{(z-w)^{2}} \, s\, \Phi_{0}^{(s),\alpha}(w)+
\frac{1}{(z-w)} \, \partial\Phi_{0}^{(s),\alpha}(w)
+\cdots, \nonu \\
G^{i}(z)\,\Phi_{2}^{(s),\alpha}(w) 
& = & 
-\frac{1}{(z-w)^{3}}\,4\Bigg[(1+2s)\alpha\,\Phi_{\frac{1}{2}}^{(s),i,\alpha}
-i\,({ T}^{ij}_{R})^{\alpha \beta}\,\Phi_{\frac{1}{2}}^{(s),j,\beta}
\Bigg](w)
\nonu \\
& - & 
\frac{1}{(z-w)^{2}}\,\Bigg[(3+2s)\Phi_{\frac{3}{2}}^{(s),i,\alpha}-2\,\alpha\,\partial\Phi_{\frac{1}{2}}^{(s),i,\alpha}
-2i\,({ T}^{ij}_{L})^{\alpha \beta}\,\Phi_{\frac{3}{2}}^{(s),j,\beta}\Bigg](w)
\nonu \\
& - & \frac{1}{(z-w)}\,\partial\Phi_{\frac{3}{2}}^{(s),i,\alpha}(w)+\cdots,
\nonu \\
G^{i}(z)\,\Phi_{\frac{3}{2}}^{(s),j,\alpha}(w) 
& = & 
\frac{1}{(z-w)^{3}}\Bigg[8\: s\:\alpha\:\delta_{ij}\,\Phi_{0}^{(s),\alpha}
-4\, i\,({ T}^{ij}_{R})^{\alpha \beta}\,\Phi_{0}^{(s),\beta}
\Bigg](w) \nonu \\
& - &  
\frac{1}{(z-w)^{2}}\,\Bigg[2(1+s)\,\Phi_{1}^{(s),ij,\alpha}+\varepsilon_{ijkl}(\alpha\,\Phi_{1}^{(s),kl,\alpha}
-2\, i\,({ T}^{jk}_{L})^{\alpha \beta}\,\Phi_{1}^{(s),jl,\beta}
)
\nonu \\
& + & 
2\,\delta_{ij}(\alpha\,\partial\Phi_{0}^{(s),\alpha}
-i\,({ T}^{ik}_{R})^{\alpha \beta}\,\Phi_{1}^{(s),ik,\beta}
)\,\Bigg](w)
-  
\frac{1}{(z-w)}\,\Bigg[\partial\Phi_{1}^{(s),ij,\alpha}+\delta_{ij}\,\Phi_{2}^{(s),\alpha}\Bigg](w) \nonu \\
& + & \cdots,
\nonu \\
G^{i}(z)\,\Phi_{1}^{(s),jk,\alpha}(w) 
& = & 
-\frac{1}{(z-w)^{2}}\,\Bigg[-2\,\delta_{ij}(\,{ \alpha}\,
  \Phi_{\frac{1}{2}}^{(s),k, \alpha}
+\varepsilon^{ikpq}\,({ T}^{ip}_{L})^{\alpha \beta}\,\Phi_{\frac{1}{2}}^{(s),q,\beta}\,)
\nonu \\
& + & 
2\,\delta_{ik}(j\leftrightarrow k)+\varepsilon^{ijkl}((1+2s)\,\Phi_{\frac{1}{2}}^{(s),l,\alpha}
+2i\,({ T}^{il}_{L})^{\alpha\beta}\,\Phi_{\frac{1}{2}}^{(s),i,\beta})\Bigg](w)
\nonu \\
& - & 
\frac{1}{(z-w)}\,\Bigg[(\delta_{ij}\,\Phi_{\frac{3}{2}}^{(s),k,\alpha}
  -\delta_{ik}\,\Phi_{\frac{3}{2}}^{(s),j,\alpha})
  +\partial\Phi_{\frac{1}{2}}^{(s),4-ijk,\alpha}\Bigg](w)+\cdots,
\nonu \\
G^{i}(z)\,\Phi_{\frac{1}{2}}^{(s),j,\alpha}(w) 
& = & 
-\frac{1}{(z-w)^{2}}\Bigg[2\,s\:\delta_{ij}\,\Phi_{0}^{(s),\alpha}-2\,i\,
  ({ T}^{ij}_{L})^{\alpha \beta}\,\Phi_{0}^{(s),\beta}\,\Bigg](w)
\nonu \\
& - & 
\frac{1}{(z-w)}\,\Bigg[\delta_{ij}\,\partial\Phi_{0}^{(s),\alpha}-\Phi_{1}^{(s),4-ij,\alpha}\Bigg](w)+\cdots,
\nonu \\
G^{i}(z) \, \Phi_{0}^{(s),\alpha}(w) & = & 
-\frac{1}{(z-w)} \, \Phi_{\frac{1}{2}}^{(s),i,\alpha}(w)
+\cdots, \nonu \\
T^{ij}(z)\,\Phi_{2}^{(s),\alpha}(w) & = & 
-\frac{1}{(z-w)^{3}}\,2\,({ \widetilde{T}}^{ij})^{\alpha\beta}\,\Phi_{0}^{(s),\beta}(w)
\nonu \\
& + &
\frac{1}{(z-w)^{2}}\Bigg[2\, i\,(s+1)\,\Phi_{1}^{(s),ij,\alpha}-({ T}^{ik})^{\alpha\beta}\,\Phi_{1}^{(s),jk,\beta}
+({ T}^{jk})^{\alpha\beta}\,\Phi_{1}^{(s),ik,\beta}\Bigg](w)
\nonu \\
& + &
\frac{1}{(z-w)}\,({ T}^{ij})^{\alpha\beta}\,\Phi_{2}^{(s),\beta}(w)+\cdots,
\nonu \\
T^{ij}(z)\,\Phi_{\frac{3}{2}}^{(s),k,\alpha}(w) & = &
\frac{1}{(z-w)^{2}}\Bigg[\varepsilon_{ijkl}(-i\,(2s+1)\,\Phi_{\frac{1}{2}}^{(s),l,\alpha}
-({ T}^{li})^{\alpha\beta}\,\Phi_{\frac{1}{2}}^{(s),i,\beta}-({ T}^{lj})^{\alpha\beta}\,\Phi_{\frac{1}{2}}^{(s),j,\beta})
\nonu \\
& + &
\varepsilon_{ijpq}(\delta_{ik}\,({ T}^{ip})^{\alpha\beta}\,\Phi_{\frac{1}{2}}^{(s),q,\beta}+\delta_{jk}\,({ T}^{jp})^{\alpha\beta}\,\Phi_{\frac{1}{2}}^{(s),q,\beta})\Bigg](w)
\nonu \\
& - &
\frac{1}{(z-w)}\,\Bigg[\delta_{ik}(i\,\Phi_{\frac{3}{2}}^{(s),j,\alpha}
-({ T}^{ij})^{\alpha\beta}\,\Phi_{\frac{3}{2}}^{(s),i,\beta})
-\delta_{jk}(i\,\Phi_{\frac{3}{2}}^{(s),i,\alpha}
-({ T}^{ij})^{\alpha\beta}\,\Phi_{\frac{3}{2}}^{(s),j,\beta})
\nonu \\
& - &
\varepsilon_{ijkl}\,({ \widetilde{T}}^{kl})^{\alpha\beta}\,\Phi_{\frac{3}{2}}^{(s),k,\beta}\Bigg](w)+\cdots,
\nonu \\
T^{ij}(z)\,\Phi_{1}^{(s),kl,\alpha}(w)
& = & 
\frac{1}{(z-w)^{2}}\Bigg[2\: i\: s\,\varepsilon_{ijkl}\,\Phi_{0}^{(s),\alpha}
+i\delta_{ik}\,({ \widetilde{T}}^{jl})^{\alpha\beta}\,\Phi_{0}^{(s),\beta}
-i\delta_{il}\,({ \widetilde{T}}^{jk})^{\alpha\beta}\,\Phi_{0}^{(s),\beta}
\nonu \\
& - & 
i\delta_{jk}\,({ \widetilde{T}}^{il})^{\alpha\beta}\,\Phi_{0}^{(s),\beta}
+i\delta_{jl}\,({ \widetilde{T}}^{ik})^{\alpha\beta}\,\Phi_{0}^{(s),\beta}\Bigg](w)
-\frac{1}{(z-w)}\,\Bigg[i\delta_{ik}\,\Phi_{1}^{(s),jl,\alpha}
\nonu \\
& - & 
i\delta_{il}\,\Phi_{1}^{(s),jk,\alpha}-i\delta_{jk}\,\Phi_{1}^{(s),il,\alpha}+i\delta_{jl}\,\Phi_{1}^{(s),ik,\alpha}-({ T}^{ij})^{\alpha\beta}\,\Phi_{1}^{(s),kl,\beta}\Bigg](w)+\cdots,
\nonu \\
T^{ij}(z)\;\Phi_{\frac{1}{2}}^{(s),k,\alpha}(w)
& = & 
\frac{1}{(z-w)}\Bigg[-i\delta^{ik}\,\Phi_{\frac{1}{2}}^{(s),j,\alpha}+i\delta^{jk}\,\Phi_{\frac{1}{2}}^{(s),i,\alpha}
+({ T}^{ij})^{\alpha\beta}\,\Phi_{\frac{1}{2}}^{(s),k,\beta}\Bigg](w)+\cdots,
\nonu \\
T^{ij}(z)\,\Phi_{0}^{(s),\alpha}(w)
& = & 
\frac{1}{(z-w)}\,({ T}^{ij})^{\alpha\beta}\,\Phi_{0}^{(s),\beta}(w)+\cdots,
\nonu \\
U(z)\,\Phi_{2}^{(s),\alpha}(w) 
& = & 
-\frac{1}{(z-w)^{3}}\,4\: s\,\Phi_{0}^{(s),\alpha}(w)+\frac{1}{(z-w)^{2}}\Bigg[2\,\partial\Phi_{0}^{(s),\alpha}+\frac{i}{2}\,({ \widetilde{T}}^{ij})^{\alpha\beta}\Phi_{1}^{(s),ij,\beta}\Bigg](w) \nonu \\
& + & \cdots,
\nonu \\
U(z)\,\Phi_{\frac{3}{2}}^{(s),i,\alpha}(w) 
& = & 
\frac{1}{(z-w)^{2}}\Bigg[\Phi_{\frac{1}{2}}^{(s),i,\alpha}
+i\,({ T}^{ij})^{\alpha\beta}\Phi_{\frac{1}{2}}^{(s),j,\beta}\Bigg](w)+\cdots,
\nonu \\
U(z)\,\Phi_{1}^{(s),ij,\alpha}(w) & = &  \frac{1}{(z-w)^{2}}\, i\:({ T}^{ij})^{\alpha\beta}\Phi_{0}^{(s),\beta}(w)+\cdots,
\nonu \\
\Gamma^{i}(z)\,\Phi_{2}^{(s),\alpha}(w) & = & -\frac{1}{(z-w)^{2}}\Bigg[i\:(2s+1)\,\Phi_{\frac{1}{2}}^{(s),i,\alpha}+({ T}^{ij})^{\alpha\beta}\,\Phi_{\frac{1}{2}}^{(s),j,\beta}\Bigg](w)
\nonu \\& + &
\frac{1}{(z-w)}\Bigg[\,i\,\partial\Phi_{\frac{1}{2}}^{(s),i,\alpha}
-({ \widetilde{T}}^{ij})^{\alpha\beta}\,\Phi_{\frac{3}{2}}^{(s),j,\beta}\Bigg](w)+\cdots,
\nonu \\
\Gamma^{i}(z)\,\Phi_{\frac{3}{2}}^{(s),j,\alpha}(w) 
& = &
\frac{1}{(z-w)^{2}}\Bigg[2\: i\: s\:\delta_{ij}\,\Phi_{0}^{(s),\alpha}+({ T}^{ij})^{\alpha\beta}\,\Phi_{0}^{(s),\beta}\Bigg](w)\nonu \\ 
& - &
\frac{1}{(z-w)}\,\Bigg[\delta_{ij}(\,i\,\partial\Phi_{0}^{(s),\alpha}-
  ({ \widetilde{T}}^{ik})^{\alpha\beta}\,\Phi_{1}^{(s),ik,\beta})
\nonu \\ 
& + &\varepsilon_{ijkl}(\,
\frac{i}{2}\,\Phi_{1}^{(s),kl,\alpha}
-({ T}^{jk})^{\alpha\beta}\,\Phi_{1}^{(s),jl,\beta}
\,)\Bigg](w)+\cdots,
\nonu \\
\Gamma^{i}(z)\,\Phi_{1}^{(s),jk,\alpha}(w) 
& = &
-\frac{1}{(z-w)}\,\Bigg[\delta_{ij}(i\,\Phi_{\frac{1}{2}}^{(s),k,\alpha}-
  \varepsilon_{iklq}({ \widetilde{T}}^{il})^{\alpha\beta}\,\Phi_{\frac{1}{2}}^{(s),q,\beta})
-\delta_{ik}(j\leftrightarrow k)
\nonu \\ 
& + &
\varepsilon_{ijkl}({ \widetilde{T}}^{il})^{\alpha\beta}\,\Phi_{\frac{1}{2}}^{(s),i,\beta}\Bigg](w)+\cdots,
\nonu \\
\Gamma^{i}(z)\,\Phi_{\frac{1}{2}}^{(s),j,\alpha}(w)
 & = & -\frac{1}{(z-w)}\,({ \widetilde{T}}^{ij})^{\alpha\beta}\Phi_{0}^{(s),\beta}(w)+\cdots, \qquad \alpha = \mbox{singlet, \,\, adjoint,\,\, vector}.
\nonu
\eea
We use the following simplified notations (\ref{alpha}) and
(\ref{threeT})
\bea
\widetilde{{ T}}^{ij} & \equiv &
\frac{1}{2!}\:\varepsilon_{ijkl}\:{ T}^{kl}, \qquad
{ T}^{ij}_{L}  \equiv 
    \frac{1}{2}({ T}^{ij})+{ \alpha}\,({ \widetilde{T}}^{ij}),
\qquad
    { T}^{ij}_{R}  \equiv 
    { \alpha}\,({ T}^{ij})+\frac{1}{2}\,({ \widetilde{T}}^{ij}),
\nonu\\
    { \alpha} & \equiv &
    \frac{1}{2}\frac{(k^{+}-k^{-})}{(k^{+}+k^{-})}, \qquad k^+
    \equiv k+1, \qquad k^- \equiv N+1.\:
\label{newdef}
\eea
Note that there are trivial OPEs
$U(z) \, \Phi_{\frac{1}{2}}^{(s),\alpha, i}(w) = + \cdots =
U(z) \, \Phi_{0}^{(s),\alpha}(w)= \Gamma^i(z) \, \Phi_{0}^{(s),\alpha}(w)$.
For fixed indices $i,j,k,l$
appearing in the left hand side, we do not sum over those
in the right hand side but we sum over other indices.
For example, in the OPE $G^{i}(z)\,\Phi_{\frac{3}{2}}^{(s),j,\alpha}(w)$,
the third term of the second order pole has indices
$i,j,k,l$. Among them, the only indices $k,l$ are summed.

\section{Partial expressions of quasi primary fields
in the component approach}

We present the various quasi primary fields appearing in section $4$
(the complete expressions can be found in $\tt{ancillary.nb}$ file)
\bea
 Q^{(2)}_{0}  & = & 
w_{1,2}\,\Phi^{(2)}_{0}
+w_{2,2}\,\tilde{L}
+w_{3,2}\,\partial U
+w_{4,2}\,U U
+w_{5,2}\,T^{ij}T^{ij}
+w_{6,2}\,T^{ij}\tilde{T}^{ij}
+w_{7,2}\,\Gamma^{i}\Gamma^{j}T^{ij}
\nonu\\
& + & 
w_{8,2}\,\Gamma^{i}\Gamma^{j} \tilde{T}^{ij}
+w_{9,2}\,\Gamma^{i}  \partial \Gamma^{i}
+\varepsilon^{ijkl}\,
w_{10,2}\,\Gamma^{i}\Gamma^{j}\Gamma^{k}\Gamma^{l},
\nonu \\
Q^{(\frac{3}{2}),i}_{\frac{1}{2}} & = &
w_{1,\frac{3}{2}}\,\tilde{G}^{i}
+w_{2,\frac{3}{2}}\,\partial\Gamma^{i}
+w_{3,\frac{3}{2}}\,\Gamma^{i}U+\varepsilon^{ijkl}(\,
w_{4,\frac{3}{2}}\,\Gamma^{j}T^{kl}
+w_{5,\frac{3}{2}}\,\Gamma^{j}\Gamma^{k}\Gamma^{l}),
\nonu \\ 
Q^{(\frac{5}{2}),i}_{\frac{1}{2}} & = &
w_{1,\frac{5}{2}}\,\Phi_{\frac{1}{2}}^{(2),i}
+w_{2,\frac{5}{2}}\,\partial\tilde{G}^{i}
+w_{3,\frac{5}{2}}\,\partial^{2}\Gamma^{i}
+w_{4,\frac{5}{2}}\,\Gamma^{i}\partial U
+w_{5,\frac{5}{2}}\,\partial\Gamma^{i}U
+w_{6,\frac{5}{2}}\,T^{ij}\tilde{G}^{i}
\nonu \\ 
&+ &
w_{7,\frac{5}{2}}\,\Gamma^{i}\Gamma^{j}\tilde{G}^{j}
+w_{8,\frac{5}{2}}\,\partial\Gamma^{j}T^{ij}
+w_{9,\frac{5}{2}}\,\Gamma^{j}\partial T^{ij}
+w_{10,\frac{5}{2}}\,\Gamma^{j}T^{ij}U
+w_{11,\frac{5}{2}}\,\Gamma^{i}\Gamma^{j}\partial\Gamma^{j}
\nonu \\ 
&
+ & w_{12,\frac{5}{2}}\,\Gamma^{i}T^{jk}T^{jk}
+w_{13,\frac{5}{2}}\,\Gamma^{i}\Gamma^{j}\Gamma^{k}T^{jk}
+w_{14,\frac{5}{2}}\,\Gamma^{j}T^{ik}T^{jk}
\nonu \\
& + & \varepsilon^{ijkl}(\,
w_{15,\frac{5}{2}}\,T^{jk}\tilde{G^{l}}
+w_{16,\frac{5}{2}}\,\Gamma^{j}\Gamma^{k}\tilde{G}^{l}
\nonu \\ 
&
+ & w_{17,\frac{5}{2}}\,\partial\Gamma^{j}T^{kl}
+w_{18,\frac{5}{2}}\,\Gamma^{j}T^{kl}U
+w_{19,\frac{5}{2}}\,\Gamma^{i}T^{ij}T^{kl}
+w_{20,\frac{5}{2}}\,\partial(\Gamma^{j}\Gamma^{k}\Gamma^{l})
+w_{21,\frac{5}{2}}\,\Gamma^{j}\Gamma^{k}\Gamma^{l}U\,),
\nonu\\
 Q^{(\frac{7}{2}),i}_{\frac{1}{2}} & = &
(M^{\alpha})^{ij}\,
(\,
w_{1,\frac{7}{2}}\,\Phi^{(3),j,\alpha}_{\frac{1}{2}}
+w_{2,\frac{7}{2}}\,\Gamma^{j}\Phi^{(3),\alpha}_{0}
\,)
+ \varepsilon^{ijkl}(M^{\alpha})^{jk}(\,
w_{3,\frac{7}{2}}\,\Phi^{(3),l,\alpha}_{\frac{1}{2}}
+w_{4,\frac{7}{2}}\,\Gamma^{l} \Phi^{(3),\alpha}_{0}
\,)
\nonu \\ 
&
+ & w_{0,\frac{7}{2}}\,\delta^{i}_{\mu}\,\Phi^{(\frac{7}{2}),\mu}_{0}
+w_{5,\frac{7}{2}}\,\Phi^{(2),i}_{\frac{3}{2}}
+w_{6,\frac{7}{2}}\, \partial \Phi^{(2),i}_{\frac{1}{2}}
+w_{7,\frac{7}{2}}\, U \Phi^{(2),i}_{\frac{1}{2}}
+w_{8,\frac{7}{2}}\, \tilde{G}^{i} \Phi^{(2)}_{0}
+\cdots
\nonu \\ 
&
+& w_{19,\frac{7}{2}}\,X^{(2),i}_{\frac{3}{2}}
+w_{20,\frac{7}{2}}\, \partial X^{(2),i}_{\frac{1}{2}}
+w_{21,\frac{7}{2}}\, U X^{(2),i}_{\frac{1}{2}}
+w_{22,\frac{7}{2}}\, \tilde{G}^{i} X^{(2)}_{0}
+\cdots
\nonu \\ 
&+ &
w_{32,\frac{7}{2}}\, \varepsilon^{ijkl}\, \Gamma^{j}\Gamma^{k}\Gamma^{l}X^{(2)}_{0}
+w_{33,\frac{7}{2}}\, \tilde{G}^{i}  \tilde{L}
+\cdots
+w_{122,\frac{7}{2}}\,\varepsilon^{ijkl}\, \Gamma^{j}T^{ij}T^{ik}T^{lj}
\nonu \\ 
&+ &
w_{123,\frac{7}{2}}\,(\varepsilon^{ijkl})^{2}\,\Gamma^{j}\Gamma^{k}\Gamma^{l}T^{ij}T^{kl},
\nonu \\
Q^{(1),ij}_{1} & = &
w_{1,1}\, T^{ij}
+w_{2,1}\,\Gamma^{i}\Gamma^{j}
+\varepsilon^{ijkl}(\,
w_{3,1}\, T^{kl}
+w_{4,1}\,\Gamma^{k}\Gamma^{l}\,)
-(i\leftrightarrow j),
\nonu \\ 
Q^{(2),ij}_{1}  & = &
w_{1,2}\,\Gamma^{i}\tilde{G}^{j}
+w_{2,2}\,\Gamma^{i}\Gamma^{j}U
+w_{3,2}\,\partial(\Gamma^{i}\Gamma^{j})
+w_{4,2}\,\Gamma^{i}\Gamma^{k}\tilde{T}^{jk}
-(i\leftrightarrow j),
\nonu \\ 
Q^{(3),ij}_{1} & = &
w_{0,3}\,(M^{\alpha})^{ij}\Phi_{0}^{(3),\alpha}
+w_{1,3}\,\Phi_{1}^{(2),ij}
+w_{5,3}\,\tilde{\Phi}_{1}^{(2),ij}
+w_{2,3}\,\Gamma^{i} \Phi_{\frac{1}{2}}^{(2),j}
+w_{3,3}\,T^{ij} \Phi_{0}^{(2)}
\nonu \\ 
&
+ & w_{4,3}\,\Gamma^{i}\Gamma^{j} \Phi_{0}^{(2)}
+\cdots
+w_{9,3}\,X_{1}^{(2),ij}
+w_{13,3}\,\tilde{X}_{1}^{(2),ij}
+w_{10,3}\,\Gamma^{i} X_{\frac{1}{2}}^{(2),j}
+w_{11,3}\,T^{ij} X_{0}^{(2)}
+\cdots
\nonu \\ 
& + & 
w_{16,3}\,\tilde{T}^{ij} X_{0}^{(2)}
+w_{17,3}\,\tilde{G}^{i}\tilde{G}^{j}
+w_{18,3}\,\Gamma^{i}\Gamma^{i}\tilde{L}+\cdots
+w_{91,3}\,(\varepsilon^{ijkl})^{2}\,\Gamma^{k}\Gamma^{l}T^{ik}T^{jl}
-(i\leftrightarrow j),
\nonu \\ 
Q^{(4),ij}_{1} & = &
(M^{\alpha})^{ik}(\, 
w_{1,4}\,\Phi_{1}^{(3),jk,\alpha}
+w_{2,4}\, T^{jk}\Phi_{0}^{(3),\alpha}
+w_{3,4}\,\tilde{T}^{jk}\Phi_{0}^{(3),\alpha}\,
+ w_{4,4}\,\Gamma^{j}\Phi_{\frac{1}{2}}^{(3),k,\alpha}
+\cdots\,)
\nonu\\ 
& + & \varepsilon^{ijkl}(\,
(M^{\alpha})^{kq}\, w_{5,4}\,\Phi_{1}^{(3),lq,\alpha}
+(M^{\alpha})^{ik}\, w_{6,4}\,\Gamma^{i}\Phi_{\frac{1}{2}}^{(3),l,\alpha}
+(M^{\alpha})^{kl}\, w_{7,4}\,\Gamma^{k}\Phi_{\frac{1}{2}}^{(3),k,\alpha}+\cdots\,)
\nonu\\
& + & \delta_{\mu}^{j}\, w_{13,4}\,\Phi_{\frac{1}{2}}^{(\frac{7}{2}),i,\mu}
+w_{14,4}\,\partial \Phi_{1}^{(2),ij}
+w_{15,4}\,\partial(\Gamma^{i}\Gamma^{j} \Phi_{0}^{(2)})
+\cdots
+w_{53,4}\,\partial X_{1}^{(2),ij}
\nonu\\
& + &
w_{54,4}\,\partial(\Gamma^{i}\Gamma^{j} X_{0}^{(2)})
+\cdots
+w_{87,4}\,\partial \tilde{X}_{1}^{(2),ij}
+w_{88,4}\,\tilde{L} U \Gamma^{i} \Gamma^{j}
+\cdots
\nonu\\
& + &
w_{303,4}\,\varepsilon^{ijkl}\,(\partial^{2} T^{ik}T^{il}+\partial^{2} T^{jk}T^{jl})
+w_{304,4}\,\varepsilon^{ijkl}\varepsilon^{mnpq}(\,T^{mn}T^{pq}U \Gamma^{k}\Gamma^{l}\,)
-(i\leftrightarrow j),
\nonu \\
Q^{(\frac{1}{2}),i}_{\frac{3}{2}}  & = & 
w_{1,\frac{1}{2}}\,\Gamma^{i},
\nonu\\
Q^{(\frac{3}{2}),i}_{\frac{3}{2}}  & = & 
w_{1,\frac{3}{2}}\,\tilde{G}^{i}
+w_{2,\frac{3}{2}}\,\partial \Gamma^{i}
+w_{3,\frac{3}{2}}\,\Gamma^{i} U
+w_{4,\frac{3}{2}}\,\Gamma^{j} T^{ij}
+\varepsilon^{ijkl}(\,
w_{5,\frac{3}{2}}\,\Gamma^{j} T^{kl}
+w_{6,\frac{3}{2}}\,\Gamma^{j}\Gamma^{k}\Gamma^{l}
\,),
\nonu\\
Q^{(\frac{5}{2}),i}_{\frac{3}{2}}   & = & 
w_{1,\frac{5}{2}}\, \Phi_{\frac{1}{2}}^{(2),i}
+w_{2,\frac{5}{2}}\,\Gamma^{i}  \Phi_{0}^{(2)}
+w_{3,\frac{5}{2}}\,X_{\frac{1}{2}}^{(2),i}
+w_{4,\frac{5}{2}}\, \tilde{G}^{i}
+\cdots
+w_{27,\frac{5}{2}}\,\varepsilon^{ijkl}\,T^{ji}\Gamma^{k}\Gamma^{i}
\Gamma^{l},
\nonu\\
Q^{(\frac{7}{2}),i}_{\frac{3}{2}} & = & 
(M^{\alpha})^{ij}(\,w_{1,\frac{7}{2}}\,\Phi^{(3),j,\alpha}_{\frac{1}{2}}
+w_{2,\frac{7}{2}}\,\Gamma^{j}\Phi^{(3),\alpha}_{0}
\,)
+ \varepsilon^{ijkl}(M^{\alpha})^{jk}(\,
w_{3,\frac{7}{2}}\,\Phi^{(3),l,\alpha}_{\frac{1}{2}}
+w_{4,\frac{7}{2}}\,\Gamma^{l} \Phi^{(3),\alpha}_{0}
\,)
\nonu \\ 
& + & w_{5,\frac{7}{2}}\,\delta^{i}_{\mu}\,\Phi^{(\frac{7}{2}),\mu}_{0}
+w_{6,\frac{7}{2}}\,\Phi^{(2),i}_{\frac{3}{2}}
+w_{7,\frac{7}{2}}\, \partial \Phi^{(2),i}_{\frac{1}{2}}
+w_{8,\frac{7}{2}}\, U \Phi^{(2),i}_{\frac{1}{2}}
+w_{9,\frac{7}{2}}\, \tilde{G}^{i} \Phi^{(2)}_{0}
+\cdots
\nonu \\ 
& + &
w_{22,\frac{7}{2}}\,X^{(2),i}_{\frac{3}{2}}
+w_{23,\frac{7}{2}}\, \partial X^{(2),i}_{\frac{1}{2}}
+w_{24,\frac{7}{2}}\, U X^{(2),i}_{\frac{1}{2}}
+w_{25,\frac{7}{2}}\, \tilde{G}^{i} X^{(2)}_{0}
+\cdots
\nonu \\ 
& + &
w_{37,\frac{7}{2}}\,\varepsilon^{ijkl}\, \Gamma^{j} \Gamma^{k} \Gamma^{l} X^{(2)}_{0}
+w_{38,\frac{7}{2}}\,\partial^{2}\tilde{G}^{i}
+w_{39,\frac{7}{2}}\,\partial^{3}\Gamma^{i}
+w_{40,\frac{7}{2}}\, UU\tilde{G}^{i}
+\cdots
\nonu \\ 
& + &
(\varepsilon^{ijkl})^{2}(\, 
w_{143,\frac{7}{2}}\, T^{ij}T^{kl}\Gamma^{j}\Gamma^{k}\Gamma^{l}
+w_{144,\frac{7}{2}}\, T^{ij}\tilde{T}{}^{kl}\Gamma^{j}\Gamma^{k}\Gamma^{l}\,),
\nonu\\
Q^{(\frac{9}{2}),i}_{\frac{3}{2}}  & = & 
w_{0,\frac{9}{2}}\,\delta_{\mu}^{i}\,\Phi_{0}^{(\frac{9}{2}),\mu}
+w_{1,\frac{9}{2}}\,\Phi_{\frac{1}{2}}^{(4),i}
+(M^{\alpha})^{ij}(\, 
w_{2,\frac{9}{2}}\,\Phi_{\frac{3}{2}}^{(3),\alpha}
+w_{3,\frac{9}{2}}\,\partial\Phi_{\frac{1}{2}}^{(3),\alpha}
+w_{4,\frac{9}{2}}\,\Gamma^{i} \Phi_{1}^{(3),ij,\alpha}
\nonu \\ 
&
+ & \cdots
+w_{22,\frac{9}{2}}\,\varepsilon^{ijkl}\,\Gamma^{i}\Gamma^{k}\Gamma^{l}\Phi_{0}^{(3),\alpha}\,)
+\cdots
+\delta_{\mu}^{l}(\,
\varepsilon^{ijkl}\,w_{49}^{1}\,\Phi_{1}^{(\frac{7}{2}),jk,\mu}
+w_{50,\frac{9}{2}}\,\varepsilon^{ijkl}\,\tilde{\Phi}_{1}^{(\frac{7}{2}),jk,\mu}
+\cdots
\nonu \\ 
&
+& w_{59,\frac{9}{2}}\,\partial \Phi_{0}^{(\frac{7}{2}),\mu}\,)
+w_{60,\frac{9}{2}}\,\Phi_{0}^{(2)}\Phi_{\frac{1}{2}}^{(2),i}
+w_{61,\frac{9}{2}}\,\tilde{L}\Phi_{\frac{1}{2}}^{(2),i}
+w_{62,\frac{9}{2}}\, U\Phi_{\frac{3}{2}}^{(2),i}+\cdots
+w_{149,\frac{9}{2}}\,\tilde{L}X_{\frac{1}{2}}^{(2),i}
\nonu \\ 
& + & w_{150,\frac{9}{2}}\, UX_{\frac{3}{2}}^{(2),i}+\cdots
+w_{236,\frac{9}{2}}\, \varepsilon^{ijkl}\,\partial T^{jk} \Gamma^{l} X_{0}^{(2)}
+w_{237,\frac{9}{2}}\,\tilde{L}UU\Gamma^{i}
+w_{238,\frac{9}{2}}\,\tilde{L}U\partial\Gamma^{i}
\nonu \\ 
& + & w_{239,\frac{9}{2}}\,\tilde{L}\tilde{G}^{i} U
+\cdots
+\varepsilon^{ijkl}(
\,
w_{379,\frac{9}{2}}\,\tilde{L} U \Gamma^{j}\Gamma^{k}\Gamma^{l}
+w_{380,\frac{9}{2}}\,\tilde{L} \tilde{G}^{j}\Gamma^{k}\Gamma^{l}
+\cdots
\nonu \\ 
&+& 
w_{668,\frac{9}{2}}\,\partial^{2}\Gamma^{i} \Gamma^{i} \Gamma^{j} \Gamma^{k} \Gamma^{l}
+w_{669,\frac{9}{2}}\,\partial^{3} \Gamma^{j} \Gamma^{k} \Gamma^{l}
+ w_{670,\frac{9}{2}}\,\partial^{2} \Gamma^{j} \partial\Gamma^{k} \partial\Gamma^{l}
+w_{671,\frac{9}{2}}\,\partial\Gamma^{j}\partial\Gamma^{k}\partial\Gamma^{l}\,
),
\nonu  \\
Q^{(1)}_{2}  & = &  w_{1,1}\,U,
\nonu\\
Q^{(2)}_{2}  & = &  
w_{1,2}\,\Phi_{0}^{(2)}
+w_{2,2}\, X_{0}^{(2)}
+w_{3,2}\,\tilde{L}
+w_{4,2}\,\partial U+w_{5}\, UU
+w_{6,2}\, T^{ij}T^{ij}
+w_{7,2}\, T^{ij}\tilde{T}^{ij}
\nonu\\
&
+& w_{8,2}\,\Gamma^{i}\Gamma^{j}T^{ij}
+w_{9,2}\,\Gamma^{i}\Gamma^{j}\tilde{T}^{ij}
+w_{10,2}\,\Gamma^{i}\tilde{G}^{i}
+w_{11,2}\,\Gamma^{i}\partial\Gamma^{i}
+\varepsilon^{ijkl}\, 
w_{12,2}\,\Gamma^{i}\Gamma^{j}\Gamma^{k}\Gamma^{l},
\nonu\\
Q^{(3)}_{2}  & = &  
w_{1,3}\, U\Phi_{0}^{(2)}
+w_{2,3}\,\Gamma^{i}\Phi_{\frac{1}{2}}^{(2),i}
+w_{3,3}\, U\tilde{L}
+w_{4,3}\, U\partial U
+w_{5,3}\, UUU
+w_{6,3}\,\Gamma^{i}\partial\tilde{G}^{i}
\nonu\\
&
+& w_{7,3}\,\partial\Gamma^{i}\tilde{G}^{i}
+w_{8,3}\,\Gamma^{i}\partial\Gamma^{i}U
+w_{9,3}\,\Gamma^{i}\partial^{2}\Gamma^{i}
+w_{10,3}\,\partial\Gamma^{i}\partial\Gamma^{i}
+w_{11,3}\,\Gamma^{i}T^{ij}\tilde{G}^{j}
\nonu \\
& + & w_{12,3}\,\Gamma^{i}\Gamma^{j}T^{ij}U
\nonu\\
&
+& w_{13,3}\, T^{ij}T^{ij}U
+w_{14,3}\,\Gamma^{i}\Gamma^{j}\partial T^{ij}
+w_{15,3}\,\partial\Gamma^{i}\Gamma^{j}T^{ij}
+w_{16,3}\,\partial^{2}T^{ij}
+\varepsilon^{ijkl}(\, 
w_{17,3}\,\Gamma^{i}T^{jk}\tilde{G}^{l}
\nonu\\
&
+& w_{18,3}\,\Gamma^{i}\Gamma^{j}\Gamma^{k}\tilde{G}^{l}
+w_{19,3}\,\Gamma^{i}\Gamma^{j}T^{kl}U
+w_{20,3}\, T^{ij}T^{kl}U
+w_{21,3}\,\Gamma^{i}\Gamma^{j}\partial T^{kl}
+w_{22,3}\,\partial\Gamma^{i}\Gamma^{j}T^{kl}
\nonu\\
&
+& w_{23,3}\,\partial(\Gamma^{i}\Gamma^{j}\Gamma^{k}\Gamma^{l})\,),
\nonu\\
Q^{(4)}_{2}  & = &  
w_{0,4}\,\Phi_{0}^{(4)}
+(M^{\alpha})^{ij}(\, 
w_{1,4}\,\Phi_{1}^{(3),ij,\alpha}
+w_{2,4}\,\tilde{\Phi}_{1}^{(3),ij,\alpha}
+\cdots
+\varepsilon^{ijkl}w_{8,4}\,\Gamma^{k}\Gamma^{l}\Phi_{0}^{(3),\alpha}\,)
\nonu\\
&
+& \delta_{\mu}^{i}(\, 
w_{9,4}\,\Phi_{\frac{1}{2}}^{(\frac{7}{2}),i,\mu}
+w_{10,4}\,\Gamma^{i}\Phi_{0}^{(\frac{7}{2}),\mu}\,)
+w_{11,4}\,\Phi_{0}^{(2)}\Phi_{0}^{(2)}
+w_{12,4}\,\Phi_{2}^{(2)}
+w_{13,4}\,\tilde{L}\Phi_{0}^{(2)}
+\cdots
\nonu\\
&
+& w_{36,4}\,\varepsilon^{ijkl}\, T^{ij}T^{kl}\Phi_{0}^{(2)}
+w_{37,4}\, X_{2}^{(2)}
+w_{38,4}\,\tilde{L} X_{0}^{(2)}
+\cdots
+w_{61,4}\,\varepsilon^{ijkl}\, T^{ij}T^{kl}X_{0}^{(2)}
\nonu\\
&
+& w_{62,4}\,\tilde{L}\tilde{L}
+w_{63,4}\,\tilde{L}UU
+\cdots+w_{168,4}^{2}\,\varepsilon^{ijkl}\,\Gamma^{i}\Gamma^{j}\partial\Gamma^{k}\partial\Gamma^{l},
\nonu\\
Q^{(5)}_{2}  & = &  
w_{1,5}\,\delta_{\mu}^{i}\,\Phi_{\frac{1}{2}}^{(\frac{9}{2}),i,\mu}
+(M^{\alpha})^{ij}(\, 
w_{2,5}\,\partial\Phi_{1}^{(3),ij,\alpha}
+w_{3,5}\,\partial\tilde{\Phi}_{1}^{(3),ij,\alpha}
+\cdots
+w_{38,5}\,\varepsilon^{ijkl}\, \partial T^{kl}\Phi_{0}^{(3),\alpha}
\nonu\\
&+ &
w_{39,5}\,\varepsilon^{ijkl}\, T^{jk}\Gamma^{l}\Phi_{\frac{1}{2}}^{(3)j,,\alpha}\,)
+\delta_{\mu}^{i}(\, 
w_{40,5}\,\Phi_{\frac{3}{2}}^{(\frac{7}{2}),i,\mu}
+w_{41,5}\,\partial\Phi_{\frac{1}{2}}^{(\frac{7}{2}),i,\mu}
+\cdots
\nonu \\
& + &
w_{52,5}\,\varepsilon^{ijk\mu}\, T^{ij}\Gamma^{k}\Phi_{0}^{(\frac{7}{2}),\mu}
\nonu\\
&
+& w_{53,5}\,\varepsilon^{ijk\mu}\,\Gamma^{i}\Gamma^{j}\Phi_{\frac{1}{2}}^{(\frac{7}{2}),k,\mu}\,)
+w_{54,5}\,\partial\Phi_{2}^{(2)}
+w_{55,5}\,\tilde{L}\partial\Phi_{0}^{(2)}
+\cdots
+w_{129,5}\,\partial X_{2}^{(2)}
\nonu\\
&+ &
w_{130,5}\,\tilde{L}\partial X_{0}^{(2)}
+\cdots
+w_{203,5}\,\varepsilon^{ijkl}\, \partial \Gamma^{i}\Gamma^{j}\Gamma^{k}\Gamma^{l} X_{0}^{(2)}
+w_{204,5}\,\partial^{3}\tilde{L}
+w_{205,5}\,\partial\tilde{L}\partial U
+\cdots
\nonu\\
&+ &
w_{348,5}\,\partial \tilde{G}^{i} \tilde{T}^{ij} T^{jk} \Gamma^{k}
+\varepsilon^{ijkl}\,
(\,
w_{349,5}\,\tilde{L} \partial \Gamma^{i} \Gamma^{j}\Gamma^{k}\Gamma^{l}
+ \cdots
+w_{426,5}\,\tilde{G}^{i}T^{jk}\partial\Gamma^{i}\Gamma^{i}\Gamma^{l}
\nonu\\
&+ &
w_{427,5}\,\tilde{G}^{i}T^{ij}\partial\Gamma^{i}\Gamma^{k}\Gamma^{l}
+w_{428,5}\,\varepsilon^{ijkl}\,\partial\tilde{G}^{i}T^{ij}\Gamma^{i}\Gamma^{k}\Gamma^{l}\,),
\label{Qcomp}
\eea
where the fields with tilde are defined in
(\ref{J4}), (\ref{Phialphaprimary}) and (\ref{newdef}).
Let us explain the notations in (\ref{Qcomp}).
Because the total number of coefficients is greater
than $2000$, we present some of the full composite fields.
In the begining of these expressions, the higher spin
currents appear and then the composite fields made of
$16$ currents appear. For example,
in the expression of $Q_{\frac{1}{2}}^{(\frac{7}{2}),i}(w)$,
there are $123$ terms where the higher spin dependent
terms arise until the thirty second term and
from the thirty third term to the last term,
the higher spin independent terms arise.


\section{Partial expressions of quasi primary fields
in the ${\cal N}=4$ superspace}

We also present the various quasi primary super fields appeared in
section $4$ (the complete expressions can be found
in $\tt{ancillary.nb}$ file)
\bea
{\bf Q}^{(2)}_{0} & = &
c_{1}^{0,2}\,{\bf \Phi}^{(2)}
+c_{2}^{0,2}\,{\bf J}^{4-0}
+c_{3}^{0,2}\,\partial^{2}{\bf J}
+c_{4}^{0,2}\,\partial {\bf J}\partial {\bf J}
+c_{5}^{0,2}\,{\bf J}^{ij}{\bf J}^{ij}
+c_{6}^{0,2}\,{\bf J}^{ij}{\bf J}^{4-ij}
\nonu \\ 
& + & 
c_{7}^{0,2}\,{\bf J}^{i}{\bf J}^{j}{\bf J}^{4-ij}
+c_{8}^{0,2}\,{\bf J}^{i}{\bf J}^{j}{\bf J}^{ij}
+c_{9}^{0,2}\,{\bf J}^{i}\partial {\bf J}^{i}
+c_{10}^{0,2}\,\varepsilon^{ijkl}\,{\bf J}^{i}{\bf J}^{j}{\bf J}^{k}{\bf J}^{l},
\nonu\\
 {\bf Q}^{(\frac{3}{2}),i}_{\frac{1}{2}}      &  = &
c_{1}^{\frac{1}{2},3}\,{\bf J}^{4-i}
+c_{2}^{\frac{1}{2},3}\,\partial {\bf J}^{i}
+c_{3}^{\frac{1}{2},3}\,{\bf J}^{i}\partial {\bf J}
+\varepsilon^{ijkl}(
c_{4}^{\frac{1}{2},3}\,{\bf J}^{j}{\bf J}^{4-kl}
+c_{5}^{\frac{1}{2},3}\,{\bf J}^{j}{\bf J}^{k}{\bf J}^{l}),
\nonu \\ 
{\bf Q}^{(\frac{5}{2}),i}_{\frac{1}{2}}  & = &
c_{1}^{\frac{1}{2},2}\,D^{i}{\bf \Phi}^{(2)}
+c_{2}^{\frac{1}{2},2}\,\partial {\bf J}^{4-i}
+c_{3}^{\frac{1}{2},2}\,\partial^{2}{\bf J}^{i}
+c_{4}^{\frac{1}{2},2}\,{\bf J}^{i}\partial^{2}{\bf J}
+c_{5}^{\frac{1}{2},2}\,\partial {\bf J}^{i}\partial {\bf J}
+c_{6}^{\frac{1}{2},2}\,{\bf J}^{4-ij}{\bf J}^{4-j}
\nonu \\ 
&
+& c_{7}^{\frac{1}{2},2}\,{\bf J}^{i}{\bf J}^{j}{\bf J}^{4-j}
+c_{8}^{\frac{1}{2},2}\,\partial {\bf J}^{j}{\bf J}^{4-ij}
+c_{9}^{\frac{1}{2},2}\,{\bf J}^{j}\partial {\bf J}^{4-ij}
+c_{10}^{\frac{1}{2},2}\,{\bf J}^{j}{\bf J}^{4-ij}\partial {\bf J}
+c_{11}^{\frac{1}{2},2}\,{\bf J}^{i}{\bf J}^{j}\partial {\bf J}^{j}
\nonu \\ 
&
+& c_{12}^{\frac{1}{2},2}\,{\bf J}^{i}{\bf J}^{jk}{\bf J}^{jk}
+c_{13}^{\frac{1}{2},2}\,{\bf J}^{i}{\bf J}^{j}{\bf J}^{k}{\bf J}^{4-jk}
+c_{14}^{\frac{1}{2},2}\,{\bf J}^{j}{\bf J}^{4-ik}{\bf J}^{jk}
+\varepsilon^{ijkl}(\,
c_{15}^{\frac{1}{2},2}\,{\bf J}^{4-jk}{\bf J}^{4-l}
+c_{16}^{\frac{1}{2},2}\,{\bf J}^{j}{\bf J}^{k}{\bf J}^{4-l}
\nonu \\ 
&
+& c_{17}^{\frac{1}{2},2}\,\partial {\bf J}^{j}{\bf J}^{4-kl}
+c_{18}^{\frac{1}{2},2}\,{\bf J}^{j}{\bf J}^{4-kl}\partial {\bf J}
+c_{19}^{\frac{1}{2},2}\,{\bf J}^{i}{\bf J}^{ij}{\bf J}^{kl}
+c_{20}^{\frac{1}{2},2}\,\partial({\bf J}^{j}{\bf J}^{k}{\bf J}^{l})
+c_{21}^{\frac{1}{2},2}\,{\bf J}^{j}{\bf J}^{k}{\bf J}^{l}\partial {\bf J}\,),
\nonu\\
    {\bf Q}^{(\frac{7}{2}),i}_{\frac{1}{2}} & = &
(M^{\alpha})^{ij}(\,
c^{\frac{1}{2},1}_{1}\,D^{j}{\bf \Phi}^{(3),\alpha}
+c^{\frac{1}{2},1}_{2}\,{\bf J}^{j}{\bf \Phi}^{(3),\alpha}\,)
+\varepsilon^{ijkl}\,(M^{\alpha})^{jk}(\,
c_{3}^{\frac{1}{2},1}\,D^{l}{\bf \Phi}^{(3),\alpha}
\nonu \\ 
&+ & c_{4}^{\frac{1}{2},1}\,{\bf J}^{l}{\bf \Phi}^{(3),\alpha}\,)
+c_{0}^{\frac{1}{2},1}\,\delta^{i}_{\mu}\,{\bf \Phi}^{(\frac{7}{2}),\mu}
+c_{5}^{\frac{1}{2},1}\,D^{4-i}{\bf \Phi}^{(2)}
+c_{6}^{\frac{1}{2},1}\,\partial D^{i}{\bf \Phi}^{(2)}
+c_{7}^{\frac{1}{2},1}\,\partial {\bf J} D^{i}{\bf \Phi}^{(2)}
\nonu \\ 
&+ & c_{8}^{\frac{1}{2},1}\,{\bf J}^{4-i}{\bf \Phi}^{(2)}
+\cdots
+c_{19}^{\frac{1}{2},1}\,D^{4-i}{\bf X}^{(2)}
+c_{20}^{\frac{1}{2},1}\,\partial D^{i}{\bf X}^{(2)}
+c_{21}^{\frac{1}{2},1}\,\partial {\bf J} D^{i}{\bf X}^{(2)}
+c_{22}^{\frac{1}{2},1}\,{\bf J}^{4-i}{\bf X}^{(2)}
\nonu \\ 
&+ &
\cdots
+c_{32}^{\frac{1}{2},1}\,\varepsilon^{ijkl}\,{\bf J}^{j}{\bf J}^{k}{\bf J}^{l}{\bf X}^{(2)}
+c_{33}^{\frac{1}{2},1}\,{\bf J}^{4-i} {\bf J}^{4-0}
+ \cdots
+c_{122}^{\frac{1}{2},1}\,\varepsilon^{ijkl}\,{\bf J}^{j}{\bf J}^{4-ij}{\bf J}^{4-ik}{\bf J}^{4-lj}
\nonu \\ 
&+ &
c_{123}^{\frac{1}{2},1}\,(\varepsilon^{ijkl})^2 \,{\bf J}^{j}{\bf J}^{k}{\bf J}^{l}{\bf J}^{4-ij}{\bf J}^{4-kl}, 
\nonu \\
{\bf Q}^{(1),ij}_{1} & = &
c_{1}^{1,4}\,{\bf J}^{4-ij}
+c_{2}^{1,4}\,{\bf J}^{i}J^{j}
+c_{3}^{1,4}\,{\bf J}^{ij}
+\varepsilon^{ijkl}\,c_{4}^{1,4}\,{\bf J}^{k}{\bf J}^{l}
-(i\leftrightarrow j),
\nonu\\
 {\bf Q}^{(2),ij}_{1} & = &
c_{1}^{1,3}\,{\bf J}^{i}J^{4-j}
+c_{2}^{1,3}\,{\bf J}^{i}{\bf J}^{j}\partial J
+c_{3}^{1,3}\,\partial({\bf J}^{i}{\bf J}^{j})
+\varepsilon^{ijkl}\,
c_{4}^{1,3}\,({\bf J}^{i}{\bf J}^{k}{\bf J}^{4-l}+{\bf J}^{j}{\bf J}^{k}{\bf J}^{4-jl})
-(i\leftrightarrow j),
\nonu\\
 {\bf Q}^{(3),ij}_{1} & = &
c_{0}^{1,2}\,(M^{\alpha})^{ij}{\bf \Phi}^{(3),\alpha}
+c_{1}^{1,2}\, D^{4-ij}{\bf \Phi}^{(2)}
+c_{5}^{1,2}\, D^{ij}{\bf \Phi}^{(2)}
+c_{2}^{1,2}\, {\bf J}^{i}D^{j}{\bf \Phi}^{(2)}
\nonu \\ 
&
+& c_{3}^{1,2}\, {\bf J}^{4-ij}{\bf \Phi}^{(2)}
+\cdots
+c_{9}^{1,2}\, D^{4-ij}{\bf X}^{(2)}
+c_{13}^{1,2}\, D^{ij}{\bf X}^{(2)}
+c_{10}^{1,2}\, {\bf J}^{i}D^{j}{\bf X}^{(2)}
\nonu \\
& + & c_{11}^{1,2}\, {\bf J}^{4-ij}{\bf X}^{(2)}
+\cdots
+ 
c_{16}^{1,2}\, {\bf J}^{ij}{\bf X}^{(2)}
+c_{17}^{1,2}\, {\bf J}^{4-i}{\bf J}^{4-j}
+c_{18}^{1,2}\, {\bf J}^{i}{\bf J}^{j}{\bf J}^{4-0}
+\cdots
\nonu \\
& + &
c_{91}^{1,2}\,(\varepsilon^{ijkl})^{2}\, {\bf J}^{k}{\bf J}^{l}{\bf J}^{4-ik}{\bf J}^{4-jl}
-(i\leftrightarrow j),
\nonu\\
 {\bf Q}^{(4),ij}_{1} & = & 
(M^{\alpha})^{ik}(\, 
c_{1}^{1,1}\, D^{4-jk}{\bf \Phi}^{(3),\alpha}
+c_{2}^{1,1}\, {\bf J}^{4-jk}{\bf \Phi}^{(3),\alpha}
+c_{3}^{1,1}\, {\bf J}^{jk}{\bf \Phi}^{(3),\alpha}
+c_{4}^{1,1}\, {\bf J}^{j}D^{k}{\bf \Phi}^{(3),\alpha}
+\cdots\,)
\nonu\\
& + & \varepsilon^{ijkl}(\,
(M^{\alpha})^{kq}c_{5}^{1,1}\, D^{4-lq}{\bf \Phi}^{(3),\alpha}
+(M^{\alpha})^{ik}c_{6}^{1,1}\, {\bf J}^{i}D^{l}{\bf \Phi}^{(3),\alpha}
+(M^{\alpha})^{kl}c_{7}^{1,1}\, {\bf J}^{k}D^{k}{\bf \Phi}^{(3),\alpha}
+\cdots\,)
\nonu\\
& +& c_{12}^{1,1}\,\delta_{\mu}^{j}\, D^{i}{\bf \Phi}^{(\frac{7}{2}),\mu}
+c_{13}^{1,1}\,\partial D^{4-ij}{\bf \Phi}^{(2)}
+c_{14}^{1,1}\,\partial({\bf J}^{i}{\bf J}^{j}{\bf \Phi}^{(2)})
+\cdots
+c_{52}^{1,1}\,\partial D^{4-ij}{\bf X}^{(2)}
\nonu\\
& + & 
c_{53}^{1,1}\,\partial({\bf J}^{i}{\bf J}^{j}{\bf X}^{(2)})
+\cdots
c_{86}^{1,1}\,\partial D^{ij}{\bf X}^{(2)}
+c_{87}^{1,1}\,{\bf J}^{4-0}\partial{\bf J}{\bf J}^{i}{\bf J}^{j}
+\cdots
\nonu\\
& + &
c_{302}^{1,1}\,\varepsilon^{ijkl}(\partial^{2}{\bf J}^{ik}{\bf J}^{il}+\partial^{2}{\bf J}^{jk}{\bf J}^{jl})
+c_{303}^{1,1}\,\varepsilon^{ijkl}\varepsilon^{mnpq}\, {\bf J}^{4-mn}{\bf J}^{4-pq}\partial {\bf J} {\bf J}^{k}{\bf J}^{l}
- (i\leftrightarrow j),
\nonu \\
{\bf Q}^{(\frac{1}{2}),i}_{\frac{3}{2}} & = &
c_{1}^{\frac{3}{2},5}\,{\bf J}^{i},
\nonu\\
 {\bf Q}^{(\frac{3}{2}),i}_{\frac{3}{2}} & = &
c_{1}^{\frac{3}{2},4}\, {\bf J}^{4-i}
+c_{2}^{\frac{3}{2},4}\,\partial {\bf J}^{i}
+c_{3}^{\frac{3}{2},4}\, {\bf J}^{i}\partial {\bf J}
+c_{4}^{\frac{3}{2},4}\, {\bf J}^{j}{\bf J}^{4-ij}
+\varepsilon^{ijkl}\,
(\,
c_{5}^{\frac{3}{2},4}\, {\bf J}^{j}{\bf J}^{4-kl}
+c_{6}^{\frac{3}{2},4}\, {\bf J}^{j}{\bf J}^{k}{\bf J}^{l}\,
),
\nonu\\
 {\bf Q}^{(\frac{5}{2}),i}_{\frac{3}{2}} & = &
c_{1}^{\frac{3}{2},3}\, D^{i}{\bf \Phi}^{(2)}
+c_{2}^{\frac{3}{2},3}\, {\bf J}^{i}{\bf \Phi}^{(2)}
+c_{3}^{\frac{3}{2},3}\, D^{i}{\bf X}^{(2)}
+c_{4}^{\frac{3}{2},3}\,\partial {\bf J}^{4-i}
+\cdots
+c_{27}^{\frac{3}{2},3}\,\varepsilon^{ijkl}\, {\bf J}^{4-ji}{\bf J}^{k}{\bf J}^{i}{\bf J}^{l},
\nonu\\
 {\bf Q}^{(\frac{7}{2}),i}_{\frac{3}{2}} &  = &
(M^{\alpha})^{ij}\,(\, 
c_{1}^{\frac{3}{2},2}\, D^{j}{\bf \Phi}^{(3),\alpha}
+c_{2}^{\frac{3}{2},2}\, {\bf J}^{j}{\bf \Phi}^{(3),\alpha}\,)
\nonu\\
&  
+ & \varepsilon^{ijkl} (M^{\alpha})^{jk}(\, 
c_{3}^{\frac{3}{2},2}\, D^{l}{\bf \Phi}^{(3),\alpha}
+c_{4}^{\frac{3}{2},2}\, {\bf J}^{l}{\bf \Phi}^{(3),\alpha}\,)
+c_{5}^{\frac{3}{2},2}\,\delta_{\mu}^{i}\,{\bf \Phi}^{(\frac{7}{2}),\mu}
+c_{6}^{\frac{3}{2},2}\, D^{4-i}{\bf \Phi}^{(2)}
\nonu\\
&  
+ & c_{7}^{\frac{3}{2},2}\,\partial D^{i}{\bf \Phi}^{(2)}
+c_{8}^{\frac{3}{2},2}\,\partial {\bf J} D^{i}{\bf \Phi}^{(2)}
+c_{9}^{\frac{3}{2},2}\, {\bf J}^{4-i}{\bf \Phi}^{(2)}
+\cdots 
+c_{22}^{\frac{3}{2},2}\, D^{4-i}{\bf X}^{(2)}
+c_{23}^{\frac{3}{2},2}\,\partial D^{i}{\bf X}^{(2)}
\nonu\\
&  
+ & c_{24}^{\frac{3}{2},2}\,\partial {\bf J} D^{i}{\bf X}^{(2)}
+c_{25}^{\frac{3}{2},2}\, {\bf J}^{4-i}{\bf X}^{(2)}
+\cdots
+c_{37}^{\frac{3}{2},2}\,\varepsilon^{ijkl} \, {\bf J}^{j}{\bf J}^{k}{\bf J}^{l}{\bf X}^{(2)}
+c_{38}^{\frac{3}{2},2}\,\partial^{2}{\bf J}^{4-i}
+c_{39}^{\frac{3}{2},2}\,\partial^{3}{\bf J}^{i}
\nonu\\
&  
+ &c_{40}^{\frac{3}{2},2}\,\partial {\bf J}\tilde{G}^{i}+\cdots
+ \varepsilon^{ijkl}(\cdots
+c_{141}^{\frac{3}{2},2}\, {\bf J}^{j}{\bf J}^{k}{\bf J}^{l}{\bf J}^{kl}{\bf J}^{4-kl}
+w_{142}^{2}\, {\bf J}^{j}{\bf J}^{k}{\bf J}^{l}{\bf J}^{ij}{\bf J}^{ij}\,)
,\nonu\\
{\bf Q}^{(\frac{9}{2}),i}_{\frac{3}{2}} & = &
c_{0}^{\frac{3}{2},1}\,\delta_{\mu}^{i}\,{\bf \Phi}^{(\frac{9}{2}),\mu}
+c_{1}^{\frac{3}{2},1}\, D^{i}{\bf \Phi}^{(4)}
+(M^{\alpha})^{ij}(\, c_{2}^{\frac{3}{2},1}\, D^{4-j}{\bf \Phi}^{(3),\alpha}
+c_{3}^{\frac{3}{2},1}\,\partial D^{j}{\bf \Phi}^{(3),\alpha}
\nonu\\
& 
+ & c_{4}^{\frac{3}{2},1}\, {\bf J}^{i}D^{4-ij}{\bf \Phi}^{(3),\alpha}
+\cdots
+c_{22}^{\frac{3}{2},1}\,\varepsilon^{ijkl}\, {\bf J}^{i}{\bf J}^{k}{\bf J}^{l}{\bf \Phi}^{(3),\alpha}\,)
+\cdots
+\delta_{\mu}^{l}(\,\varepsilon^{ijkl}\, c_{49}^{\frac{3}{2},1}\, D^{4-jk}{\bf \Phi}^{(\frac{7}{2}),\mu}
\nonu\\
& 
+ &
c_{50}^{\frac{3}{2},1}\,\varepsilon^{ijkl}\, D^{jk}{\bf \Phi}^{(\frac{7}{2}),\mu}+\cdots
+c_{59}^{\frac{3}{2},1}\,\partial{\bf \Phi}^{(\frac{7}{2}),\mu}\,)
+c_{60}^{\frac{3}{2},1}\,{\bf \Phi}^{(2)}D^{i}{\bf \Phi}^{(2)}
+c_{61}^{\frac{3}{2},1}\, {\bf J}^{4-0}D^{i}{\bf \Phi}^{(2)}
\nonu\\
& 
+ & c_{62}^{\frac{3}{2},1}\,\partial {\bf J} D^{4-i}{\bf \Phi}^{(2)}
+\cdots
+c_{149}^{\frac{3}{2},1}\, {\bf J}^{4-0}D^{i}{\bf X}^{(2)}
+c_{150}^{\frac{3}{2},1}\,\partial {\bf J} D^{4-i}{\bf X}^{(2)}
+\cdots
+c_{236}^{\frac{3}{2},1}\, \partial {\bf J}^{ij}{\bf J}^{j} {\bf X}^{(2)}
\nonu\\
& + & 
c_{237}^{\frac{3}{2},1}\, {\bf J}^{4-0}\partial {\bf J}\partial {\bf J} {\bf J}^{i}
+c_{238}^{\frac{3}{2},1}\, {\bf J}^{4-0}\partial {\bf J}\partial {\bf J}^{i}
+c_{239}^{\frac{3}{2},1}\, {\bf J}^{4-0}{\bf J}^{4-i}\partial {\bf J}+\cdots
+\varepsilon^{ijkl}(\, c_{379}^{\frac{3}{2},1}\, {\bf J}^{4-0}\partial {\bf J} {\bf J}^{j}{\bf J}^{k}{\bf J}^{l}
\nonu\\
& + & 
c_{380}^{\frac{3}{2},1}\, {\bf J}^{4-0}{\bf J}^{4-j}{\bf J}^{k}{\bf J}^{l}
+\cdots
+c_{668}^{\frac{3}{2},1}\,\partial^{2}{\bf J}^{i}{\bf J}^{i}{\bf J}^{j}{\bf J}^{k}{\bf J}^{l}
+c_{669}^{\frac{3}{2},1}\,\partial^{3}{\bf J}^{j}{\bf J}^{k}{\bf J}^{l}
+c_{670}^{\frac{3}{2},1}\,\partial^{2}{\bf J}^{j}\partial {\bf J}^{k}\partial {\bf J}^{l}
\nonu\\
& + & 
c_{671}^{\frac{3}{2},1}\,\partial {\bf J}^{j}\partial {\bf J}^{k}\partial {\bf J}^{l}\,),
\nonu\\
{\bf Q}^{(1)}_{2} & = &
c_{1}^{2,5}\,\partial {\bf J},
\nonu\\
{\bf Q}^{(2)}_{2} &  = &
c_{1}^{2,4}\,{\bf \Phi}^{(2)}
+c_{2}^{2,4}\,{\bf X}^{(2)}
+c_{3}^{2,4}\, {\bf J}^{4-0}
+c_{4}^{2,4}\,\partial^{2}{\bf J}
+c_{5}^{2,4}\,\partial {\bf J}\partial {\bf J}
+c_{6}^{2,4}\, {\bf J}^{i}{\bf J}^{4-i}
\nonu\\
&  
+ & c_{7}^{2,4}\, {\bf J}^{i}\partial {\bf J}^{i}
+c_{8}^{2,4}\, {\bf J}^{ij}{\bf J}^{ij}
+c_{9}^{2,4}\, {\bf J}^{ij}{\bf J}^{4-ij}
+c_{10}^{2,4}\, {\bf J}^{i}{\bf J}^{j}{\bf J}^{4-ij}
+c_{11}^{2,4}\, {\bf J}^{i}{\bf J}^{j}{\bf J}^{ij}
+c_{12}^{2,4}\,\varepsilon^{ijkl}\, {\bf J}^{i}{\bf J}^{j}{\bf J}^{k}{\bf J}^{l},
\nonu\\
{\bf Q}^{(3)}_{2}  & = &
c_{1}^{2,3}\,\partial J{\bf \Phi}^{(2)}
+c_{2}^{2,3}\, {\bf J}^{i}D^{i}{\bf \Phi}^{(2)}
+c_{3}^{2,3}\,\partial {\bf J}{\bf J}^{4-0}
+c_{4}^{2,3}\,\partial {\bf J}\partial^{2}{\bf J}
+c_{5}^{2,3}\,\partial {\bf J}\partial {\bf J}\partial {\bf J}
\nonu\\
& 
+ & c_{6}^{2,3}\, {\bf J}^{i}\partial {\bf J}^{4-i}
+c_{7}^{2,3}\,\partial {\bf J}^{i}{\bf J}^{4-i}
+c_{8}^{2,3}\, {\bf J}^{i}\partial {\bf J}^{i}\partial {\bf J}
+c_{9}^{2,3}\, {\bf J}^{i}\partial^{2}{\bf J}^{i}
+c_{10}^{2,3}\,\partial {\bf J}^{i}\partial {\bf J}^{i}
+c_{11}^{2,3}\, {\bf J}^{i}{\bf J}^{4-ij}{\bf J}^{4-j}
\nonu\\
& 
+ & c_{12}^{2,3}\, {\bf J}^{i}{\bf J}^{j}{\bf J}^{4-ij}\partial {\bf J}
+c_{13}^{2,3}\, {\bf J}^{4-ij}{\bf J}^{4-ij}\partial {\bf J}
+c_{14}^{2,3}\, {\bf J}^{i}{\bf J}^{j}\partial {\bf J}^{4-ij}
+c_{15}^{2,3}\,\partial {\bf J}^{i}{\bf J}^{j}{\bf J}^{4-ij}
+c_{16}^{2,3}\,\partial^{2}{\bf J}^{ij}
\nonu\\
& 
+ & \varepsilon^{ijkl}(\, 
c_{17}^{2,3}\, {\bf J}^{i}{\bf J}^{4-jk}{\bf J}^{4-l}
+c_{18}^{2,3}\, {\bf J}^{i}{\bf J}^{j}{\bf J}^{k}{\bf J}^{4-l}
+c_{19}^{2,3}\, {\bf J}^{i}{\bf J}^{j}{\bf J}^{4-kl}\partial {\bf J}
+c_{20}^{2,3}\, {\bf J}^{4-ij}{\bf J}^{4-kl}\partial {\bf J}
\nonu\\
& 
+ & c_{21}^{2,3}\, {\bf J}^{i}{\bf J}^{j}\partial {\bf J}^{4-kl}
+c_{22}^{2,3}\,\partial {\bf J}^{i}{\bf J}^{j}{\bf J}^{4-kl}
+c_{23}^{2,3}\,\partial({\bf J}^{i}{\bf J}^{j}{\bf J}^{k}{\bf J}^{l})\,
),
\nonu\\
{\bf Q}^{(4)}_{2} & = &
c_{0}^{2,2}\,{\bf \Phi}^{(4)}
+(M^{\alpha})^{ij}(\, 
c_{1}^{2,2}\, D^{ij}{\bf \Phi}^{(3),\alpha}
+c_{2}^{2,2}\, D^{4-ij}{\bf \Phi}^{(3),\alpha}
+\cdots
+\varepsilon^{ijkl}c_{8}^{2,2}\, {\bf J}^{k}{\bf J}^{l}{\bf \Phi}^{(3),\alpha}\,)
\nonu\\
& 
+ & \delta_{\mu}^{i}(\, c_{9}^{2,2}\, D^{i}{\bf \Phi}^{(\frac{7}{2}),\mu}
+c_{10}^{2,2}\, {\bf J}^{i}{\bf \Phi}^{(\frac{7}{2}),\mu}\,)
+c_{11}^{2,2}\,{\bf \Phi}^{(2)}{\bf \Phi}^{(2)}
+c_{12}^{2,2}\, D^{4-0}{\bf \Phi}^{(2)}
+c_{13}^{2,2}\, {\bf J}^{4-0}{\bf \Phi}^{(2)}+\cdots
\nonu\\
& 
+ & c_{36}^{2,2}\,\varepsilon^{ijkl}\, {\bf J}^{4-ij}{\bf J}^{4-kl}{\bf \Phi}^{(2)}
+c_{37}^{2,2}\, D^{4-0}{\bf X}^{(2)}
+c_{38}^{2,2}\, {\bf J}^{4-0}{\bf X}^{(2)}+\cdots
+c_{61}^{2,2}\,\varepsilon^{ijkl}\, {\bf J}^{4-ij}{\bf J}^{4-kl}{\bf X}^{(2)}
\nonu\\
& 
+ & c_{62}^{2,2}\, {\bf J}^{4-0}{\bf J}^{4-0}
+c_{63}^{2,2}\, {\bf J}^{4-0}\partial {\bf J}\partial {\bf J}
+\cdots
+c_{169}^{2,2}\,\varepsilon^{ijkl}\, {\bf J}^{i}{\bf J}^{j}\partial {\bf J}^{k}\partial {\bf J}^{l},
\nonu\\
 {\bf Q}^{(5)}_{2} &  = &
c_{1}^{2,1}\,\delta_{\mu}^{i}\, D^{i}{\bf \Phi}^{(\frac{9}{2}),\mu}
+(M^{\alpha})^{ij}(\, 
c_{2}^{2,1}\,\partial D^{4-ij}{\bf \Phi}^{(3),\alpha}
+c_{3}^{2,1}\,\partial D^{ij}{\bf \Phi}^{(3),\alpha}
+\cdots
\nonu\\
& 
+ & c_{38}^{2,1}\,\varepsilon^{ijkl}\,\partial {\bf J}^{4-kl}{\bf \Phi}^{(3),\alpha}
+c_{39}^{2,1}\,\varepsilon^{ijkl}\, {\bf J}^{4-jk}{\bf J}^{l}D^{j}{\bf \Phi}^{(3),\alpha}\,)
+\delta_{\mu}^{i}(\, 
c_{40}^{2,1}\, D^{4-i}{\bf \Phi}^{(\frac{7}{2}),\mu}
+c_{41}^{2,1}\,\partial D^{i}{\bf \Phi}^{(\frac{7}{2}),\mu}
\nonu\\
& 
+ & \cdots
+c_{52}^{2,1}\,\varepsilon^{jjk\mu}\, {\bf J}^{4-ij}J^{k}{\bf \Phi}^{(\frac{7}{2}),\mu}
+c_{53}^{2,1}\,\varepsilon^{ijk\mu}\,\Gamma^{i}\Gamma^{j}D^{k}{\bf \Phi}^{(\frac{7}{2}),\mu}\,)
+c_{54}^{2,1}\,\partial D^{4-0}{\bf \Phi}^{(2)}
\nonu\\
& 
+ & c_{55}^{2,1}\, {\bf J}^{4-0}\partial{\bf \Phi}^{(2)}
+\cdots
+c_{129}^{2,1}\,\partial D^{4-0}{\bf X}^{(2)}
+c_{130}^{2,1}\, {\bf J}^{4-0}\partial{\bf X}^{(2)}
+\cdots
+c_{204}^{2,1}\,\varepsilon^{ijkl}\, \partial{\bf J}^{i}{\bf J}^{j}{\bf J}^{k}{\bf J}^{l}{\bf X}^{(2)}
\nonu\\
& + &
c_{205}^{2,1}\,\partial^{3}{\bf J}^{4-0}
+ c_{206}^{2,1}\,\partial {\bf J}^{4-0}\partial {\bf J}
+\cdots
+c_{351}^{2,1}\,\partial {\bf J}^{4-i}{\bf J}^{ij}{\bf J}^{4-jk}{\bf J}^{k}
\nonu \\
& + & \varepsilon^{ijkl}\,(\, 
c_{352}^{2,1}\, {\bf J}^{4-0}\partial
{\bf J}^{i}{\bf J}^{j}{\bf J}^{k}{\bf J}^{l}
+\cdots
\nonu\\
& 
+ &
c_{428}^{2,1}\, {\bf J}^{4-i}{\bf J}^{ij}\partial {\bf J}^{j}{\bf J}^{k}{\bf J}^{l}
+c_{429}^{2,1}\, {\bf J}^{4-i}{\bf J}^{4-ij}\partial {\bf J}^{i}{\bf J}^{k}{\bf J}^{l}
+c_{430}^{2,1}\,\varepsilon^{ijkl}\,\partial {\bf J}^{4-i}{\bf J}^{4-ij}{\bf J}^{i}{\bf J}^{k}{\bf J}^{l}\,).
\label{quasisuper}
\eea
These can be obtained from Appendix $B$ by taking the procedure
in (\ref{comptosuper}). The coefficients appearing in
(\ref{quasisuper}) are proportional to the ones in (\ref{Qcomp})
according to (\ref{comptosuper}) in most of the terms. The number
of terms are little different from each other. For example, see
$Q_1^{(4),ij}(w)$ and ${\bf Q}_1^{{\bf (4)}, ij}(Z_2)$ and other cases.

\section{The single ${\cal N}=4$ OPE in different order }

By applying the replacements in (\ref{comptosuper}), we
obtain the following intermediate expression 
\bea
&&{\bf \Phi^{(2)}}(Z_{1})\,{\bf \Phi^{(2)}}(Z_{2}) =
\frac{\theta_{12}^{4-0}}{z_{12}^{6}}\, 8 \, \al \, c^{0,4}_{0}
+\frac{\theta_{12}^{4-0}}{z_{12}^{5}}\,{\bf Q}^{(1)}_{2}(Z_2)
+\frac{\theta_{12}^{4-0}}{z_{12}^{4}}\,\Bigg[\,
\frac{3}{2}\,\partial {\bf Q}^{(1)}_{2}
+{\bf Q}^{(2)}_{2}
+{\bf R}^{(2)}_{2}\,\Bigg](Z_{2})
\nonu\\
&&
+\frac{\theta_{12}^{4-0}}{z_{12}^{3}}\,\Bigg[\,
\partial^{2} {\bf Q}^{(1)}_{2}
+\partial {\bf Q}^{(2)}_{2}
+{\bf Q}^{(3)}_{2}
+{\bf R}^{(3)}_{2}\,\Bigg](Z_{2})
\nonu\\
&&
+\frac{\theta_{12}^{4-0}}{z_{12}^{2}}\,\Bigg[\,
\frac{5}{12}\,\partial^{3} {\bf Q}^{(1)}_{2}
+\frac{1}{2}\, \partial {\bf Q}^{(2)}_{2}
+\frac{5}{6}\,{\bf Q}^{(3)}_{2}
+{\bf Q}^{(4)}_{2}
+{\bf R}^{(4)}_{2}
\,\Bigg](Z_{2})
\nonu\\
&&
+\frac{\theta_{12}^{4-0}}{z_{12}}\,\Bigg[\,
\frac{1}{8}\,\partial^{4} {\bf Q}^{(1)}_{2}
+\frac{1}{6}\,\partial^{3} {\bf Q}^{(2)}_{2}
+\frac{5}{14}\,\partial^{2} {\bf Q}^{(3)}_{2}
+\frac{3}{4}\,\partial {\bf Q}^{(4)}_{2}
+ {\bf Q}^{(5)}_{2}\,
\Bigg](Z_{2})
\nonu\\
&&
+\frac{\theta_{12}^{4-i}}{z_{12}^{5}}\,
{\bf Q}^{(\frac{1}{2}),i}_{\frac{3}{2}} \, (Z_{2})
+\frac{\theta_{12}^{4-i}}{z_{12}^{4}}\,\Bigg[\,
2\,\partial {\bf Q}^{(\frac{1}{2}),i}_{\frac{3}{2}}+{\bf Q}^{(\frac{3}{2}),i}_{\frac{3}{2}}
+{\bf R}^{(\frac{3}{2}),i}_{\frac{3}{2}}
\,\Bigg](Z_{2})
\nonu\\
&&
+\frac{\theta_{12}^{4-i}}{z_{12}^{3}}\,\Bigg[\,
\frac{3}{2}\,\partial^{2} {\bf Q}^{(\frac{1}{2}),i}_{\frac{3}{2}}
+\partial {\bf Q}^{(\frac{3}{2}),i}_{\frac{3}{2}}
+{\bf Q}^{(\frac{5}{2}),i}_{\frac{3}{2}}
+{\bf R}^{(\frac{5}{2}),i}_{\frac{3}{2}}
\,\Bigg](Z_{2})
\nonu\\
&&
+\frac{\theta_{12}^{4-i}}{z_{12}^{2}}\,\Bigg[\,
\frac{2}{3}\,\partial^{3} {\bf Q}^{(\frac{1}{2}),i}_{\frac{3}{2}}
+\frac{1}{2}\,\partial^{2} {\bf Q}^{(\frac{3}{2}),i}_{\frac{3}{2}}
+\frac{4}{5}\,\partial {\bf Q}^{(\frac{5}{2}),i}_{\frac{3}{2}}
+ {\bf Q}^{(\frac{7}{2}),i}_{\frac{3}{2}}
+{\bf R}^{(\frac{7}{2}),i}_{\frac{3}{2}}
\,\Bigg](Z_{2})
\nonu\\
&&
+\frac{\theta_{12}^{4-i}}{z_{12}}\,\Bigg[\,
\frac{5}{24}\,\partial^{4} {\bf Q}^{(\frac{1}{2}),i}_{\frac{3}{2}}
+\frac{1}{6}\,\partial^{3} {\bf Q}^{(\frac{3}{2}),i}_{\frac{3}{2}}
+\frac{1}{3}\,\partial^{2} {\bf Q}^{(\frac{5}{2}),i}_{\frac{3}{2}}
+\frac{5}{7}\,\partial {\bf Q}^{(\frac{7}{2}),i}_{\frac{3}{2}}
+ {\bf Q}^{(\frac{9}{2}),i}_{\frac{3}{2}}
\,\Bigg](Z_{2})
\nonu\\
&&
+\frac{\theta_{12}^{4-ij}}{z_{12}^{4}}\,
{\bf Q}^{(1),ij}_{1}\,(Z_{2})
+\frac{\theta_{12}^{4-ij}}{z_{12}^{3}}\,\Bigg[\,
\partial{\bf Q}^{(1),ij}_{1}
+{\bf Q}^{(2),ij}_{1}
\,\Bigg](Z_{2})
\nonu\\
&&
+\frac{\theta_{12}^{4-ij}}{z_{12}^{2}}\,\Bigg[\,
\frac{1}{2}\,\partial^{2} {\bf Q}^{(1),ij}_{1}
+\frac{3}{4}\,\partial {\bf Q}^{(2),ij}_{1}
+{\bf Q}^{(3),ij}_{1}
\,\Bigg](Z_{2})
\nonu\\
&&
+\frac{\theta_{12}^{4-ij}}{z_{12}}\,\Bigg[\,
\frac{1}{6}\,\partial^{3} {\bf Q}^{(1),ij}_{1} 
+\frac{3}{10}\,\partial^{2} {\bf Q}^{(2),ij}_{1}
+\frac{2}{3}\,\partial {\bf Q}^{(3),ij}_{1}
+{\bf Q}^{(4),ij}_{1}
\,\Bigg](Z_{2})
+\frac{\theta_{12}^{i}}{z_{12}^{3}}\,
{\bf Q}^{(\frac{3}{2}),i}_{\frac{1}{2}}\, (Z_{2})
\nonu\\
&&
+\frac{\theta_{12}^{i}}{z_{12}^{2}}\,\Bigg[\,
\frac{2}{3}\,\partial  {\bf Q}^{(\frac{3}{2}),i}_{\frac{1}{2}}
+ {\bf Q}^{(\frac{5}{2}),i}_{\frac{1}{2}}
\,\Bigg](Z_{2})
+\frac{\theta_{12}^{i}}{z_{12}}\,\Bigg[\,
\frac{1}{4}\,\partial^{2}  {\bf Q}^{(\frac{3}{2}),i}_{\frac{1}{2}} +
\frac{3}{5}\,\partial {\bf Q}^{(\frac{5}{2}),i}_{\frac{1}{2}}
+ {\bf Q}^{(\frac{7}{2}),i}_{\frac{1}{2}}
\,\Bigg](Z_{2})
\nonu\\
&&
+\frac{1}{z_{12}^{4}}\, c^{0,4}_{0}
+\frac{1}{z_{12}^{2}}\,
{\bf Q}^{(2)}_{0}\, (Z_{2})
+\frac{1}{z_{12}}\,
\frac{1}{2} \, \partial \, {\bf Q}^{(2)}_{0}\,(Z_{2})
+\cdots.
\label{intermediate}
\eea
By rearranging this (\ref{intermediate}) in the order of
increasing spin, it is easy to see that we can
write down this expression as (\ref{singleOPE}).

\section{ The expression of $X_0^{(2)}$ for $N=5$ }

We present the higher spin-$2$ current (the coset indices
ar given by $1, 2, \cdots, 10$, the $SO(4)$ indices are
$11$, $12$ and $13$ and the remaining indices
$14, \cdots, 18$ are $SO(5)$ ones) 
appearing in (\ref{Q22})
\bea
X^{(2)}_0 & = &
c_1\,(\, 
Q^1 Q^2 Q^{19} Q^{20}
+ Q^1 Q^3 Q^{19} Q^{21}
+ Q^1 Q^4 Q^{19}  Q^{21}
+ Q^1 Q^5 Q^{19} Q^{23}
+ Q^2 Q^3 Q^{20} Q^{21} \nonu\\
& + &
 Q^2 Q^4 Q^{20} Q^{22}
+ Q^2 Q^5 Q^{20} Q^{23} 
+ Q^{3} Q^{4} Q^{21} Q^{22}
+ Q^{3} Q^{5} Q^{21} Q^{23}
+ Q^{4} Q^{5} Q^{22} Q^{23} \nonu\\
& + & Q^{6} Q^{7} Q^{24} Q^{25}
+ Q^{6} Q^{8} Q^{24} Q^{26}
+ Q^{6} Q^{9} Q^{24} Q^{27}
+ Q^{6} Q^{10} Q^{24} Q^{28}
+ Q^{7} Q^{8} Q^{25} Q^{26} \nonu\\
& + &
 Q^{7} Q^{9} Q^{25} Q^{27}
+ Q^{7} Q^{10} Q^{25} Q^{28}
+ Q^{8} Q^{9} Q^{26} Q^{27}
+ Q^{8} Q^{10} Q^{26} Q^{28}
+ Q^{9} Q^{10} Q^{27} Q^{28}
\,) 
\nonu\\
& + &
c_2\,(\, 
 Q^{1} Q^{6} Q^{20} Q^{25}
+ Q^{1} Q^{6} Q^{21} Q^{26}
+ Q^{1} Q^{6} Q^{22} Q^{27}
+ Q^{1} Q^{6} Q^{23} Q^{28}
+ Q^{2} Q^{7} Q^{19} Q^{24} \nonu\\
& + &
 Q^{2} Q^{7} Q^{22} Q^{27}
+ Q^{2} Q^{7} Q^{23} Q^{28}
+ Q^{3} Q^{8} Q^{19} Q^{24}
+ Q^{3} Q^{8} Q^{22} Q^{27}
+ Q^{3} Q^{8} Q^{23} Q^{28} \nonu\\
& + &
 Q^{4} Q^{9} Q^{19} Q^{24}
+ Q^{4} Q^{9} Q^{20} Q^{25}
+ Q^{4} Q^{9} Q^{21} Q^{26}
+ Q^{5} Q^{10} Q^{19} Q^{24}
+ Q^{5} Q^{10} Q^{20} Q^{25} \nonu\\
& + &
 Q^{5} Q^{10} Q^{21} Q^{26}
\,)
\nonu\\
& + &
c_3\,(\,
Q^{1} Q^{7} Q^{19} Q^{25} 
+ Q^{1} Q^{8} Q^{19} Q^{26} 
+ Q^{1} Q^{9} Q^{19} Q^{27} 
+ Q^{1} Q^{10} Q^{19} Q^{28} 
+ Q^{2} Q^{6} Q^{20} Q^{24} 
 \nonu\\
& + &
 Q^{2} Q^{9} Q^{20} Q^{27} 
+ Q^{2} Q^{10} Q^{20} Q^{28} 
+ Q^{3} Q^{6} Q^{21} Q^{24} 
+ Q^{3} Q^{9} Q^{21} Q^{27}
+ Q^{3} Q^{10} Q^{21} Q^{28} 
 \nonu\\
& + &
  Q^{4} Q^{6} Q^{22} Q^{24} 
+ Q^{4} Q^{7} Q^{22} Q^{25} 
+  Q^{4} Q^{8} Q^{22} Q^{26}
+ Q^{5} Q^{6} Q^{23} Q^{24} 
+ Q^{5} Q^{7} Q^{23} Q^{25} 
\nonu\\
& + &
 Q^{5} Q^{8} Q^{23} Q^{26}
\,) 
\nonu\\
& + &
c_4\,(\,
Q^{1} Q^{7} Q^{21} Q^{24} 
+ Q^{1} Q^{8} Q^{20} Q^{24} 
+ Q^{1} Q^{9} Q^{23} Q^{24} 
+Q^{1} Q^{10} Q^{22} Q^{24} 
+ Q^{2} Q^{6} Q^{19} Q^{26} 
\nonu\\
& + &
 Q^{2} Q^{9} Q^{23} Q^{26} 
+ Q^{2} Q^{10} Q^{22} Q^{26}
+Q^{3} Q^{6} Q^{19} Q^{25} 
+ Q^{3} Q^{9} Q^{23} Q^{25} 
+ Q^{3} Q^{10} Q^{22} Q^{25} 
\nonu\\
& + &
  Q^{4} Q^{6} Q^{19} Q^{28} 
+ Q^{4} Q^{7} Q^{21} Q^{28} 
+ Q^{4} Q^{8} Q^{20} Q^{28} 
+ Q^{5} Q^{6} Q^{19} Q^{27} 
+ Q^{5} Q^{7} Q^{21} Q^{27} 
\nonu\\
& + &
 Q^{5} Q^{8} Q^{20} Q^{27}
\,)  
\nonu\\
& + &
c_5\,(\, 
Q^{2} Q^{7} Q^{20} Q^{25}
+Q^{3} Q^{8} Q^{21} Q^{26} 
+ Q^{4} Q^{9} Q^{22} Q^{27} 
+ Q^{5} Q^{10} Q^{23} Q^{28}\,) 
\nonu\\
& + &
 c_6\,(\,
Q^{2} Q^{7} Q^{21} Q^{26} 
+ Q^{3} Q^{8} Q^{20} Q^{25}
+ Q^{4} Q^{9} Q^{23} Q^{28} 
+ Q^{5} Q^{10} Q^{22} Q^{27}\,)
\nonu\\
& + &
c_7\,(\,
Q^{3} Q^{7} Q^{21} Q^{25} 
+ Q^{2} Q^{8} Q^{20} Q^{26} 
+ Q^{4} Q^{10} Q^{22} Q^{28} 
+ Q^{5} Q^{9} Q^{23} Q^{27}\,)
  \nonu\\
& + &
c_8\, Q^{1} Q^{6} Q^{19} Q^{24}
\nonu\\
& + &
 c_9\,(\,
Q^{1} Q^{19} V^{12} 
+ Q^{2} Q^{20} V^{12} 
+ Q^{3} Q^{21} V^{12} 
+ Q^{4} Q^{22} V^{12} 
+ Q^{5} Q^{23} V^{12} 
  \nonu\\
& - &
  Q^{6} Q^{24} V^{12} 
- Q^{7} Q^{25} V^{12} 
- Q^{8} Q^{26} V^{12} 
- Q^{9} Q^{27} V^{12} 
- Q^{10} Q^{28} V^{12}\,)  
\nonu\\
& + &
c_{10}\,(\,
Q^{1} Q^{19} V^{30} 
 + Q^{2} Q^{20} V^{30} 
+ Q^{3} Q^{21} V^{30} 
+ Q^{4} Q^{22} V^{30} 
+ Q^{5} Q^{23} V^{30}\,) 
\nonu\\
& + &
c_{11}\,(\,
Q^{1} Q^{20} V^{17} 
+ Q^{1} Q^{22} V^{18} 
+ Q^{2} Q^{19} V^{35} 
+ Q^{2} Q^{23} V^{34} 
+ Q^{3} Q^{23} V^{32} 
+  Q^{4} Q^{19} V^{36} 
\nonu\\
& + & 
 Q^{5} Q^{20} V^{16} 
+Q^{5} Q^{21} V^{14} 
+ Q^{6} Q^{26} V^{17} 
+ Q^{6} Q^{28} V^{18} 
+ Q^{7} Q^{27} V^{32} 
+ Q^{8} Q^{24} V^{35} 
  \nonu\\
& + &
 Q^{8} Q^{27} V^{34} 
+ Q^{9} Q^{25} V^{14} 
+ Q^{9} Q^{26} V^{16} 
+ Q^{10} Q^{24} V^{36} 
- Q^{1} Q^{21} V^{35} 
- Q^{1} Q^{23} V^{36} 
  \nonu\\
& - & 
Q^{2} Q^{22} V^{14} 
- Q^{3} Q^{19} V^{17} 
- Q^{3} Q^{22} V^{16} 
- Q^{4} Q^{20} V^{32} 
- Q^{4} Q^{21} V^{34} 
- Q^{5} Q^{19} V^{18} 
  \nonu\\
& - & 
Q^{6} Q^{25} V^{35} 
- Q^{6} Q^{27} V^{36} 
 - Q^{7} Q^{24} V^{17} 
- Q^{7} Q^{28} V^{16} 
- Q^{8} Q^{28} V^{14} 
- Q^{9} Q^{24} V^{18} 
  \nonu\\
& - & 
Q^{10} Q^{26} V^{32} 
- Q^{10} Q^{25} V^{34}\,) 
  \nonu\\
& + &
c_{12}\,(\,
Q^{1} Q^{24} V^{11} 
+ Q^{2} Q^{26} V^{11} 
+  Q^{3} Q^{25} V^{11} 
+ Q^{4} Q^{28} V^{11} 
+ Q^{5} Q^{27} V^{11} 
+ Q^{6} Q^{19} V^{29} 
 \nonu\\
& + & Q^{7} Q^{21} V^{29} 
+ Q^{8} Q^{20} V^{29}
+ Q^{9} Q^{23} V^{29}
 + Q^{10} Q^{22} V^{29}\,)
 \nonu\\
& + &
c_{13}\,(\,
Q^{2} Q^{20} V^{15} + Q^{2} Q^{20} V^{33}
 + Q^{8} Q^{26} V^{15} 
+ Q^{8} Q^{26} V^{33}
- Q^{3} Q^{21} V^{15} 
- Q^{3} Q^{21} V^{33} 
 \nonu\\
&- & Q^{7} Q^{25} V^{15} 
- Q^{7} Q^{25} V^{33}\,)
 \nonu\\
& + &
c_{14}\,(\,
Q^{4} Q^{22} V^{15} 
- Q^{4} Q^{22} V^{33}
- Q^{5} Q^{23} V^{15}
+ Q^{5} Q^{23} V^{33} 
- Q^{9} Q^{27} V^{15}
+ Q^{9} Q^{27} V^{33} 
 \nonu\\
& + &
 Q^{10} Q^{28} V^{15} 
- Q^{10} Q^{28} V^{33 }\,) 
\nonu\\
& + &
c_{15}\,(\,
Q^{6} Q^{24} V^{30}
+ Q^{7} Q^{25} V^{30} 
+ Q^{8} Q^{26} V^{30} 
+ Q^{9} Q^{27} V^{30} + Q^{10} Q^{28} V^{30}\,)
\nonu\\
& + &
c_{16}\,(\,
Q^{1}\partial Q^{19} 
+ Q^{2}\partial Q^{20} 
+ Q^{3}\partial Q^{21} 
+ Q^{4}\partial Q^{22} 
+ Q^{5}\partial Q^{23} 
+ Q^{6}\partial Q^{24} 
+ Q^{7}\partial Q^{25} 
\nonu\\
& + &
 Q^{8}\partial Q^{26} 
+ Q^{9}\partial Q^{27} 
+ Q^{10}\partial Q^{28} 
- \partial Q^{1} Q^{19} 
- \partial Q^{2} Q^{20} 
- \partial Q^{3} Q^{21} 
- \partial Q^{4} Q^{22} 
- \partial Q^{5} Q^{23} 
\nonu\\
& - &
 \partial Q^{6} Q^{24} 
- \partial Q^{7} Q^{25} 
- \partial Q^{8} Q^{26} 
- \partial Q^{9} Q^{27} 
- \partial Q^{10}Q^{28} \,)
\nonu\\
& + &
 c_{17}\, V^{11} V^{29} 
+ c_{18}\, (V^{12} V^{12} - V^{30} V^{30}) 
+ c_{19}\, V^{12} V^{30} 
\nonu\\
& + &
 c_{20} (
V^{14} V^{32} 
+ V^{15} V^{33} 
+ V^{16} V^{34} 
+ V^{17} V^{35} 
+ V^{18} V^{36})
+c_{21}\,\partial V^{12} 
+c_{22}\, \partial V^{30}
\nonu\\
& + &
c_{23}\,\partial V^{15} 
+ c_{24}\, \partial V^{33},
\nonu
\eea
where the
$k$ dependent coefficients
are 
\bea
c_ {1} & = & \frac{126 k (5 + 2 k) (19 + 13 k)^2 (23 + 12 k + 
      k^2)} {25 (-1 + k) (2 + k) (7 + k)^6 (241 + 132 k + 11 k^2)}, 
\nonu\\
c_ {2} & = &
\frac{54 k (3 + k) (9 + k) (5 + 2 k) (19 + 13 k)^2} {25 (-1 +
       k) (2 + k) (7 + k)^6 (241 + 132 k + 11 k^2)},
\nonu\\
c_ {3} & = & -\frac{72 k (10 + k) (5 + 2 k) (19 + 13 k)^2}{
  25 (-1 + k) (7 + k)^6 (241 + 132 k + 11 k^2)}, 
\nonu\\
c_ {4} & = & -\frac{18 k (5 + 2 k) (19 + 13 k)^2}{
  25 (-1 + k) (2 + k) (7 + k)^6},
\nonu\\
c_ {5} & = &
-\frac{18 k (5 + 2 k) (19 + 13 k)^2 (-1 + 12 k + k^2)}{25 (-1 + 
       k) (2 + k) (7 + k)^6 (241 + 132 k + 11 k^2)}, 
\nonu\\
c_ {6} & = & -\frac{(144 k (10 + k) (5 + 2 k) (19 + 13 k)^2}{
  25 (-1 + k) (7 + k)^6 (241 + 132 k + 11 k^2)}, 
\nonu\\
c_ {7} & = &
-\frac{54 k (5 + 2 k) (19 + 13 k)^2 (107 + 60 k + 
       5 k^2)}{25 (-1 + k) (2 + k) (7 + k)^6 (241 + 132 k + 11 k^2)},
\nonu\\
c_ {8} & = & -\frac{216 k (10 + k) (5 + 2 k) (19 + 13 k)^2}{
  25 (-1 + k) (7 + k)^6 (241 + 132 k + 11 k^2)}, 
\nonu\\
c_ {9} & = &
-\frac {(\frac{252}{25} - \frac{252 i}{25}) \sqrt{2} (10 + k) (5 + 2 k) (19 + 13 k)^2}{(-1 + k) (7 + k)^5 (241 + 
      132 k + 11 k^2)}, 
\nonu\\
c_ {10} & = &
-\frac {(\frac{252}{25}+ \frac{252 i}{25}) \sqrt{2} (10 + k) (5 + 2 k) (19 + 13 k)^2}{(-1 + k) (7 + k)^5 (241 + 
      132 k + 11 k^2)}, 
\nonu\\
c_ {11} & = &
-\frac {378 (3 + k) (9 + k) (5 + 2 k) (19 + 13 k)^2}{25 (-1 + 
      k) (2 + k) (7 + k)^5 (241 + 132 k + 11 k^2)}, 
\nonu\\
c_ {12} & = & -\frac {1008 (10 + k) (5 + 2 k) (19 + 13 k)^2}{
  25 (-1 + k) (7 + k)^5 (241 + 132 k + 11 k^2)}, 
\nonu\\
c_ {13} & = & \frac {189 \sqrt{
     2} (3 + k) (9 + k) (5 + 2 k) (19 + 13 k)^2)}{25 (-1 + k) (2 + 
      k) (7 + k)^5 (241 + 132 k + 11 k^2)}, 
\nonu\\
c_ {14} & = & \frac {189 i \sqrt{
     2} (3 + k) (9 + k) (5 + 2 k) (19 + 13 k)^2}{25 (-1 + k) (2 + 
      k) (7 + k)^5 (241 + 132 k + 11 k^2}, 
\nonu\\
c_ {15} & = & \frac {(\frac{252}{25} + \frac{252 i}{25}) \sqrt{
     2} (10 + k) (5 + 2 k) (19 + 13 k)^2}{(-1 + k) (7 + k)^5 (241 + 
      132 k + 11 k^2)}, 
\nonu\\
c_ {16} & = & \frac {756 k (5 + 2 k) (19 + 13 k)^2}{
 25 (-1 + k) (2 + k) (7 + k)^5 (241 + 132 k + 11 k^2)}, 
\nonu\\
c_ {17} & = & -\frac {1008 (10 + k) (5 + 2 k) (19 + 13 k)^2}{
  5 (-1 + k) (2 + k) (7 + k)^4 (241 + 132 k + 11 k^2)}, 
\nonu\\
c_ {18} & = & \frac {252 i (10 + k) (5 + 2 k) (19 + 13 k)^2}{
 5 (-1 + k) (2 + k) (7 + k)^4 (241 + 132 k + 11 k^2)}, 
\nonu\\
c_ {19} & = & -\frac {504 (10 + k) (5 + 2 k) (19 + 13 k)^2}{
  5 (-1 + k) (2 + k) (7 + k)^4 (241 + 132 k + 11 k^2}, 
\nonu\\
c_ {20} & = &
\frac {1512 (9 + k) (5 + 2 k) (19 + 13 k)^2}{ 25 (-1 + k) (2 + k) (7 + k)^4 (241 + 132 k + 11 k^2)}, 
\nonu\\
c_ {21} & = &
-\frac {(\frac{252}{5 }- \frac{252 i}{5}) \sqrt{
   2} (10 + k) (5 + 2 k) (19 + 13 k)^2)}{(-1 + k) (2 + k) (7 + k)^4 (241 + 132 k + 11 k^2)}, 
\nonu\\
c_ {22} & = &
-\frac {(\frac{252}{5} + \frac{252 i}{5}) \sqrt{2} (10 + k) (5 + 2 k) (19 + 13 k)^2}{(-1 + k) (2 + k) (7 + 
     k)^4 (241 + 132 k + 11 k^2}, 
\nonu\\
c_ {23} & = &
-\frac {(\frac{378}{25} + \frac{1134i}{25}) \sqrt{2} (9 + k) (5 + 2 k) (19 + 13 k)^2}{(-1 + k) (2 + k) (7 + 
     k)^4 (241 + 132 k + 11 k^2)}, 
\nonu\\
c_ {24} & = &
\frac {(\frac{378}{25} - \frac{1134i}{25}) \sqrt{2} (9 + k) (5 + 2 k) (19 + 13 k)^2}{(-1 + k) (2 + k) (7 + 
    k)^4 (241 + 132 k + 11 k^2)}.
\nonu
\eea
The lowest power of $\frac{1}{k}$ of $X_0^{(2)}$
under the large $k$ limit is given by $4$ while the corresponding
value in (\ref{T2ansatz}) is given by
$1$, $2$ or $0$. Therefore, the higher spin-$2$ current
$X_0^{(2)}$ will vanish under the large $k$ limit for fixed $N$.


\end{document}